\documentclass[12pt]{article}
\usepackage[english]{babel}
\usepackage[utf8]{inputenc}
\usepackage{bm}
\usepackage{nicefrac,xfrac}
\usepackage{amsmath}
\usepackage[linesnumbered,ruled,vlined]{algorithm2e}
\usepackage[colorlinks=true,linkcolor=blue, allcolors=blue,hyperfootnotes=false]{hyperref}
\usepackage{xcolor}
\usepackage{braket}
\usepackage{amsthm}
\usepackage{amssymb}
\usepackage{siunitx}
\usepackage[version=4]{mhchem}
\usepackage{url}
\usepackage{float}
\usepackage{placeins} 
\DeclareSIUnit \Ha {Ha}

\usepackage{booktabs}
\usepackage{multirow}
\usepackage{soul}  %
\usepackage{csquotes}
\usepackage{cite}
\usepackage{orcidlink}

\usepackage{subcaption}

\usepackage{cleveref}

\usepackage{geometry}
\SetKwInput{KwHardware}{Hardware Resources}

\DeclareSIUnit\angstrom{\text {Å}}

\DeclareMathOperator*{\argmin}{arg\,min}
 
\usepackage{setspace}

\usepackage[at]{easylist}
\usepackage{versions}
\usepackage{authblk}
\usepackage{etoc}

\makeatletter
\def\@fnsymbol#1{\ensuremath{\ifcase#1\or \dagger\or \ddagger\or
   \mathsection\or \mathparagraph\or \|\or **\or \dagger\dagger
   \or \ddagger\ddagger \else\@ctrerr\fi}}
    \makeatother

\newif\ifPrintOutline
\PrintOutlinefalse

\excludeversion{commentdone}
\ifPrintOutline
\includeversion{comment}
\else
\excludeversion{comment}
\fi

\title{Fast gradient-free optimization of excitations in variational quantum eigensolvers}

\author{
Jonas Jäger\thanks{\textcolor{blue}{jojaeger@cs.ubc.ca}}$~^{1,2,3,4}$\orcidlink{0000-0001-7631-8689},
Thierry N.~Kaldenbach\thanks{\textcolor{blue}{thierry.kaldenbach@dlr.de}}$~^{1}$\orcidlink{0009-0008-5607-4427},
Max Haas*$^{1}$\orcidlink{0009-0004-9414-5442} and
Erik Schultheis*$^{1}$\orcidlink{0009-0007-4728-7124}}

\affil{
$^{1}$German Aerospace Center (DLR), Institute of Materials Research, Cologne, Germany\\
$^{2}$University of British Columbia (UBC), Department of Computer Science, Vancouver, BC, Canada\\
$^{3}$University of British Columbia (UBC), Institute of Applied Mathematics, Vancouver, BC, Canada\\
$^{4}$Stewart Blusson Quantum Matter Institute, Vancouver, BC, Canada}

\date{\today}
\footnotetext{Both authors contributed equally to this work.
\\Corresponding authors:}

\begin{document}

    \maketitle
    
    \begin{abstract}
    Finding molecular ground states and energies with variational quantum eigensolvers is central to chemistry applications on quantum computers. 
         Physically motivated ansätze based on excitation operators respect physical symmetries, but existing quantum-aware optimizers, such as Rotosolve, have been limited to simpler operator types.
         To fill this gap, we introduce ExcitationSolve, a fast quantum-aware optimizer that is globally-informed, gradient-free, and hyperparameter-free.
         ExcitationSolve extends these optimizers to parameterized unitaries with generators $G$ of the form $G^3=G$ exhibited by excitation operators in approaches such as unitary coupled cluster.
         ExcitationSolve determines the global optimum along each variational parameter using the same quantum resources that gradient-based optimizers require for one update step. We provide optimization strategies for both fixed and adaptive variational ansätze, along with generalizations for simultaneously selecting and optimizing multiple excitations.
         On molecular ground state energy benchmarks, ExcitationSolve outperforms state-of-the-art optimizers by converging faster, achieving chemical accuracy for equilibrium geometries in a single parameter sweep, yielding shallower adaptive ansätze and remaining robust to real hardware noise.
         By uniting physical insight with efficient optimization, ExcitationSolve paves the way for scalable quantum chemistry calculations.
    \end{abstract}

    \clearpage
    
    \onehalfspacing
    \section{Introduction}
        
\begin{figure}[!ht]
            \centering
                \begin{subfigure}[t]{0.95\textwidth}
                    \centering
                   \includegraphics[width=\textwidth]{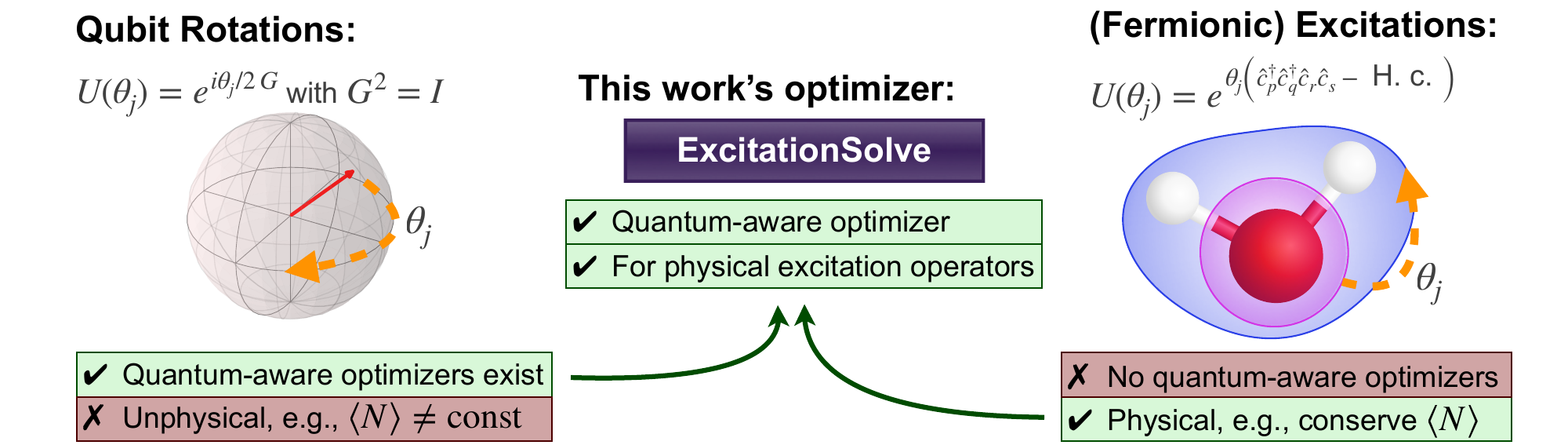}
                \end{subfigure}
                \\\vspace{.45cm}\hrule\vspace{.3cm}
                \begin{subfigure}[t]{0.9\textwidth}
                    \centering
                     \includegraphics[width=\textwidth]{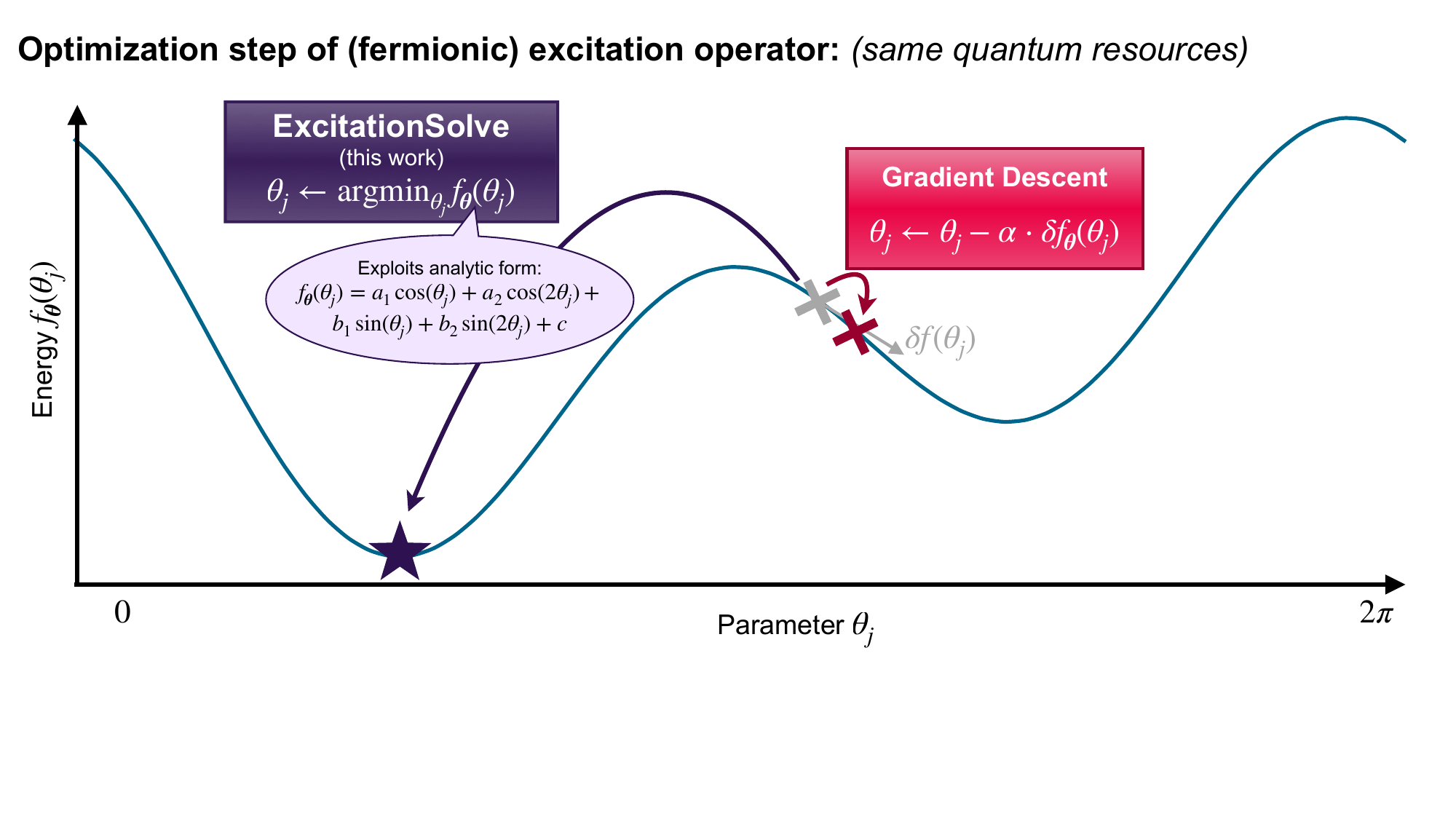}
                \end{subfigure}
            \caption{\textbf{Schematic overview of our work.} \textit{Top:}~while hardware-efficient ansätze, typically composed of parameterized rotations, allow for fast quantum-aware optimization \cite{ostaszewski_StructureOptimizationParameterized_2021,nakanishi_SequentialMinimalOptimization_2020,parrish_JacobiDiagonalizationAnderson_2019,vidal_CalculusParameterizedQuantum_2018}, they do not preserve physical properties \cite{gard2020efficient}, e.g., vary the average particle number $\langle N \rangle$, the opposite is true for physically-motivated ansätze such as those assembled from fermionic excitation operators \cite{peruzzo_VariationalEigenvalueSolver_2014}. Our new optimizer, \emph{ExcitationSolve}, fills this gap and combines fast optimization with physical guarantees. \textit{Bottom:}~ExcitationSolve (purple) relies on the same quantum resources, i.e., same number of energy measurements, to jump to the global energy minimum along a single parameter $\theta_j$, as a gradient-based optimizer (red) evaluating and following the (partial) derivative in $\theta_j$. The latter does not consider global information of the energy landscape, thus being limited to a local parameter region. Note that since gradient descent is based on the full gradient evaluated over $N$ parameters, ExcitationSolve in fact performs $N$ update steps while gradient descent updates locally once.}
            \label{fig:graphical_abstract}
        \end{figure}

        The choice of ansatz for the parameterized or variational quantum circuit plays a crucial role in the variational quantum eigensolver (VQE) \cite{peruzzo_VariationalEigenvalueSolver_2014}, which aims to prepare the ground state of a Hamiltonian and find the corresponding ground state energy. The Hamiltonian can describe, e.g., an electronic structure problem in a molecule or material \cite{peruzzo_VariationalEigenvalueSolver_2014,tilly2022variational,ma2020quantum, bauer2020quantum}. Physically-motivated ansätze, such as a composition of excitation operators like single and double fermionic excitations in the \emph{Unitary Coupled Cluster} (UCCSD) ansatz \cite{peruzzo_VariationalEigenvalueSolver_2014}, are particularly relevant because of their guarantees of producing physically plausible states. By design, relevant physical properties of an initial reference state, typically the Hartree-Fock (HF) state, are conserved, such as the number of electrons or spin symmetries. Furthermore, number-conserving, yet hardware-efficient approaches, such as \emph{qubit-excitation based} (QEB) ansätze \cite{yordanov2021qubitexcitationbased} exist, most prominently appearing in the \emph{Qubit Coupled Cluster Singles Doubles} (QCCSD) ansatz \cite{xia2020qubit}. In contrast, problem-agnostic ansätze like generic hardware-efficient ansätze might yield physically implausible states and energies by, e.g., not conserving the number of particles \cite{gard2020efficient}. These ansätze are composed of parameterized qubit rotations. The implications of ansatz choice are visualized in Fig.~\ref{fig:graphical_abstract}.
        
        After specifying the variational ansatz $U(\bm{\theta})$, its $N$ parameters $\bm{\theta} \in (-\pi,\pi]^N$ have to be optimized to prepare the desired ground state.
        This optimization happens iteratively in a hybrid loop involving a quantum computer to evaluate the energy and a classical computer to optimize the parameters. On the quantum computer we evaluate the expectation value $\braket{\psi(\bm{\theta}) | H | \psi(\bm{\theta})}$ of the Hamiltonian $H$ with respect to prepared $n$-qubit state $\ket{\psi(\bm{\theta})} = U(\bm{\theta}) \ket{\psi_0}$, as a function of the current parameters. The energy landscape $f(\bm{\theta})$ we want to minimize can be written as 
        \begin{equation}\label{eq:energy_func_expectation} 
            f(\bm{\theta}) = \braket{H} = \braket{\psi_0 | U^\dagger(\bm{\theta}) H U(\bm{\theta}) | \psi_0}.
        \end{equation} 
        However, the VQE optimization problem is generally challenging because the energy landscape is a $N$-dimensional trigonometric function \cite{vidal_CalculusParameterizedQuantum_2018}, leading to a large number of local minima, many of which are sub optimal \cite{anschuetz_QuantumVariationalAlgorithms_2022,bittel_TrainingVariationalQuantum_2021,wang_TrainabilityEnhancementParameterized_2023,you_ExponentiallyManyLocal_2021}.
        Therefore, gradient-based optimizers (e.g., gradient descent, Adam \cite{kingma_AdamMethodStochastic_2017} or BFGS \cite{broyden1970convergence, fletcher1970new, goldfarb1970family, shanno1970conditioning}) as well as a gradient-free black-box optimizers (e.g., COBYLA \cite{powell1994direct}, SPSA \cite{spall1987stochastic, spall1992multivariate}) struggle to navigate the complex energy landscape as for larger molecules or materials.
        Note that, instead of a fixed ansatz $U$, an adaptive ansatz can also be employed, where operators are iteratively added to the ansatz during the optimization. For VQE, this concept was first introduced as ADAPT-VQE \cite{grimsley_AdaptiveVariationalAlgorithm_2019}.
        
        \emph{Quantum-aware} optimizers, defined as those that leverage problem-specific knowledge and properties of the quantum system to be optimized, pose a promising alternative. In contrast to optimizers that treat the quantum system as a \emph{black box} (e.g., the ones listed above), they utilize this information to navigate the energy landscape more efficiently.
        A prominent example is the \emph{Rotosolve} optimization method \cite{ostaszewski_StructureOptimizationParameterized_2021}, which was simultaneously proposed under the term \emph{Sequential Minimal Optimization} (SMO) \cite{nakanishi_SequentialMinimalOptimization_2020}%
        , as well as analogously %
        mentioned in Refs. \cite{parrish_JacobiDiagonalizationAnderson_2019,vidal_CalculusParameterizedQuantum_2018}. Rotosolve globally optimizes parameterized operators individually based on the closed form of the energy landscape slice, thus leveraging the operator-specific properties to significantly reduce the required quantum resources \cite{ostaszewski_StructureOptimizationParameterized_2021}. This provides an efficient alternative to gradient-based optimization.
        However, the type of parameterized operators or gates incorporated in the ansatz must be compatible with the quantum-aware optimizer. The applicability of Rotosolve is limited to unitaries with self-inverse generators, e.g., (Pauli) rotation gates, although generalizations were suggested \cite{wierichs_GeneralParametershiftRules_2022}. While the more complicated unitaries relevant for quantum chemistry applications can be decomposed into fixed entangling gates and parameterized rotations \cite{barkoutsos_QuantumAlgorithmsElectronic_2018}, Rotosolve's performance degrades as it then overestimates the required number of energy evaluations. While these gradient-free optimizers, like their gradient-based counterparts, are generally only guaranteed to converge to a local optimum, empirical evidence demonstrates their effectiveness across various applications \cite{nakanishi_SequentialMinimalOptimization_2020,ostaszewski_StructureOptimizationParameterized_2021,parrish_JacobiDiagonalizationAnderson_2019,vidal_CalculusParameterizedQuantum_2018}. Variations of these optimizers, such as Free-Axis or Free-Quaternion Selection \cite{watanabe_OptimizingParameterizedQuantum_2021,watanabe_OptimizingParameterizedQuantum_2023,wada_SimulatingTimeEvolution_2022,wada_SequentialOptimalSelections_2024,kurogi_OptimizingParameterizedControlled_2024} and the Unitary Block Optimization Scheme \cite{slattery_UnitaryBlockOptimization_2022} further support their utility.
        
        In this work, we combine the advantages of incorporating excitation operators in physically-motivated VQE ansätze with the effectiveness of quantum-aware optimizers. We introduce \emph{ExcitationSolve}, a fast globally-informed gradient-free optimizer for ansätze composed of excitation operators. ExcitationSolve is quantum-aware since we know the analytical form of the energy landscape in one parameter (or a subset of parameters) for excitation operators, which is a (multi-dimensional) \emph{second-order Fourier series}. This applies to fermionic excitations \cite{peruzzo_VariationalEigenvalueSolver_2014}, qubit excitations \cite{yordanov2021qubitexcitationbased,xia2020qubit}, and Givens rotations \cite{arrazola_UniversalQuantumCircuits_2022} -- not limited to single and double excitations.
        ExcitationSolve can be applied to fixed and adaptive VQE ansätze as in the UCCSD ansatz \cite{peruzzo_VariationalEigenvalueSolver_2014} and ADAPT-VQE \cite{grimsley_AdaptiveVariationalAlgorithm_2019}, respectively.
        Within the literature on such optimizers, ExcitationSolve can be characterized as an extension of Rotosolve \cite{nakanishi_SequentialMinimalOptimization_2020,ostaszewski_StructureOptimizationParameterized_2021,parrish_JacobiDiagonalizationAnderson_2019,vidal_CalculusParameterizedQuantum_2018} and Greedy Gradient-free Adaptive VQE  (GGA-VQE) \cite{feniou_GreedyGradientfreeAdaptive_2023} for VQE and ADAPT-VQE, respectively.
        A schematic summary of our work is provided in Fig.~\ref{fig:graphical_abstract}.

        In Sec.~\ref{sec:ExcitationSolve} we introduce the framework of the ExcitationSolve algorithm, followed by a description of operator classes covered by the algorithm in Sec.~\ref{sec:excitation_ops}. In Sec.~\ref{sec:adapt_cirterion} we describe how ExcitationSolve can be utilized to globally inform the selection criterion in ADAPT-VQE. 
        Generalizations of ExcitationSolve to a multi-dimensional case (Sec.~\ref{sec:multi_param_gen}) and to ansätze with multiple occurrences of the same variational parameters (Sec.~\ref{sec:multi_occ}) are introduced.
        Section \ref{sec:results} provides simulated experiments comparing ExcitationSolve to optimizers commonly used in VQE literature for fixed- and adaptive ansätze exemplified by ground state preparation of molecules. Here, Sec.~\ref{subsec:fixedAnsatz} focuses on the setting of a fixed ansatz scenario, while Sec.~\ref{subsec:ADAPT} targets adaptive ansätze. In Sec.~\ref{subsec:StrongCorrelations} we explore the utility of ExcitationSolve in the realm of strongly correlated systems and further employ 2D optimization to accelerate convergence. Last, Sec.~\ref{subsec:NISQ} demonstrates how the different optimizers perform under the influence of real hardware noise. Section \ref{sec:discussion} concludes our work with a discussion. Details on the used methodology can be found in Sec.~\ref{sec:methods}, Appendix \ref{app:experimental_setup} and Appendix \ref{app:algorithms}, while Appendix \ref{app:supp_experiments} provide further experimental examinations and Appendix \ref{app:proofs} provides mathematical derivations and proofs.

    \section{ExcitationSolve algorithm} \label{sec:ExcitationSolve}
    
        In this section, we introduce the quantum-aware optimization algorithm \emph{ExcitationSolve}, which readily extends Rotosolve-type optimizers \cite{ostaszewski_StructureOptimizationParameterized_2021, nakanishi_SequentialMinimalOptimization_2020} to excitation operators, which obey the following more general form.  %
        Throughout this work, we assume variational ansätze $U(\bm{\theta})$ consisting of a product of unitary operators $U(\theta_j)$ of the generic form 
        \begin{equation}\label{eq:excitation_operator_herm_form}
            U(\theta_j) = \exp( -i \theta_j G_j),
        \end{equation}
        depending on a single parameter $\theta_j$ each (the $j$-th component of $\bm \theta$). Most importantly, with the Hermitian generators $G_j$ fulfilling $G_j^3 = G_j$. Note that any generator with $G_j^2=I$ (the prerequisite of Rotosolve) fits into this description.
        However, this work is concerned with the class of excitation operators because their generators fulfill $G_j^3=G_j$ and, importantly, $G_j^2\neq I$. 
        
        In the following, we first present the analytic form of the energy landscape when varying a single parameter in an operator of the aforementioned structure, and, second, how this is exploited to derive an optimization algorithm.
        The analytic form of the energy with respect to a single parameter $\theta_j$ associated with some generator $G_j$ %
        is a finite Fourier series (also known as a trigonometric polynomial) of second-order with period $2\pi$ and has the form
        \begin{equation} \label{eq:analytic_form_fourier}
            f_{\bm{\theta}}(\theta_j) = a_1 \cos(\theta_j) + a_2 \cos(2\theta_j) + b_1 \sin(\theta_j) + b_2 \sin(2\theta_j) + c. %
        \end{equation}
        Here, the notation $f_{\bm \theta}(\theta_j)$ refers to the energy landscape $f(\bm \theta)$ from Eq.~\eqref{eq:energy_func_expectation} with all parameters being fixed except $\theta_{j}$. The five coefficients $a_1, a_2, b_1, b_2, c$ are independent of the parameter $\theta_j$ but may depend on the remaining parameters $\theta_{i \neq j}$, which is detailed in the constructive proof in Appendix~\ref{app:analy_energy_func_single_param}. In order to determine these five coefficients, we need energy values in at least five distinct configurations of the parameter $\theta_j$. For five evaluations, the coefficients are the solution to the linear equation system, whereas for more than five evaluations, the overdetermined equation system can be solved using either the least square method or truncated (fast) Fourier transform. Appendix~\ref{sec:noise_robustness} discusses how this relates to noise robustness.

\begin{figure}[ht]
            \centering
            \includegraphics[width=\textwidth]{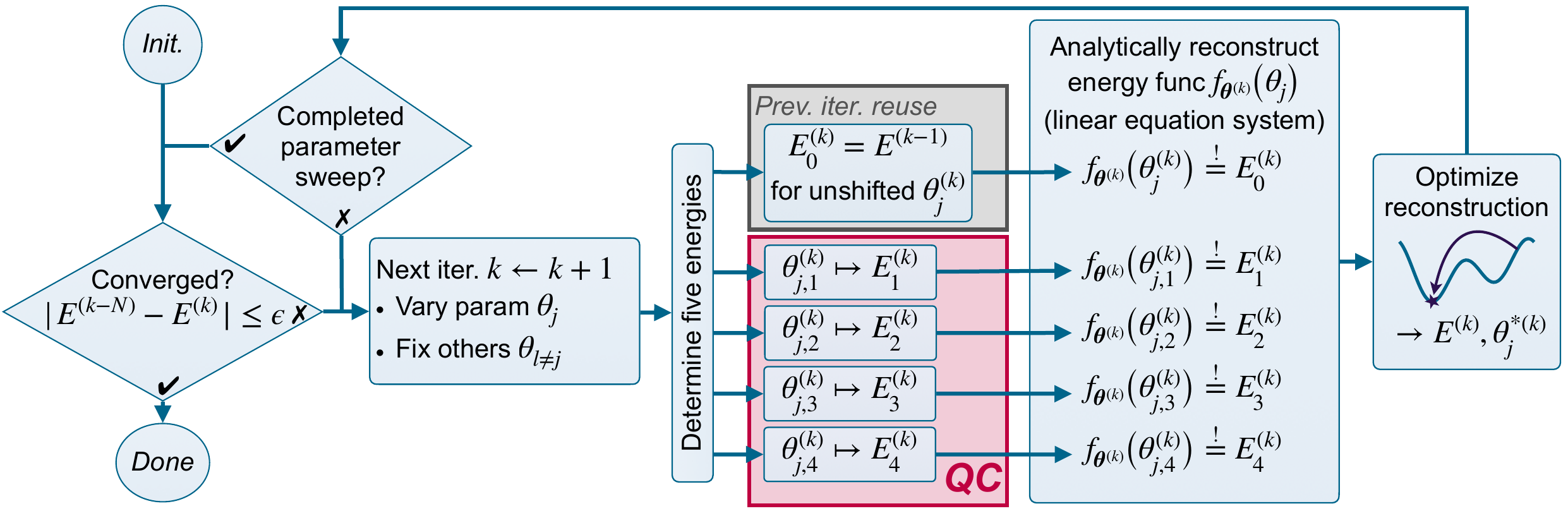}
            \caption{\textbf{Flow chart for ExcitationSolve for fixed ansätze.}
            In this iterative algorithm, the $k$-th iteration updates a single parameter $\theta_j$ through repeated sweeps over all $N$ parameters until convergence. To reflect the flexibility in the sweep order, we use separate indices $j$ and $k$.
            Per iteration, the parameter is shifted to four different positions $\theta^{(k)}_{j,1}, \theta^{(k)}_{j,2}, \theta^{(k)}_{j,3}, \theta^{(k)}_{j,4}$, and the quantum computer (QC) is used to obtain the corresponding energy values. This is the only part requiring the quantum hardware (purple). All remaining steps are efficiently computed classically. The energy associated with the unshifted current parameter value $\theta_j^{(k)}$ is re-used from the previous iteration $k-1$.}
            \label{fig:algorithm_diagram_fixed}
        \end{figure}

        The proposed optimization algorithm ExcitationSolve (cf.~Fig.~\ref{fig:algorithm_diagram_fixed}) iteratively sweeps through the $N$ parameters $\bm{\theta}$, reconstructs the energy landscape per parameter $\theta_j$ analytically,
        globally minimizes the reconstructed function classically, and assigns the parameter $\theta_j$ to the value where the global minimum is attained. Hence, each parameter sweep consists of $N$ updates, and the order in which the $N$ parameters are optimized can be chosen freely. This process is repeated until convergence, which is defined by a threshold criterion on the absolute or relative energy reduction of the last parameter sweep. In this algorithm, the quantum computer is used exclusively to obtain energy evaluations while the reconstruction and subsequent minimization of the energy landscape is performed on a classical computer. 
        To determine the minimum energy and corresponding parameter classically, we utilize a companion-matrix method \cite{boyd_ComputingZerosMaxima_2006}, which is a direct numerical method detailed in Sec.~\ref{sec:classical_min_analytic_energy_function}.
        Importantly, in each optimization step, the previously determined minimum energy can be reused, requiring only an additional \emph{four} parameter shifts to reconstruct the energy landscape along the next parameter. Appendix \ref{app:algorithms:fixed} describes the exact algorithmic details.
        It should be emphasized that for the specific analytic form in Eq.~\eqref{eq:analytic_form_fourier} each parameter $\theta_j$ must occur only once in the ansatz in terms of parameterized excitations, which is a commonly satisfied assumption. This number of occurrences should not be confused with the number of single-qubit rotations arising when further decomposing an excitation into basic gates, which in general will be more than one. We yet further generalize ExcitationSolve to multiple occurrences of the same parameter in Sec.~\ref{sec:multi_occ}, e.g., making it compatible with ansätze constituted of higher-order product formulas or multiple Trotter steps.

        In essence, ExcitationSolve performs a gradient-free \emph{coordinate descent} with (efficient) \emph{exact line search}, i.e., independently optimizing each parameter $\theta_j$ iteratively until convergence. For the exact line search, it leverages an efficient analytic reconstruction of the energy landscape in a \emph{single} parameter to determine its \emph{global} optimum directly while the other parameters $\theta_{i\neq j}$ remain fixed. Most importantly, the effective resource demands on the quantum hardware per parameter are equivalent to gradient-based optimizers. %

        \subsection{Supported types of excitation operators} \label{sec:excitation_ops} %
        
        Excitation operators are one class of operators whose generators satisfy $G^3=G$. Such operators appear for example as generators in UCC theory \cite{tilly2022variational, bartlett1989alternative}, which is a post Hartree-Fock method that unitarily evolves the Hartree-Fock ground state based on fermionic excitations. 
        For a fermionic excitation of $m$ electrons, the $m$-excitation generator reads
        
            \begin{equation}
                \tau^{(m)}_{\bm{o}, \bm{v}} =  \prod_{l=1}^m a^\dagger_{v_l} a_{o_l} - \text{H.c.},
                \label{eq:excitation_generator}
            \end{equation}
            
        where $a^\dagger$/$a$ are the standard fermionic creation/annihilation operators and the
        $m$-component vectors $\bm{o}$/$\bm{v}$ entail the involved occupied/virtual orbitals, respectively. The product $\prod_l$ is intuitively taken right-to-left, however, changing this order conveniently leaves the expression invariant.
        The corresponding $m$-electron unitary excitation operator is defined as 
        \begin{equation}
            U^{(m)}_{\bm{o},\bm{v}}(\theta) = \exp\left(\theta~\tau^{(m)}_{\bm{o},\bm{v}}\right).
            \label{eq:excitation_operator}
        \end{equation}
        To incorporate all eligible $m$-electron excitations from occupied to virtual orbitals, the $m$-th cluster operator $T^{(m)}=\sum_{\bm{o}, \bm{v}} \theta_{\bm{o},\bm{v}}~\tau^{(m)}_{\bm{o}, \bm{v}}$ is introduced.
        The truncated cluster operator, including all excitations of $M$ or less electrons, is defined as $T = \sum_{m=1}^M T^{(m)}$ and serves as the generator of the variational unitary. Typically, the unitary $ \exp(T)$ is then approximated through a first-order Trotter-Suzuki \cite{suzuki1976generalized} decomposition
            \begin{equation}
                U(\bm{\theta}) = \prod_{m=1}^M \prod_{\bm{o}, \bm{v}} U^{(m)}_{\bm{o},\bm{v}}(\theta_{\bm{o},\bm{v}}),
            \end{equation}
        where $\bm{\theta}$ contains all the variational parameters $\theta_{\bm{o},\bm{v}}$.
        In Appendix \ref{app:excitation_operators} we show that the anti-Hermitian fermionic excitation operators (Eq.~\eqref{eq:excitation_generator}) obey the equation $\tau^3=-\tau$ for arbitrary excitation-orders. Consequently, we can define the Hermitian generator $G=i\tau$ with $G^3=G$, such that the excitation operators from Eq.~\eqref{eq:excitation_generator} comply with the form in Eq.~\eqref{eq:excitation_operator_herm_form}. Thus, the energy landscape $f(\bm{\theta})=\braket{U^\dagger(\bm{\theta}) H U(\bm{\theta})}$ in a single parameter takes the form from Eq.~\eqref{eq:analytic_form_fourier}.
        
        In practice, the truncated cluster operator is often restricted to only include single-electron- and double-electron excitations ($M=2$), resulting in the \emph{Unitary Coupled Cluster Singles Doubles} unitary (UCCSD). We note, that ExcitationSolve is applicable for arbitrary truncation orders $M$ and, importantly, that the required energy evaluations for the optimization stays constant regardless of the order of the excitation $m$ the energy always obeys a second-order Fourier series. 
        In contrast, in order to optimize excitation operators using Rotosolve/SMO, one applies a fermionic mapping, e.g., Jordan-Wigner (JW)\cite{jordan1993paulische} or Bravyi-Kitaev (BK)\cite{bravyi_FermionicQuantumComputation_2002,tranter_BravyiKitaevTransformation_2015, tranter2018comparison}, to decompose the operation into compatible Pauli rotations. 
        The order of the Fourier series predicted by SMO scales \emph{exponentially} in the order of the excitation operator \cite{kottmann_FeasibleApproachAutomatically_2021, seeley2012bravyi}. %
        We further emphasize that this approach works for any fermion-to-qubit mapping. 
            Analogously to fermionic excitations, one can use other types of excitations such as \emph{qubit excitations} used in QEB ansätze such as QCCSD \cite{yordanov2021qubitexcitationbased,xia2020qubit} (recently explored in Ref.~\cite{feniou_GreedyGradientfreeAdaptive_2023}) also sometimes referred to as \emph{Givens rotations}, or \emph{controlled} excitations \cite{arrazola_UniversalQuantumCircuits_2022}. The latter are universal for particle-number preserving unitaries. We further note that ExcitationSolve can be readily applied to the recently introduced couple exchange operators (CEOs) consisting of linear combinations of excitations \cite{ramôa2024reducingresourcesrequiredadaptvqe}. In particular, the linear combination of two distinct excitations acting on the same (spin-) orbitals with one shared variational parameter (OVP-CEOs) satisfies the requirements. As a final note, our algorithm is readily applicable to generalized excitations \cite{nooijen2000can, lee2018generalized}, which do not distinguish between occupied and virtual orbitals.

        \subsection{ExcitationSolve for ADAPT-VQE: Globally-informed selection} \label{sec:adapt_cirterion}

            \begin{figure}[tb]
                \centering
                \includegraphics[width=\textwidth]{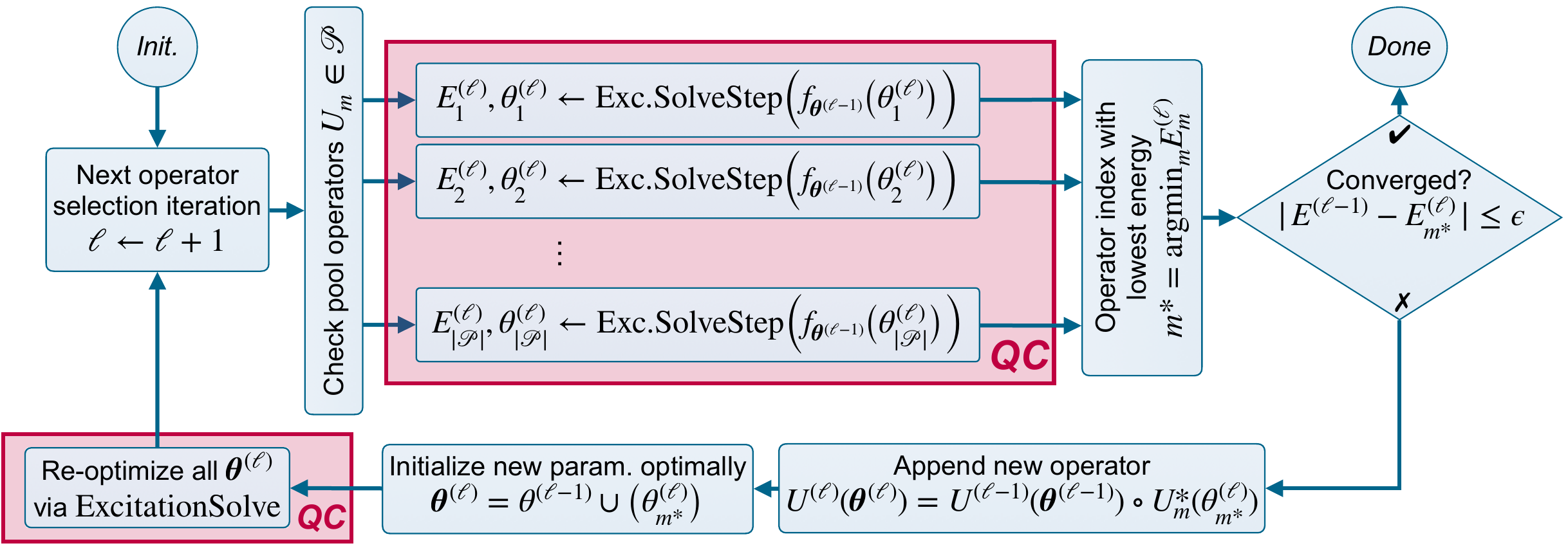}
                \caption{\textbf{Flow chart for ExcitationSolve for ADAPT-VQE (adpative ansätze).} ExcitationSolve is integrated into ADAPT-VQE in two parts. First, for the selection criterion from the operator pool $\mathcal{P}$ by determining the immediate energy improvements via a single ExcitationSolve iteration when appending each of the operator candidates $U_m$ separately. Second, to re-optimize all parameter $\bm{\theta}^{(\ell)}$ at the end of each ADAPT iteration.
                The usage of the quantum computer (QC, red) solely happens in $\mathrm{ExcitationSolve(Step)}$ (Fig.~\ref{fig:algorithm_diagram_fixed}) when invoked as sub-routines.
                Note that ADAPT iteration $\ell$ denotes how many operators have been appended to the ansatz.}
                \label{fig:algorithm_diagram_adapt}
            \end{figure}

            When optimizing adaptive ansätze, e.g., ADAPT-VQE \cite{grimsley_AdaptiveVariationalAlgorithm_2019}, a scoring criterion is needed to select an operator from the pool to append to the ansatz. The goal of this criterion is to assess the effectiveness of this operator selection in producing the ground state and energy. 
            Naturally, we apply ExcitationSolve to ADAPT-VQE (cf.~Fig.~\ref{fig:algorithm_diagram_adapt}) to obtain a globally-informed ADAPT-VQE operator selection criterion by leveraging analytic energy reconstructions for each operator candidate separately when added to the current ansatz. 
            We select the operator that achieves the strongest \emph{immediate} decrease in energy to be appended to the current ansatz (and parameters) and initialize it in its optimal value. Given the potential for a stronger energy decrease by adjusting the preceding parameters, we use ExcitationSolve to re-optimize all parameters in the typical fixed ansatz VQE manner before extending the ansatz further. We only proceed to the next ADAPT(-VQE) iteration if the threshold criterion for convergence has not yet been met. The details are described in Appendix \ref{app:algorithms:adaptive}.
            Note that a similar approach, known as \emph{Greedy Gradient-free Adaptive VQE (GGA-VQE)}, %
            was recently proposed \cite{feniou_GreedyGradientfreeAdaptive_2023}, but it was limited to parameterized qubit excitations and rotations and neglected the re-optimization of the intermediate parameters. In contrast we extend it to fermionic excitation operators and include effective re-optimization. We further note that the energy ranking of the operator pool can be utilized to append the top two (or more) operators at once, as recently proposed in Ref.~\cite{feniou_GreedyGradientfreeAdaptive_2023}. This is only a heuristic for the most impactful pair (or subset) of operators, since the largest individual impact does not necessarily imply the largest simultaneous impact. However, an efficient initialization in their simultaneous optimum can still be achieved via the multi-parameter extension of ExcitationSolve (cf.~Sec.~\ref{sec:multi_param_gen}).

            \begin{figure}[tb]
                \centering
                \includegraphics{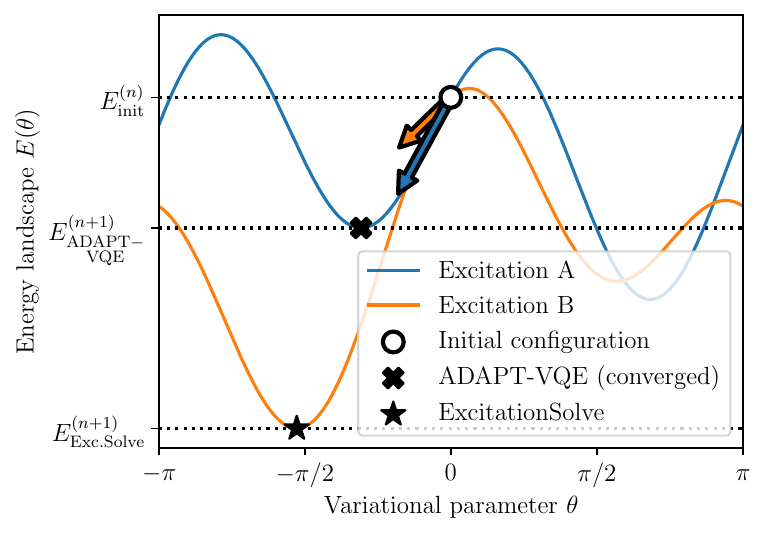}
                \caption{\textbf{ADAPT-VQE vs.~ExcitationSolve.} We consider the selection among two excitation operator candidates A and B in the adaptive setting. In the original ADAPT-VQE approach \cite{grimsley_AdaptiveVariationalAlgorithm_2019}, excitation A is selected based on the gradient criterion, i.e.~the steepest gradient at $\theta=0$. This is then converged with a gradient descent towards a (potentially only local) minimum, typically requiring multiple gradient evaluations. ExcitationSolve chooses excitation B (despite the smaller gradient at $\theta=0$) based on the energy criterion, i.e., the attainable global energy minimum, and already initializes $\theta$ in its optimal configuration.}
                \label{fig:ADAPT-VQE_vs_ExcitationSolve}
            \end{figure}

                Figure \ref{fig:ADAPT-VQE_vs_ExcitationSolve} demonstrates the advantage of our globally-informed selection criterion by comparing it with the original local ADAPT-VQE criterion \cite{grimsley_AdaptiveVariationalAlgorithm_2019}, which selects operators based on the magnitude of their partial derivative at zero $\left|f'_{\bm{\theta}}(0) \right|$ (details in Sec.~\ref{sec:method_adapt_vqe}).
                In contrast, ExcitationSolve assesses the potential impact of each operator on a global scale, identifying the operator that provides the greatest immediate improvement.
                Note that the example provided in Fig.~\ref{fig:ADAPT-VQE_vs_ExcitationSolve} is fabricated through a random three-qubit Hamiltonian to showcase the motivation and does not correspond to an actual scenario observed in numerical simulations.
                From a theoretical point of view, a valuable insight can be made by considering that the original ADAPT-VQE criterion approximates the energy landscape in the selected operator's parameter $f_{\bm{\theta}}(\theta_{N+1})$ by a \emph{first-order} Taylor expansion around $\theta_{N+1} = 0$, while ExcitationSolve utilizes the exact energy landscape or, analogously, the \emph{full} Taylor series, i.e., 
                \begin{equation}
                    f_{\bm{\theta}}(\theta_{N+1}) = \overbrace{\underbrace{f_{\bm{\theta}}(0) 
                    + \frac{1}{1!}f'_{\bm{\theta}}(0)\theta_{N+1}}_{\text{Original ADAPT-VQE}}
                    + \frac{1}{2!}f''_{\bm{\theta}}(0)\theta_{N+1}^2
                    + \frac{1}{3!}f'''_{\bm{\theta}}(0)\theta_{N+1}^3
                    + \ldots}^{\text{ExcitationSolve ADAPT-VQE}}\ .
                \end{equation}
                Given that the first-order Taylor approximation is a linear function, the minimum energy is trivially attained at either boundary, $\theta_{N+1} = \pm\pi$, with energy $f_{\bm{\theta}}(0) \pm f'_{\bm{\theta}}(0)\pi$. Thus, the original ADAPT-VQE selects the operator from the pool that decreases the energy the most in the first-order Taylor approximation (in $\theta_{N+1} = 0$) of the energy landscape\footnote{The approximation error is expected to be high at the boundaries $\theta_{N+1} = \pm\pi$, which is why the initialization of $\theta_{N+1} = 0$ \cite{grimsley_AdaptiveVariationalAlgorithm_2019} remains a meaningful choice in this theoretical picture.}, whereas ExcitationSolve does so based on the full Taylor series, i.e., the exact energy landscape. 
                Especially for such trigonometric functions, a first-order Taylor approximation falls short in faithfully capturing the up to four optima and identifying the global optimum, highlighting the effectiveness of the global criterion in ExcitationSolve. 
                This underscores the importance of such global criteria for operators that induce higher-degree trigonometric functions, unlike rotations corresponding to simple sine curves where local minima trivially coincide with global minima.

        \subsection{Multi-parameter generalization} \label{sec:multi_param_gen}
\begin{figure}[!hbt]
            \centering
            \includegraphics{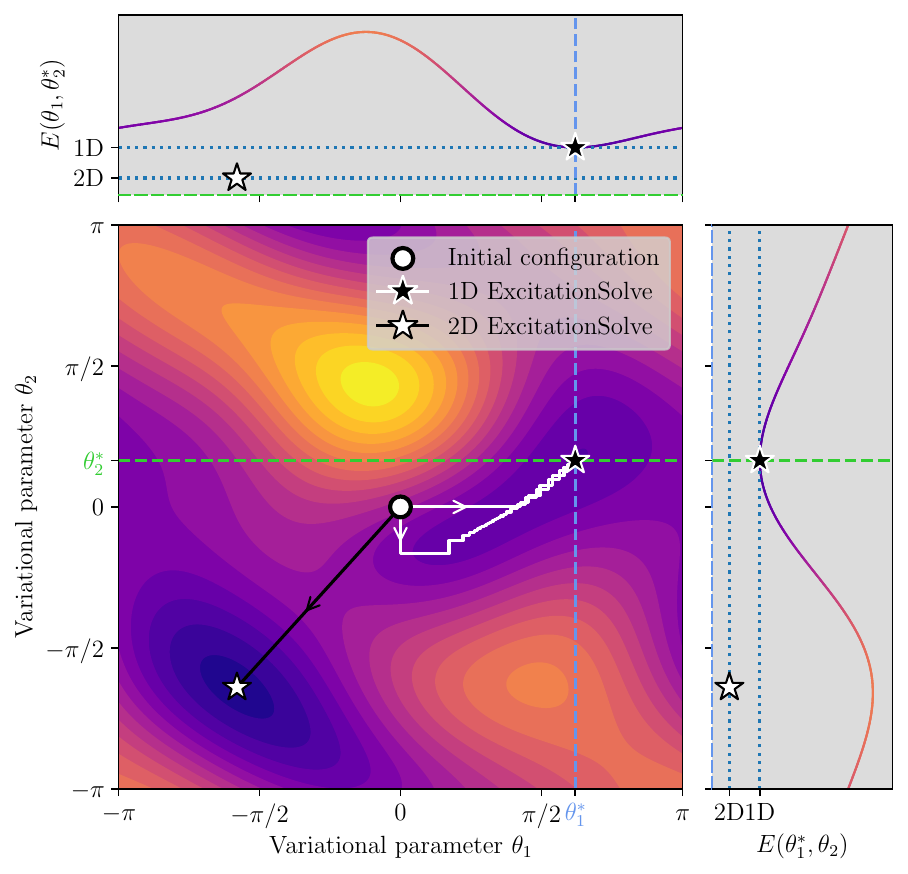}
            \caption{\textbf{1D ExcitationSolve with Coordinate Descent vs.~2D ExcitationSolve}. The simultaneous optimization of two parameters can be achieved either by effectively reducing it to a 1D optimization task using coordinate descent or employing a true 2D optimization based on the energy landscape from Eq.~\eqref{eq:analytic_form_fourier_d_dim}. Color indicates energy linearly from low (violet) to high (yellow). In this example, no matter which parameter is tuned first, the 1D coordinate descent approach (white arrows) converges only to a local minimum (black star marker). Also, this convergence takes up multiple iterations. Meanwhile, in the proper 2D case, the global reconstruction of the 2D second-order Fourier series permits an immediate jump (black arrow) to the global minimum (white star marker). 
            }
            \label{fig:1D+CD_vs_2D}
        \end{figure}

        In this section, we generalize the one-dimensional optimization to multiple dimensions, i.e.~independent parameters. As illustrated in Fig.~\ref{fig:1D+CD_vs_2D}, this enables ExcitationSolve to potentially avoid and escape local minima in the energy landscape because a improved local or global optimum may be unveiled in a higher-dimensional space.
        The multi-parameter generalization can be used for both the optimization of fixed and adaptive ansätze. In the context of rotation operators, this has already been explored \cite{nakanishi_SequentialMinimalOptimization_2020}: The energy varied in $D$ parameters is analytically described by a $D$-dimensional first-order Fourier series.
        Analogously, we show in Appendix \ref{app:analy_energy_func_multi_param} that a simultaneous variation of $D$ excitation operators can be described through a $D$-dimensional \emph{second-order} Fourier series. The energy landscape can thus be expressed as
         
        \begin{equation}
                f_{\bm \theta}(\bm{\theta}_{\mathcal M}) = 
                \bm c \cdot \left[\bigotimes_{i\in \mathcal M}\begin{pmatrix}
                    \cos(\theta_i)  \\
                    \cos(2\theta_i) \\
                    \sin(\theta_i) \\
                    \sin(2\theta_i) \\
                    1
                \end{pmatrix}\right],
                \label{eq:analytic_form_fourier_d_dim}
        \end{equation}
        where $\bm c$ is a $5^D$-dimensional real-valued vector and $\mathcal M$ denotes the index set of the $|\mathcal{M}|=D$ simultaneously varied parameters. Consequently, the full reconstruction of the energy landscape requires a total of $5^D-1$ new energy evaluations. Once reconstructed, we can classically find the minimum of the energy landscape, our method of choice is detailed in Sec.~\ref{sec:classical_min_analytic_energy_function}.
        
        The exponential number of energy evaluations in the number of parameters hinders multi-parameter ExcitationSolve from being always blindly employed. Nonetheless, it offers a useful tool when employed in the right place. Figure \ref{fig:1D+CD_vs_2D} demonstrates an example of when a single application of a 2D optimization not only requires significantly fewer resources than the individual 1D optimization to converge, but also finds the global minimum instead of a local one. Again, this specific example is fabricated through a random three-qubit Hamiltonian. Generally, it can possibly set the optimizer on a more profitable path any time the 1D optimization reaches a local minimum that it cannot escape.
        To complete the scope of applicability of ExcitationSolve, Appendix~\ref{sec:multi_occ_multi_param} covers the most generic case of a multi-parameter optimization where each parameter may occur multiple times.
        
\section{Experiments} \label{sec:results}

    In this section, we assess the performance of ExcitationSolve on both fixed and adaptive ansätze. We compare it to other optimizers commonly found in VQE literature: Gradient Descent (GD), Constrained Optimization By Linear Approximation (COBYLA) \cite{powell1994direct}, Adam \cite{kingma_AdamMethodStochastic_2017}, Simultaneous Perturbation Stochastic Approximation (SPSA) \cite{spall1987stochastic, spall1992multivariate} and the Broyden-Fletcher-Goldfarb-Shannon (BFGS) algorithm \cite{broyden1970convergence, fletcher1970new, goldfarb1970family, shanno1970conditioning}. 
    The VQE is initialized with the Hartree-Fock (HF) state and parameters set to zero such that the initial energy is the HF energy $E_{\text{HF}}$. To evaluate the experiments, we consider the absolute error between the VQE energy $E_{\text{VQE}}$ with respect to the exact Full Configuration Interaction (FCI) energy $E_{\text{FCI}}$, i.e., $|E_{\text{VQE}}-E_{\text{FCI}}|$. This approach helps us determine when the error falls below the desirable \emph{chemical accuracy} of $\SI{e-3}{\Ha}$ \cite{peruzzo_VariationalEigenvalueSolver_2014}. Note that, although the term chemical accuracy is commonly used in quantum computing literature, it should more precisely be referred to as chemical precision \cite{tilly2022variational,elfving_HowWillQuantum_2020}. The quantum resource demand of the optimizers is tracked in the number of \emph{energy evaluations}, which refers to obtaining the expectation value of the Hamiltonian and is proportional to the actual number of measurements and terms of the Hamiltonian. Gradients for GD and Adam are computed using the four-term parameter-shift rule, requiring four energy evaluations per partial derivative. This equivalence in terms of energy evaluations provides a common basis for directly comparing all optimizers. By reusing previous energies, ExcitationSolve matches the cost of one iteration (parameter sweep) to that of one partial derivative (gradient) computation. The presented results are for optimizers with tuned hyperparameters. Full details on the experiments, their implementation and evaluation can be found in Appendix \ref{app:experimental_setup}.
    
    \subsection{Fixed ansatz (UCCSD) comparison with other optimizers} \label{subsec:fixedAnsatz}

\begin{figure}[!hpt]
                \centering
                \begin{subfigure}[t]{0.49\textwidth}
                    \centering
                    \includegraphics[width=\textwidth]{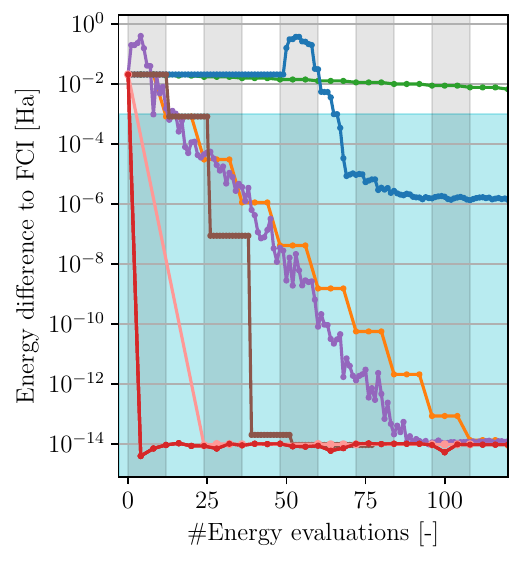}
                    \vspace{-0.75cm}
                    \caption{\ce{H2}, $4$ qubits. %
                    }
                    \label{fig:fixed_ansatz_H2_gd_adam_spsa_cobyla}
                    \vspace{0.4cm}
                \end{subfigure}
                \hfill
                \begin{subfigure}[t]{0.49\textwidth}
                    \centering
                    \includegraphics[width=\textwidth]{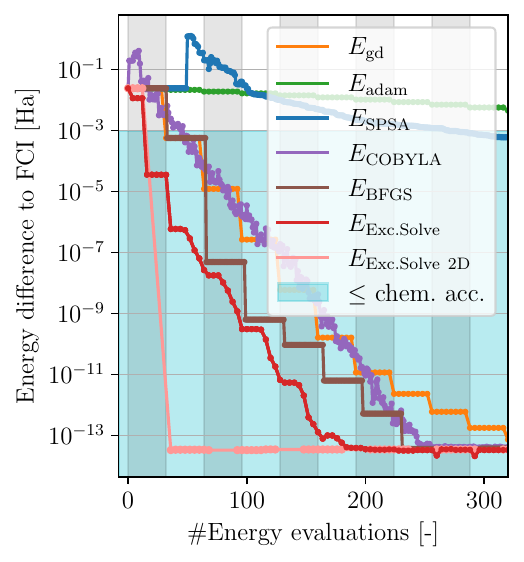}
                    \vspace{-0.75cm}
                    \caption{\ce{H3+}, $6$ qubits.
                    }
                    \label{fig:fixed_ansatz_H3_gd_adam_spsa_cobyla}
                \end{subfigure}
                \hfill
                \begin{subfigure}[t]{0.49\textwidth}
                    \centering
                    \includegraphics[width=\textwidth]{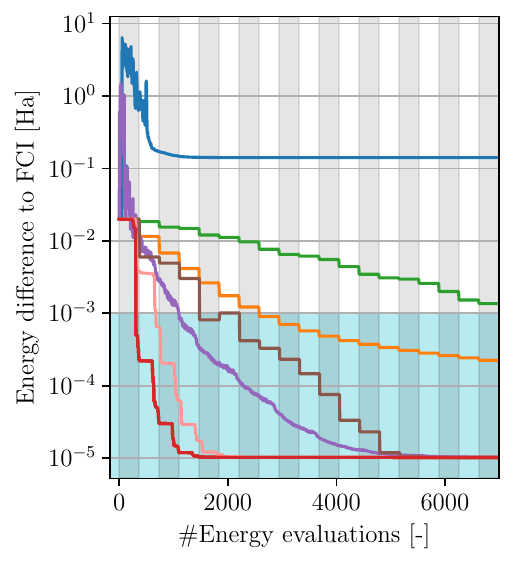}
                    \vspace{-0.75cm}
                    \caption{\ce{LiH}, $12$ qubits. %
                    }
                    \label{fig:fixed_ansatz_LiH_gd_adam_spsa_cobyla}
                \end{subfigure}
                \hfill
                \begin{subfigure}[t]{0.49\textwidth}
                    \centering
                    \includegraphics[width=\textwidth]{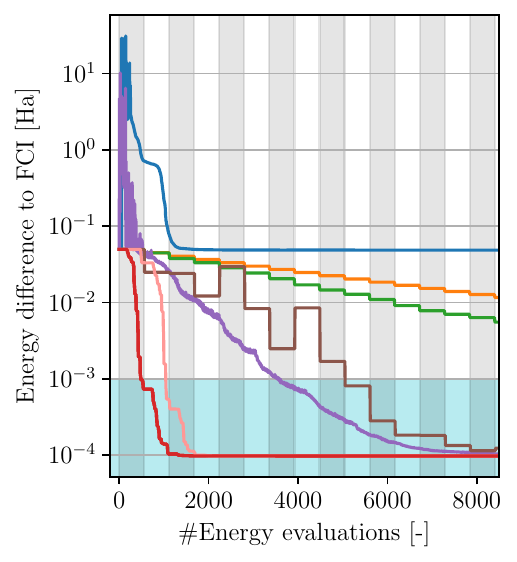}
                    \vspace{-0.75cm}
                    \caption{\ce{H2O}, $14$ qubits.
                    }
                    \label{fig:fixed_ansatz_H2O_gd_adam_spsa_cobyla}
                \end{subfigure}
                \caption{\textbf{Comparison of optimizers for fixed UCCSD ansätze.} The optimizers under consideration are ExcitationSolve (red), COBYLA (purple), Gradient descent (yellow), Adam (green), SPSA (blue) and BFGS (brown). The plots show the error of the VQE with respect to the FCI solution
                ${|E_{\text{VQE}}-E_{\text{FCI}}|}$ over the number of energy evaluations for the molecules a) \ce{H2}, b) \ce{H3+}, c) \ce{LiH} and d) \ce{H2O} in their respective equilibrium geometries.  The light blue region signifies the chemical accuracy ($\SI{e-3}{\Ha}$), while alternating vertical shadings mark full sweeps over all parameters. As each sweep incurs the same cost as a single gradient computation via the parameter-shift rule, gradient-based optimizers exhibit piecewise constant progress in comparison. The BFGS optimizer needs additional energy evaluations per update step to approximate the Hessian. Therefore the BFGS updates do not align with the vertical shadings.}
                \label{fig:combined_fixed_ansatz}
            \end{figure}

            To compare the optimizers we use a fixed ansatz where the tunable parameters and the order of the parameters are exactly the same for all optimizers. We choose the UCCSD ansatz in its first-order Trotter-approximation where we apply a single layer of first all double and then all single excitations.
            Figure \ref{fig:combined_fixed_ansatz} presents the results where we studied the ground state energies of the molecules \ce{H2} ($4$ qubits, Fig.~\ref{fig:fixed_ansatz_H2_gd_adam_spsa_cobyla}), \ce{H3+} ($6$ qubits, Fig.~\ref{fig:fixed_ansatz_H3_gd_adam_spsa_cobyla}), \ce{LiH} ($12$ qubits, Fig.~\ref{fig:fixed_ansatz_LiH_gd_adam_spsa_cobyla}) and \ce{H2O} ($14$ qubits, Fig.~\ref{fig:fixed_ansatz_H2O_gd_adam_spsa_cobyla}), each in their equilibrium geometry \cite{Utkarsh2023ChemistryComb} in the STO-3G basis.

            We find that ExcitationSolve does not only take fewer evaluations to reach chemical accuracy but also achieves this within a single sweep over the parameters. ExcitationSolve finds the exact ground state energy faster than all other optimizers with the most prominent speedup for larger molecules. 
            For the \ce{H2} molecule, ExcitationSolve even converges to the FCI energy in just one VQE iteration. This efficiency is attributed to the ground state being a superposition of the Hartree-Fock state and one doubly-excited state. Therefore, optimizing only the parameter in the UCCSD ansatz corresponding to the relevant double excitation is sufficient for convergence, which ExcitationSolve accomplishes optimally in a single VQE iteration. Similarly, for the \ce{H3+} molecule, ExcitationSolve with a 2D optimization strategy converges to the FCI energy after one 2D optimization, since the FCI ground state is a superposition of the HF state and two excited states. When using ExcitationSolve for both \ce{H2} and \ce{H3+} we note energy increases that appear around $5$ and $170$ energy evaluations, respectively. We attribute these increases in energy to numerical inaccuracies since the energy differences to the exact FCI energy are below $\SI{e-13}{\Ha}$ in both cases. 
            For \ce{H2O} ExcitationSolve reaches both the chemical accuracy and exact solution 7 times faster than the next best optimizer (COBYLA and BFGS, respectively). In contrast, the slowest, yet converging, optimizer (GD) takes 46 and 86 times longer to reach these accuracies. 
            It is worth noting for larger molecules that all optimizers consistently converge with a higher error, likely due to the absence of higher-order excitations in the UCCSD ansatz and the limited expressivity of a single first-order Trotter step. ExcitationSolve readily applies to higher-order excitations, and Sec.~\ref{sec:multi_occ} provides an extension to ansätze with repeated parameters resulting from higher-order Trotterization.
            As both SPSA and Adam have a very slow convergence and in some cases do not even manage to reach the chemical accuracy, we disregard them for further studies.

    \subsection{ADAPT-VQE} \label{subsec:ADAPT}
    
        As often suggested in recent literature on variational algorithms \cite{larocca_ReviewBarrenPlateaus_2024}, fixed ansätze may not be the way forward. We therefore probe ExcitationSolve in an adaptive setting where not only the parameters are optimized using ExcitationSolve, but also the choice of the next operator to append to the ansatz is made using the same strategy. We compare it to the original ADAPT-VQE implementation \cite{grimsley_AdaptiveVariationalAlgorithm_2019} where the operator selection is made by the gradient criterion and the re-optimization is performed with GD. Both are initialized in the HF state. 
        
\begin{figure}[!hpt]
                \centering
                \begin{subfigure}[t]{0.49\textwidth}
                    \centering
                    \includegraphics[width=\textwidth]{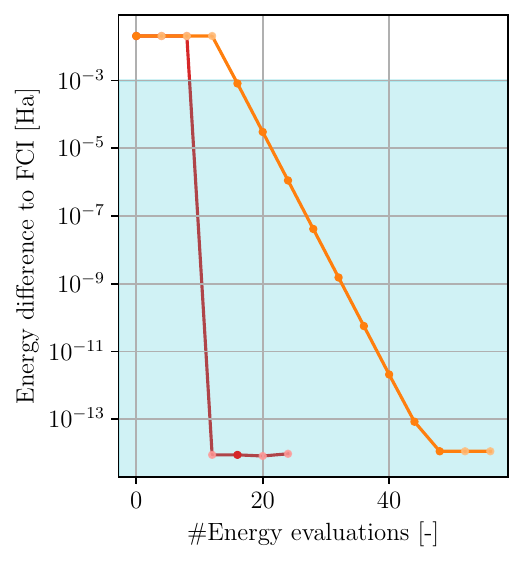}
                    \vspace{-0.75cm}
                    \caption{\ce{H2}, $4$ qubits. %
                    }
                    \label{fig:adapt_H2}
                    \vspace{0.4cm}
                \end{subfigure}
                \hfill
                \begin{subfigure}[t]{0.49\textwidth}
                    \centering
                    \includegraphics[width=\textwidth]{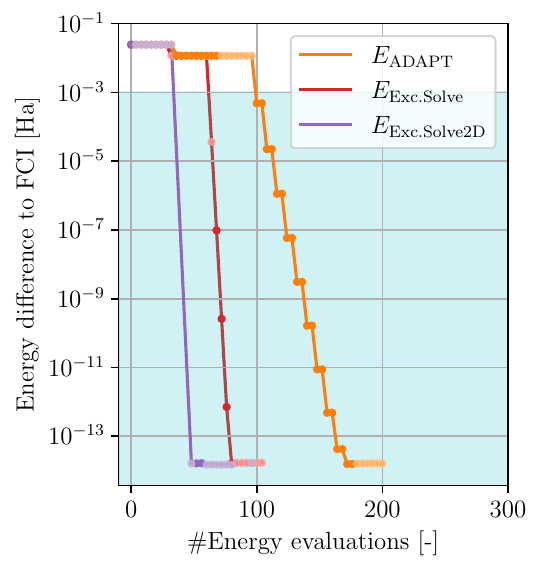}
                    \vspace{-0.75cm}
                    \caption{\ce{H3+}, $6$ qubits.
                    }
                    \label{fig:adapt_H3+}
                \end{subfigure}
                \hfill
                \begin{subfigure}[t]{0.49\textwidth}
                    \centering
                    \includegraphics[width=\textwidth]{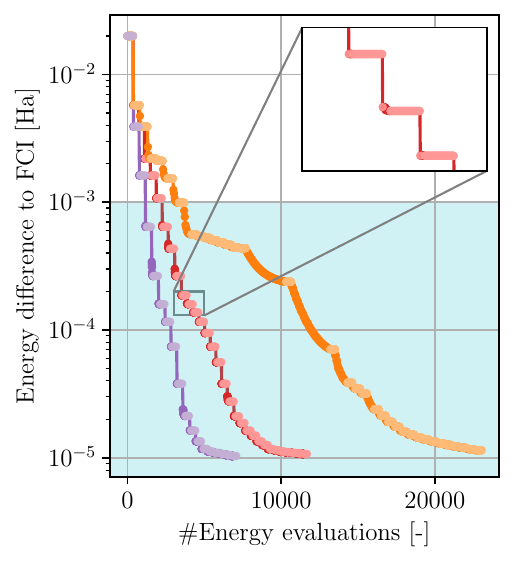}
                    \vspace{-0.75cm}
                    \caption{\ce{LiH}, $12$ qubits. %
                    }
                    \label{fig:adapt_LiH}
                \end{subfigure}
                \hfill
                \begin{subfigure}[t]{0.49\textwidth}
                    \centering
                    \includegraphics[width=\textwidth]{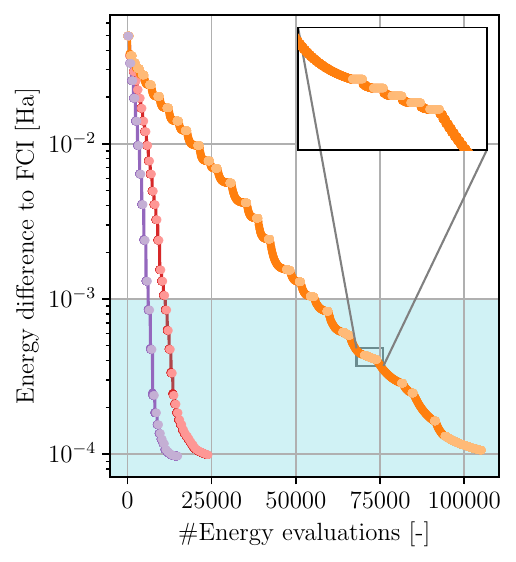}
                    \vspace{-0.75cm}
                    \caption{\ce{H2O}, $14$ qubits.
                    }
                    \label{fig:adapt_H2O}
                \end{subfigure}
                \caption{\textbf{Comparison between ExcitationSolve and original ADAPT-VQE with the UCCSD operator pool.} 
                The plots show the error of the VQE with respect to the FCI solution
                ${|E_{\text{VQE}}-E_{\text{FCI}}|}$ over the number of energy evaluations for the molecules a) \ce{H2}, b) \ce{H3+}, c) \ce{LiH} and d) \ce{H2O}. Lighter plot colors signal evaluations needed for operator selection, darker plot colors mark optimization steps, with insets magnifying regions of interest. The light blue region signifies the chemical accuracy ($\SI{e-3}{\Ha}$).}
                \label{fig:Adapt_Benchmark_combined}
            \end{figure}

        Figure \ref{fig:Adapt_Benchmark_combined} shows the convergence of the adaptive optimizations to the ground states of molecules \ce{H2}, \ce{H3+}, \ce{LiH} and \ce{H2O} in their equilibrium geometry. We compare ADAPT-VQE with GD as optimizer to ExcitationSolve and to a 2D variant of ExcitationSolve. Here in each adapt step not only one, but the two best operators are selected, 2D optimized with respect to both parameters $\theta_i,\ \theta_j$ and then appended to the ansatz. All further optimization is performed using 1D ExcitationSolve. The total number of evaluations is composed of the evaluations to select a new operator and the evaluations to re-optimize the parameters already present in the ansatz. The former lead to plateaus, during which the energy remains unchanged.
        For all four molecules ExcitationSolve reaches faster convergence than ADAPT-VQE for both the chemical accuracy and the limit within the UCCSD ansatz. The reason for the faster convergence stems mainly from two key advantages that ExcitationSolve features:
        First, using ExcitationSolve to select new operators leads to fewer operators being added to the ansatz. This results in a shallower circuit and a cheaper re-optimization.
        This can be seen for the larger molecules, i.e., \ce{LiH} in Fig.~\ref{fig:adapt_LiH}, where ADAPT-VQE requires 34 operators, while ExcitationSolve only needs 30 to converge. For \ce{H2O}, ADAPT-VQE needs 48 operators, while ExcitationSolve only requires 42. Note that the operator reduction mostly but not only becomes significant beyond chemical accuracy.
        Second, initializing the new operators with ExcitationSolve at their optimal values offers a beneficial warm start for the intermediate parameter optimization, further leading to a convergence within fewer iterations in the re-optimization of the parameters. A special case can be seen for \ce{H2} in Fig.~\ref{fig:adapt_H2}, where only a single excitation contributes to the ground state and ExcitationSolve immediately initializes it with its optimal parameter value. The most significant reduction in energy evaluations can be observed for \ce{H2O} where ExcitationSolve reaches the chemical accuracy approximately 15 times faster. Appendix \ref{app:adapt_vqe_resource_comp} adds resource comparisons.
        One might wonder whether the reduction in selected operators is due to the ExcitationSolve parameter optimizer or the ExcitationSolve operator selection -- or perhaps even a joint effort. An in-depth analysis in Appendix~\ref{app:mixing} shows that the reduction stems from the ExcitationSolve selection. The fastest convergence is still achieved by employing ExcitationSolve for both tasks.

    \subsection{Dissociation curves} \label{subsec:StrongCorrelations}
    
            \begin{figure}[!hpt]
                \centering
                \begin{subfigure}[t]{0.49\textwidth}
                    \centering
                    \includegraphics[width=\textwidth]{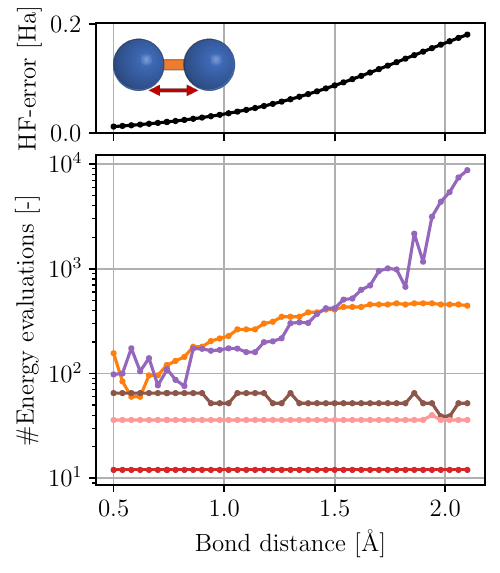}
                    \vspace{-0.75cm}
                    \caption{\ce{H2}, 4 qubits.%
                    }
                    \label{fig:diss_curve_H2_gd_cobyla}
                    \vspace{0.4cm}
                \end{subfigure}
                \hfill
                \begin{subfigure}[t]{0.49\textwidth}
                    \centering
                    \includegraphics[width=\textwidth]{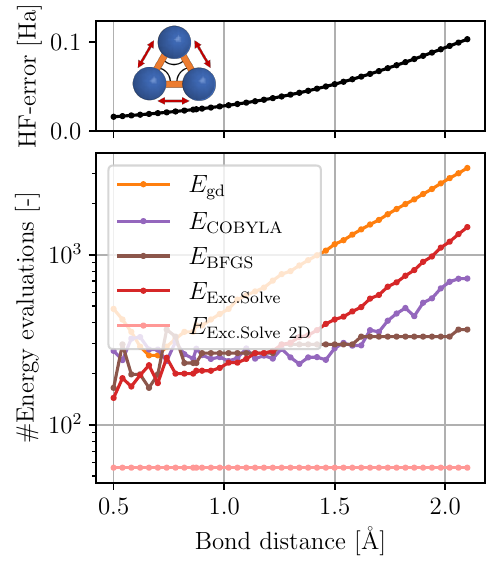}
                    \vspace{-0.75cm}
                    \caption{\ce{H3+}, $6$ qubits. %
                    }
                    \label{fig:diss_curve_H3_gd_cobyla}
                \end{subfigure}
                \hfill
                \begin{subfigure}[t]{0.49\textwidth}
                    \centering
                    \includegraphics[width=\textwidth]{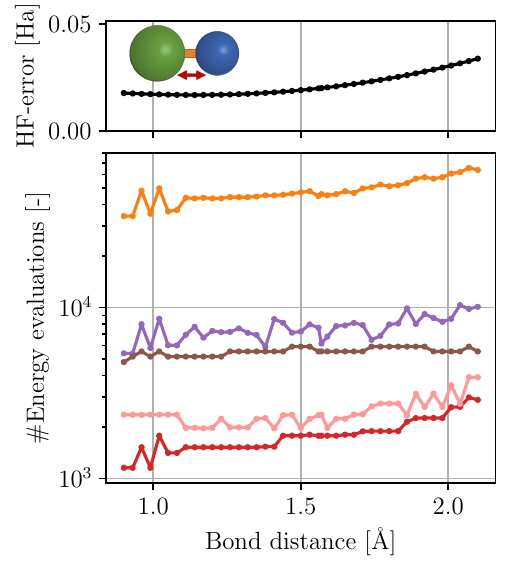}
                    \vspace{-0.75cm}
                    \caption{\ce{LiH}, 12 qubits.%
                    }
                    \label{fig:diss_curve_LiH_gd_cobyla}
                \end{subfigure}
                \hfill
                \begin{subfigure}[t]{0.49\textwidth}
                    \centering
                    \includegraphics[width=\textwidth]{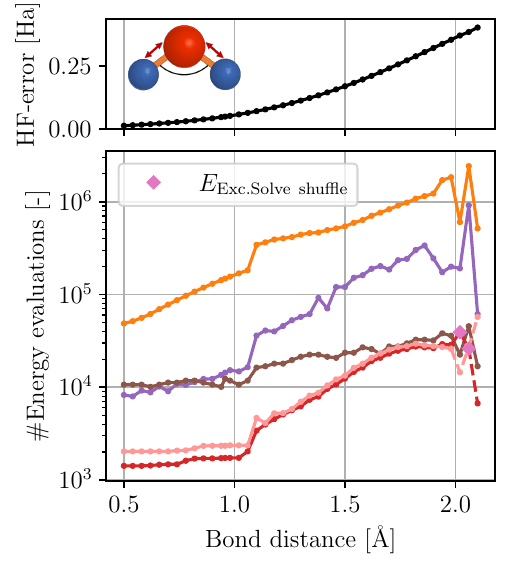}
                    \vspace{-0.75cm}
                    \caption{\ce{H2O}, 14 qubits.%
                    }
                    \label{fig:diss_curve_H2O_gd_cobyla}
                \end{subfigure}
                \caption{\textbf{Comparison of optimizers using fixed UCCSD ansätze for non-equilibrium geometries.} Hartree-Fock (HF) error (black) and energy evaluations until VQE convergence in dependence of the bond length for optimizers ExcitationSolve (red), COBYLA (purple) and Gradient descent (yellow) for molecules a) \ce{H2}, b) \ce{H3+}, c) \ce{LiH} and d) \ce{H2O}.  For \ce{H2O} with high bond lengths, random shuffling of the parameter order in ExcitationSolve is additionally utilized to achieve convergence (diamond marker). The HF error is the absolute difference between the FCI energy and HF energy. %
                }
                \label{fig:combined_diss_curves}
            \end{figure}

            We further analyze how a deviation of the HF state from the actual ground state influences the performance of ExcitationSolve compared to GD, BFGS and COBYLA when provided as the initial state in the fixed UCCSD ansatz VQE. Concretely, by varying the inter-atomic distances in the molecules, we affect how closely the HF state approximates the true ground state: The further the bond is stretched, the larger the initial HF error $|E_{\text{HF}}-E_{\text{FCI}}|$ of the HF energy $E_{\text{HF}}$ to the energy of the FCI solution $E_{\text{FCI}}$, signaling the emergence of \emph{strong correlations} \cite{giner_BasissetErrorCorrection_2020}. This HF error dependence on the bond distance for all studied molecules is shown in Fig.~\ref{fig:combined_diss_curves}.
            
            Figure \ref{fig:combined_diss_curves} shows how many energy evaluations are needed for different bond distances to reach convergence for each molecule. See Appendix \ref{app:diss_curves_energy_errors} about the final energy differences to FCI when convergence is reached.
            Among all molecules it becomes apparent that the higher the bond distance gets, i.e., the higher the HF error since the initial HF state deviates more from the ground state solution, the more executions of the circuit are necessary to find the ground state. We see only two exceptions: for \ce{H2} (Fig.~\ref{fig:diss_curve_H2_gd_cobyla}) the ExcitationSolve optimizer finds the ground state with a constant number of evaluations and for \ce{H3+} (Fig.~\ref{fig:diss_curve_H3_gd_cobyla}) ExcitationSolve also needs a constant number of evaluations when optimizing two parameters at the same time. These two cases are special, because (2D-)ExcitationSolve can set the parameters to the global optimum within the first VQE iteration because the ground state of \ce{H2} and \ce{H3+} are superpositions of the HF state with one and two excited states, respectively.
            Overall, it can be observed that ExcitationSolve outperforms the other optimizers for each bond distance. Although, the relative difference in the number of energy evaluations needed for the optimizers to reach convergence is mostly independent of the bond distance. Indeed the scaling with the bond length for the two larger molecules \ce{LiH} (Fig.~\ref{fig:diss_curve_LiH_gd_cobyla}) and \ce{H2O} (Fig.~\ref{fig:diss_curve_H2O_gd_cobyla}) is almost identical for all optimizers, with BFGS seeming least impacted by the bond distance. 
            Finally, see Appendix~\ref{app:local_min_diss_curves} for a closer look at certain large \ce{H2O} bond lengths, where ExcitationSolve faces convergence issues in local minima and flat regions, and how strategies like ExcitationSolve2D and parameter shuffling help overcome these challenges.

    \subsection{NISQ hardware benchmarks}\label{subsec:NISQ}
    
        To conclude the experimental evaluation of ExcitationSolve, we examine the near-term applicability of ExcitationSolve by repeating previous experiments on the IBM quantum computer \texttt{ibm\_quebec}, which is referred to as IBM-Q henceforth. This quantum processor features the \textit{Eagle r3} architecture with 127 qubits and classifies as a noisy intermediate-scale quantum (NISQ) device.
        Decoherence over time significantly impedes the execution of quantum circuits of increased depth, primarily limited by the two-qubit gate count. Therefore, the implementation of excitation operators on current NISQ devices is a challenge due to their decomposition into rather deep circuits over a hardware-native gate set \cite{barkoutsos_QuantumAlgorithmsElectronic_2018}. This is a limitation of these ansätze rather than ExcitationSolve or any other optimizer per se.

        \begin{figure}[htbp]
            \centering
            \begin{subfigure}[t]{0.49\textwidth}
                \centering
                \includegraphics[width=\textwidth]{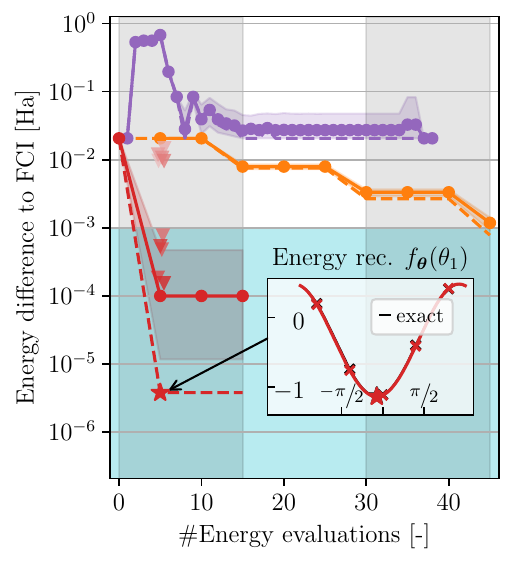}
                \vspace{-0.75cm}
                \caption{\ce{H2}, $4$ qubits. %
                }
                \label{fig:ibmq_H2}
            \end{subfigure}
            \hfill
            \begin{subfigure}[t]{0.49\textwidth}
                \centering
                \includegraphics[width=\textwidth]{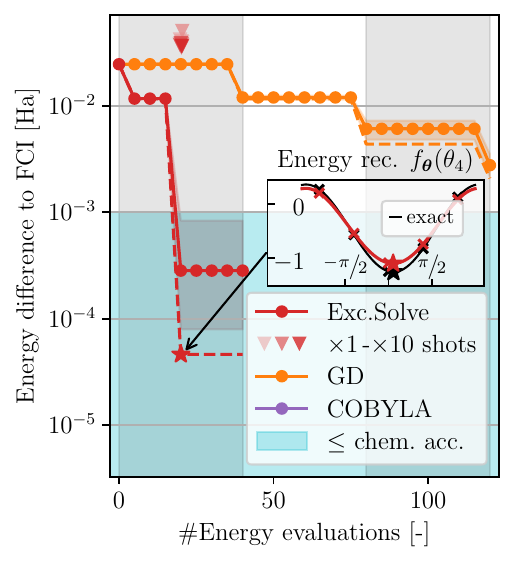}
                \vspace{-0.75cm}
                \caption{\ce{H3+}, $6$ qubits.
                }
                \label{fig:ibmq_H3}
            \end{subfigure}
            \caption{\textbf{Benchmarks on NISQ (\texttt{ibm\_quebec}) quantum processor for fixed UCCSD ansätze.}
            The molecules a) \ce{H2} and b) \ce{H3+} are studied, analogous to the simulated experiments in Figs.~\ref{fig:fixed_ansatz_H2_gd_adam_spsa_cobyla} and \ref{fig:fixed_ansatz_H3_gd_adam_spsa_cobyla}, respectively.
            The optimizers considered are ExcitationSolve (red), Gradient descent (yellow), and COBYLA (purple), where the latter was discarded from the \ce{H3+} experiments. Per optimizer, five experiments are performed out of which the best run (dashed), along with the mean (solid) and 95\% confidence intervals (bands), are presented in terms of the VQE energy with respect to the FCI solution ${|E_{\text{VQE}}-E_{\text{FCI}}|}$. 
            While all presented optimizations exclusively rely on energy value (and gradient) information extracted from the the NISQ device \texttt{ibm\_quebec}, the FCI errors shown are based on exact re-evaluations via state vector simulation for a clear quality assessment. Triangles instead represent such \texttt{ibm\_quebec} energy estimates in the final ExcitationSolve parameter configuration (star), with shot counts from 1$\times$ (light red) to 10$\times$ (dark red) of the 8192 default.
            The inset plots also show the noisy energy values (red crosses) and compare the resulting energy reconstructions $f_{\bm{\theta}}(\cdot)$ (red) as used by ExcitationSolve with the exact energy function (black).
            The light blue background signifies chemical accuracy. Vertical lines mark the iterations in ExcitationSolve and GD.
            }
            \label{fig:combined_ibmq}
        \end{figure}
        
        Nevertheless, we select \ce{H2} and \ce{H3+} for fixed UCCSD ansatz IBM-Q experiments, analogous to Sec.~\ref{subsec:fixedAnsatz}. The results are depicted in Fig.~\ref{fig:combined_ibmq} and discussed in the following. For a comparison, simulated experiments with pure shot noise are presented in Appendix \ref{subsec:Noise}. 
        In Appendix \ref{app:nisq_adapt}, \ce{LiH} serves to study adaptive ansätze by focusing on the robustness of the operator selection criterion of ExcitationSolve compared to the original ADAPT-VQE on a NISQ device.

        Over the five repeated experiments for both, \ce{H2} (Fig.~\ref{fig:ibmq_H2}) and \ce{H3+} (Fig.~\ref{fig:ibmq_H3}), ExcitationSolve demonstrates strong robustness against hardware noise by reproducing the convergence within a single iteration as observed in both exact and shot noise simulation. Convergence means that the parameters that were found by the optimizers by solely using energy (or gradient) information from the IBM-Q prepare the ground state within chemical accuracy when re-evaluating the associated energy \emph{via exact simulation}. ExcitationSolve hereby surpasses GD in both speed and quality: in all five repetitions of the \ce{H2} experiment, ExcitationSolve reaches chemical accuracy within the first iteration, even within the 95\% confidence intervals. The GD step size was tuned in preliminary \ce{H2} IBM-Q experiments and chosen as the largest non-diverging step size tested. Even for the more challenging \ce{H3+}, ExcitationSolve still achieves chemical accuracy within the first iteration on average, which was no longer observed for GD in the given time.
        COBYLA was not examined for \ce{H3+} after it proved incapable of handling the hardware noise in the smaller \ce{H2} case where no improvement over the HF energy was achieved. COBYLA does not preserve sparsity in the iterates, i.e., does not keep most parameters zero, which leads to deeper and thus noisier transpiled circuits. This is in contrast to ExcitationSolve and GD, which owe some of their success under hardware noise to this sparsity, which allows smaller quantum circuits to be executed on the IBM-Q.
        As a conclusion, the successful convergence especially via ExcitationSolve within a single iteration is striking and rather unexpected given the relatively high errors in the energy estimates obtained from the IBM-Q device, which the energy reconstructions were derived from. 
        To put this into perspective, not only is the deviation typically outside the chemical accuracy, but the absolute error may even exceed the initial HF error. Nevertheless, the parameters that ExcitationSolve optimized based on these noisy inputs prove to prepare the ground states within chemical accuracy. A reason why the reconstruction remains useful is that the hardware noise introduces a systematic error. This error may manifest as a rescaled amplitude or an offset shift of the reconstructed finite Fourier series, as depicted for \ce{H3+} in the inset plot of Fig.~\ref{fig:ibmq_H3}. Importantly, these transformations do not affect the location of the minimum, leading to parameter values that still prepare the ground state accurately. Estimating the energy for the final ExcitationSolve parameters on IBM-Q with a shot budget exceeding 8192, which was otherwise used for energy evaluations, reveals that an at least five-fold shot increase achieves chemical accuracy for \ce{H2}, whereas no such improvement -- not even below the HF error -- is observed for the more complex \ce{H3+}. 
        Beyond \ce{H3+}, we observed hardware noise to obscure energy information due to an increased circuit depth from multiple excitations being active in the ansatz. Therefore, the error is not systematic enough to allow for sufficient energy reconstructions.

        \FloatBarrier

    \section{Discussion}
\label{sec:discussion}

        The main motivation behind ExcitationSolve is to unite the benefits of quantum-aware optimization and physically-motivated ansätze composed of excitation operators. Due to the quantum-awareness in particular, ExcitationSolve improves over common gradient-based optimizers by informing updates globally instead of being limited to the local vicinity -- without imposing any resource overhead and, on top of that, circumventing any hyperparameter tuning. Specifically, reconstructing the energy landscape for a single parameter requires four energy evaluations, optimized by reusing the final minimal energy from the previous step, which matches the resource requirements for the four-term parameter-shift rule. Interestingly, the same resource-efficiency already can be observed with simpler rotations in Rotosolve.  
        When considering real wave functions (real-valued quantum states), as we did for all examples in Sec.~\ref{sec:results}, a more efficient approach which manages with only two shifts exists \cite{kottmann_FeasibleApproachAutomatically_2021} This approach is, however, at the cost of circuit modifications as detailed in Appendix~\ref{sec:param_shift_rules}. 
        Table \ref{tab:resource_comparison} provides a resource comparison.
        
        In the adaptive setting, we globally inform the operator selection criterion with ExcitationSolve by using the same quantum resources more effectively compared to the original gradient criterion in ADAPT-VQE \cite{grimsley_AdaptiveVariationalAlgorithm_2019}.
        We remark that any double excitation can be decomposed into a product of 8 Pauli rotations of the same angle \cite{barkoutsos_QuantumAlgorithmsElectronic_2018}. A naive application of SMO/Rotosolve to the Pauli decomposition leads to an overestimation of the Fourier order by a factor of 4. This further worsens for higher-order excitations where SMO exponentially overestimates the Fourier order. However, the more sophisticated decomposition of excitation operators into \emph{two} commuting self-inverse operators (cf.~Appendix~\ref{app:proofs} and Ref.~\cite{kottmann_FeasibleApproachAutomatically_2021}) reveals that ExcitationSolve can be interpreted as the double-occurrence case of SMO.

        \begin{table}[ht]
            \centering
            \caption{Number of shifts comparison for the full single-parameter energy landscape reconstruction vs.~the partial derivative via a parameter-shift rule. Reconstruction numbers distinguish the \emph{theoretical} minimum and the \emph{effective} number after reusing previously evaluated energies. Note that the number of shifts for excitations on real-valued wavefunctions ($^*$) does not exactly correspond to pure energy evaluations \cite{kottmann_FeasibleApproachAutomatically_2021}.}
            \label{tab:resource_comparison}
            \begin{tabular}[t]{@{}lcccc@{}}
            \toprule
                                           & \multicolumn{2}{c}{\multirow{2}{*}{Function reconstruction}} & \multicolumn{2}{c}{Partial derivative} \\
                                           & \multicolumn{2}{c}{}        & \multicolumn{2}{c}{via parameter-shift rule} \\ 
                                           & \textit{Theoretical}            & \textit{Effective} & \textit{Complex WF} & \textit{Real WF} \\ \midrule
            Rotations ($G^2 = I$)   & 3 \cite{ostaszewski_StructureOptimizationParameterized_2021,nakanishi_SequentialMinimalOptimization_2020}                   & 2 \cite{ostaszewski_StructureOptimizationParameterized_2021,nakanishi_SequentialMinimalOptimization_2020}                   & 2 \cite{mitarai_QuantumCircuitLearning_2018,schuld_EvaluatingAnalyticGradients_2019} & -                \\
            Excitations ($G^3 = G$) & 5                    & 4                    & 4 \cite{anselmetti_LocalExpressiveQuantumnumberpreserving_2021}  & 2* \cite{kottmann_FeasibleApproachAutomatically_2021}                  \\\bottomrule
            \end{tabular}
        \end{table}
        
        Moreover, the relevance of ExcitationSolve is intrinsically linked to the advantages of employing physical ansätze using excitation operators. The significance of such ansätze lies in their ability to preserve essential physical quantities and symmetries. It could be argued that \emph{qubit tapering} \cite{bravyi_TaperingQubitsSimulate_2017, setia_ReducingQubitRequirements_2020} for hardware-efficient or problem-agnostic ansätze, i.e., composed of rotation gates, could achieve the same advantages, however qubit tapering conserves the particle numbers and spin symmetries only up to their parity.
        Note that ExcitationSolve natively handles excitation operators, independent of the actual decompositions, fermion-to-qubit mappings and simplifications of the ansatz to the quantum circuit.  %
        In addition, organizing the parameters in a problem-informed way via such physical excitation operators could lead to a simpler, more suggestive optimization landscape as it has been observed in other contexts \cite{wiersema_HereComesMathrmSU_2023}.
        Alternatively, the choice of physically-motivated ansätze can be interpreted as encoding an \emph{inductive bias}, and, hence, could effectively restrict the exponentially growing underlying Hilbert space. This potentially counteracts the emergence of so-called \emph{barren plateaus}, which would obstruct the practical use in realistic, large-scale problem sizes \cite{larocca_ReviewBarrenPlateaus_2024}.
        
        Our experimental results demonstrate multiple advantages of ExcitationSolve over previous state-of-the-art optimizers, including gradient descent (GD), COBYLA \cite{powell1994direct}, Adam \cite{kingma_AdamMethodStochastic_2017}, SPSA \cite{spall1987stochastic, spall1992multivariate}, and BFGS \cite{broyden1970convergence, fletcher1970new, goldfarb1970family, shanno1970conditioning}:
        Firstly, for a fixed UCCSD ansatz ExcitationSolve generally takes fewer iterations to converge to the ground state than any other tested optimizer. How significant the advantage is depends on the molecule.
        Secondly, ExcitationSolve has no hyperparameters that need to be tuned and needs no calibration. %
        Thirdly, in an adaptive setting, like ADAPT-VQE \cite{grimsley_AdaptiveVariationalAlgorithm_2019}, ExcitatonSolve can be used to choose the next operator to append based on its highest impact on the energy value outperforming the locally informed selection such as the original ADAPT-VQE gradient criterion. The newly picked operator is also already initialized in its optimal parameter value, which significantly warm-starts the intermediate optimization of all present parameters before extending the ansatz further. 
        In NISQ experiments on the IBM quantum device, ExcitationSolve shows robustness against both shot- and hardware noise, and preserves the advantages over other optimizers that were observed in simulation.
        Decoherence errors ultimately impose a limit but are natural for the depth of transpiled circuits from ansätze composed of excitation operators, which is independent of the chosen optimizer.
        
        While ExcitationSolve offers promising features, it faces certain challenges, such as the potential for getting trapped in a local minimum. This issue becomes particularly significant for larger molecules, where the proliferation of local minima poses a critical challenge. To address this, we proposed a multi-dimensional extension of ExcitationSolve, though this approach introduces new complexities, such as increased computational cost, as the number of evaluations grows exponentially with the number of parameters. Moreover, determining which parameters to pair for optimization is complex due to the combinatorial nature of the problem. One effective heuristic is to pair operators with the strongest impact from single-parameter optimization, as demonstrated with the \ce{H3+} molecule. In the case of \ce{H2O}, we observe that this heuristic is not always sufficient to avoid local minima. Fortunately, we find that parameter shuffling provides another, complementary approach to avoid local minima -- and unlike the 2D optimizer does not even require additional computational effort.
        We did not find a systematic way of deciding when to apply shuffling to avoid local minima. But, what we find is that the order in which the parameters in the UCCSD ansatz are optimized can, at least for ExcitationSolve, play a crucial role in overall convergence. One could investigate if turning parameter shuffling on and off during optimization can increase convergence speed or even avoid local minima. Furthermore, one needs to analyze what causes the 1D optimization to settle in a local minimum and why parameter shuffling is able to circumvent it.
        In general, ExcitationSolve cannot be guaranteed to find a global minimum. Hence, identifying strong heuristics for beneficial parameter orderings and for how and when multi-parameter optimization can be employed to navigate the energy landscape faster and more reliably remains an area for future investigation.

        Besides the showcased application of ExcitationSolve to ground state preparations, it can be analogously employed to perform projected variational quantum dynamics (pVQD) \cite{Barison2021efficientquantum}. The utility of gradient-free optimization for time evolution has already been demonstrated \cite{Benedetti2021HardwareEfficient} via Rotosolve for hardware-efficient ansätze to replicate a Trotter step. The applicability of ExcitationSolve to excitation-based ansätze for pVQD becomes apparent when reformulating the overlap maximization with the zero-state $\ket{\bm 0}$ as the minimization of the Hermitian zero-state projector $P=\ket{\bm 0}\bra{\bm 0}$ \cite{Jaderberg_2020}, taking the role of the Hamiltonian $H$ in this work.

    \section{Methods} %
\label{sec:methods}

        \subsection{ExcitationSolve for multiple occurrences of parameters} \label{sec:multi_occ}
        
        For the practical use of UCCSD, it often suffices to employ a single time step in first-order Trotterization. However, one might encounter scenarios where the expressivity of such type of ansatz is no longer sufficient and thus requires a refinement -- e.g., through a higher order product formula or simply multiple time steps. In a higher order product formula, at least one or more parameters appear multiple times throughout the corresponding quantum circuit. Concerning the use of multiple time steps, there is the degree of freedom %
        of using different parameters for every time step or sharing these parameters between multiple steps. The latter is motivated by the physical point of view of a discretized time evolution of an adiabatic process with equal time steps where the strength of the free fermionic problem, i.e.~the single-excitations, is kept constant. For $S$ excitations sharing the same variational parameter $\theta$,
        
        \begin{equation}
            f_{\bm\theta}(\theta) = \sum_{s = 1}^{2S} a_s \cos(s \theta) +  \sum_{s = 1}^{2S} b_s \sin(s \theta) + c,
            \label{eq:energy_multi_occ}
        \end{equation}
        
        i.e., it is a one-dimensional Fourier series of order $2S$. The $4S+1$ coefficients can be determined through $4S$ energy evaluations. This result can be straightforwardly inferred from the multi-parameter generalization from Eq.~\eqref{eq:analytic_form_fourier_d_dim} by setting multiple parameters equal and reducing the trigonometric form. \Cref{eq:energy_multi_occ} is closely related to the result for multiple occurrences of a single parameter in Rotosolve/SMO, where the energy landscape is given by a Fourier series of order $S$ with $2S+1$ coefficients \cite{nakanishi_SequentialMinimalOptimization_2020} (cf.~Sec.~\ref{sec:rotosolve_smo}). For a detailed derivation, refer to Appendix \ref{app:multi_occ_single_param}. Note that unlike the exponential growth in energy evaluations for a multi-parameter optimization, we have a linear growth for the multi-occurrence case. Appendix~\ref{sec:multi_occ_multi_param} presents the most generic energy landscape form when considering multiple such repeated parameters simultaneously.

        \subsection{Classical minimization of analytic energy reconstructions} \label{sec:classical_min_analytic_energy_function}
        
            After reconstructing the energy function, in order to determine the global minimum in the single parameter (or set of parameters) classically, we suggest the utilization of the following approaches.
        
            \paragraph{Single-parameter case (companion-matrix method).}
            
                The first important realization is that the derivative of the finite Fourier series determining the energy function in one parameter as in Eq.~\eqref{eq:analytic_form_fourier}, is again a Fourier series of the same order, i.e., $\tfrac{d}{d\theta} b_s\sin(s\theta) = b_s s\cos(s\theta)$, $\tfrac{d}{d\theta} a_s\cos(s\theta) = -a_s s\sin(s\theta)$ and zero constant. Of this derivative function we can determine the zeros and evaluate the analytic energy function (classically) at these points. The smallest of the resulting energy values must be the global minimum.
            
                Finding the zeros of the derivative can be achieved efficiently through the so-called companion-matrix method. In the following we provide a brief review of this technique introduced in Ref.~\cite{boyd2006computing} for a general (finite) Fourier series. Note that the constant term $c=0$ as we deal with minima, i.e., zeros of derivatives:
                \begin{equation}
                    f(\theta) = \sum_{s = 1}^S a_s \cos(s \theta) +  \sum_{s = 1}^S b_s \sin(s \theta).
                \end{equation}
                By employing the Euler identity, this finite Fourier series may be recast into the complex form
                \begin{align}
                    f(\theta) &= \sum_{s=1}^S \left(\frac{a_s - i b_s}{2} e^{i s \theta} + \frac{a_s + i b_s}{2} e^{-i s \theta}\right) \nonumber \\
                    &= e^{-iS\theta} \sum_{s=1}^S \left(\frac{a_s - i b_s}{2} e^{i (S+s) \theta} + \frac{a_s + i b_s}{2} e^{i (S-s) \theta}\right) \nonumber \\
                    &= e^{-iS \theta} \left(\sum_{s=S+1}^{2S} \frac{a_{s-S} - i b_{s-S}}{2} e^{i s\theta} + \sum_{s=0}^{S-1} \frac{a_{S-s} + i b_{S-s}}{2} e^{i s\theta}  \right).
                \end{align}
                By introducing the transformation $z=e^{i\theta}$, we rewrite the Fourier series as
                
                \begin{equation}
                    f(\theta) = \frac{z^{-S}}{2} \sum_{s=0}^{2S} h_s z^s =: \frac{z^{-S}}{2} h(z),
                \end{equation}
                where
                \begin{equation}
                h_s =
                    \begin{cases}
                        a_{S-s} + i b_{S-s}, &s=0,1,\dots, S-1 \\
                        0, &s=S \\
                        a_{s-S} - i b_{s-S}, &s=S, S+1,\dots, 2S.
                    \end{cases}
                \end{equation}
                and $h(z)$ is referred to as the \emph{associated polynomial}\footnote{Since $\bar h_s = h_{2S-s}$ holds, $h(z)$ is a complex \textit{self-reciprocal} or \textit{palindromic} polynomial.}. Note that the problem of finding the real zeros of $f(\theta)$ has now been transformed to the task of determining the zeros of $h(z)$ on the complex unit circle. To solve for the roots of the associated polynomial, the $2S \times 2S$ companion matrix/Frobenius matrix $ B$ is constructed with the component in the $s$-th row and $t$-th column
                \begin{equation}
                    B_{st} = 
                    \begin{cases}
                        \delta_{s,t-1}, &s=1,2,\dots, 2S-1 \\
                        - \frac{h_{t-1}}{a_S-ib_S}, &s=2S,
                    \end{cases}
                \end{equation}
                where $\delta$ denotes the Kronecker delta.
                The characteristic polynomial of the companion matrix is precisely the associated polynomial from before. In the case of ExcitationSolve, where $S=2$, the companion matrix takes the form
                \begin{equation}
                    B = \begin{pmatrix}
                        0 & 1 & 0 & 0 \\
                        0 & 0 & 1 & 0 \\
                        0 & 0 & 0 & 1 \\
                        - \frac{a_2+ib_2}{a_2-ib_2} & - \frac{a_1+ib_1}{a_2-ib_2} & 0 & - \frac{a_1-ib_1}{a_2-ib_2}
                    \end{pmatrix}
                \end{equation}
                The roots $\theta_k$ of $f$ are then obtained as
                \begin{equation}
                    \theta_k = \text{arg}(z_k) + 2\pi m - i \log(|z_k|),
                \end{equation}
                where $z_k$ is the $k$-th eigenvalue of $B$ and $m$ is some integer. Here, it becomes clear that $\theta_k$ is real iff $z_k$ lies on the complex unit circle. 
            
            \paragraph{Multi-parameter case (Nyquist initialization and local optimization).}
            
                In the case of an analytic energy function in multiple parameters to be optimized as in Eq.~\eqref{eq:analytic_form_fourier_d_dim}, we can no longer employ the companion-matrix method.

                One naive way to find the minimum of a multiple-dimensional energy landscape lies in rasterizing of the parameter space. For low dimensions, such a brute-force evaluation can be easily performed on a classical computer (keep in mind that the energy function has already been faithfully reconstructed and can be evaluated at arbitrary positions in parallel). For this type of approach, however, the precision of the result is limited by the resolution of the grid, which is not desirable as the precision  then depends precisely on the position of the minimum and cannot be assumed to be constant. 
                
                Inspired by the Nyquist-Shannon sampling theorem \cite{shannon1949communication}, we conjecture that it is sufficient to evaluate the energy function with a lattice spacing of $\Delta = 2\pi/(2\omega_\mathrm{max} + 1)$, where $\omega_\mathrm{max}=2S$ is the highest frequency in the system along an $S$-fold occurring parameter, i.e., we sample with at least twice the highest frequency of the system (Nyquist frequency), which is $2\omega_\mathrm{max} + 1$ equidistant samples within the period. Each of those lattice points is then taken as an initial guess for a local optimization scheme such as gradient descent. This can be implemented on classical hardware very efficiently for the following reasons. Firstly, all runs for the different initial guesses can be performed in parallel. Secondly, since the function to be minimized is known analytically, analytical gradients are also readily available. Thirdly, through an optimal gradient descent step size $\alpha$, the convergence is guaranteed and the convergence speed can be improved significantly with $\alpha = 1/L$ where $L$ denotes a Lipschitz constant of the gradient \cite{beck_IntroductionNonlinearOptimization_2014}, which can be determined given the analytic form as a multi-dimensional Fourier series.
                Bear in mind that, while this technique performs much more efficiently and precisely than a naive high-resolution grid evaluation, its computational costs still scales exponentially in the number of (different) parameters.

    \section*{Data availability}
    Data sets generated and analyzed during the current study are available from the corresponding author upon reasonable request.
    
    \section*{Code availability}
    The source code of the ExcitationSolve algorithm is available on GitHub: \url{https://github.com/dlr-wf/ExcitationSolve}. This includes the standard algorithm for fixed and adaptive ansätze, as well as the two-dimensional variant. Some minimal examples are provided to make the experiments conducted in this work reproducible.
    
    \section*{Acknowledgments}
    We thank Jakob S. Kottmann and David Wierichs for fruitful discussions. We would also like to thank Thomas Plehn and Gabriel Breuil for helpful comments on the manuscript. Furthermore, we would like to extend our gratitude to the anonymous referees for their helpful suggestions throughout the peer review process to enhance the presentation and evaluation of our work.
    This project was made possible by the DLR Quantum Computing Initiative and the Federal Ministry for Economic Affairs and Climate Action; \url{https://qci.dlr.de/quanticom} (J.J., T.N.K., M.H., and E.S.). 
    We further acknowledge the NSERC CREATE in Quantum Computing Program, grant number 543245 (J.J.). 
    Finally, we acknowledge the use of IBM Quantum services for this work via the IBM Quantum System One Access Program funded by the Quantum Algorithms Institute (QAI) (J.J.). The views expressed are those of the authors, and do not reflect the official policy or position of IBM or the IBM Quantum team. 
    
    \section*{Author contributions}
    J.J. initiated and managed the project, and devised the main concepts and initial proofs. 
    T.N.K. worked out the large majority of the theoretical proofs. 
    All authors conceived and planned the experiments. 
    J.J. implemented the methods, while M.H., and E.S. wrote the simulation code. 
    M.H. and E.S. carried out the statevector simulation experiments. J.J. and T.N.K. carried out the quantum hardware experiments. All authors analyzed the data and contributed to the interpretation of the results. 
    All authors contributed to writing the manuscript, with J.J. and T.N.K. leading the effort.
    
    \section*{Competing interests}
    A patent application filed by the German Aerospace Center (Deutsches Zentrum für Luft- und Raumfahrt e.V., DLR), currently pending with the German Patent and Trade Mark Office (Deutsches Patent- und Markenamt, DPMA), covers aspects of this work. It specifically includes, but is not limited to, the ExcitationSolve method for fermionic excitations and fixed ansätze. The listed inventors are identical to the authors of this work. The application number is DE 10 2024 115 387.3, with the German title "Verfahren zur Bestimmung von Energien und Energiezuständen".
    The authors declare no other financial or non-financial competing interests.

    \singlespacing
    \bibliographystyle{unsrt}
    \bibliography{main}

\begin{thebibliography}{10}

\bibitem{peruzzo_VariationalEigenvalueSolver_2014}
Alberto Peruzzo, Jarrod McClean, Peter Shadbolt, Man-Hong Yung, Xiao-Qi Zhou,
  Peter~J. Love, Al{\'a}n {Aspuru-Guzik}, and Jeremy~L. O'Brien.
\newblock A variational eigenvalue solver on a photonic quantum processor.
\newblock {\em Nature Communications}, 5(1):4213, July 2014.

\bibitem{tilly2022variational}
Jules Tilly, Hongxiang Chen, Shuxiang Cao, Dario Picozzi, Kanav Setia, Ying Li,
  Edward Grant, Leonard Wossnig, Ivan Rungger, George~H Booth, et~al.
\newblock The variational quantum eigensolver: a review of methods and best
  practices.
\newblock {\em Physics Reports}, 986:1--128, 2022.

\bibitem{ma2020quantum}
He~Ma, Marco Govoni, and Giulia Galli.
\newblock Quantum simulations of materials on near-term quantum computers.
\newblock {\em npj Computational Materials}, 6(1):85, 2020.

\bibitem{bauer2020quantum}
Bela Bauer, Sergey Bravyi, Mario Motta, and Garnet Kin-Lic Chan.
\newblock Quantum algorithms for quantum chemistry and quantum materials
  science.
\newblock {\em Chemical Reviews}, 120(22):12685--12717, 2020.

\bibitem{yordanov2021qubitexcitationbased}
Yordan~S. Yordanov, V.~Armaos, Crispin H.~W. Barnes, and David R.~M.
  {Arvidsson-Shukur}.
\newblock Qubit-excitation-based adaptive variational quantum eigensolver.
\newblock {\em Communications Physics}, 4(1):1--11, October 2021.

\bibitem{xia2020qubit}
Rongxin Xia and Sabre Kais.
\newblock Qubit coupled cluster singles and doubles variational quantum
  eigensolver ansatz for electronic structure calculations.
\newblock {\em Quantum Science and Technology}, 6(1):015001, 2020.

\bibitem{gard2020efficient}
Bryan~T Gard, Linghua Zhu, George~S Barron, Nicholas~J Mayhall, Sophia~E
  Economou, and Edwin Barnes.
\newblock Efficient symmetry-preserving state preparation circuits for the
  variational quantum eigensolver algorithm.
\newblock {\em npj Quantum Information}, 6(1):10, 2020.

\bibitem{ostaszewski_StructureOptimizationParameterized_2021}
Mateusz Ostaszewski, Edward Grant, and Marcello Benedetti.
\newblock Structure optimization for parameterized quantum circuits.
\newblock {\em Quantum}, 5:391, 2021.

\bibitem{nakanishi_SequentialMinimalOptimization_2020}
Ken~M. Nakanishi, Keisuke Fujii, and Synge Todo.
\newblock Sequential minimal optimization for quantum-classical hybrid
  algorithms.
\newblock {\em Physical Review Research}, 2(4):043158, October 2020.

\bibitem{parrish_JacobiDiagonalizationAnderson_2019}
Robert~M. Parrish, Joseph~T. Iosue, Asier Ozaeta, and Peter~L. McMahon.
\newblock A {{Jacobi Diagonalization}} and {{Anderson Acceleration Algorithm
  For Variational Quantum Algorithm Parameter Optimization}}, April 2019.

\bibitem{vidal_CalculusParameterizedQuantum_2018}
Javier~Gil Vidal and Dirk~Oliver Theis.
\newblock Calculus on parameterized quantum circuits, December 2018.

\bibitem{anschuetz_QuantumVariationalAlgorithms_2022}
Eric~R. Anschuetz and Bobak~T. Kiani.
\newblock Quantum variational algorithms are swamped with traps.
\newblock {\em Nature Communications}, 13(1):7760, December 2022.

\bibitem{bittel_TrainingVariationalQuantum_2021}
Lennart Bittel and Martin Kliesch.
\newblock Training variational quantum algorithms is {{NP-hard}}.
\newblock {\em Physical Review Letters}, 127(12):120502, September 2021.

\bibitem{wang_TrainabilityEnhancementParameterized_2023}
Yabo Wang, Bo~Qi, Chris Ferrie, and Daoyi Dong.
\newblock Trainability {{Enhancement}} of {{Parameterized Quantum Circuits}}
  via {{Reduced-Domain Parameter Initialization}}, March 2023.

\bibitem{you_ExponentiallyManyLocal_2021}
Xuchen You and Xiaodi Wu.
\newblock Exponentially {{Many Local Minima}} in {{Quantum Neural Networks}}.
\newblock In {\em Proceedings of the 38th {{International Conference}} on
  {{Machine Learning}}}, pages 12144--12155. PMLR, July 2021.

\bibitem{kingma_AdamMethodStochastic_2017}
Diederik~P. Kingma and Jimmy Ba.
\newblock Adam: {{A Method}} for {{Stochastic Optimization}}, January 2017.

\bibitem{broyden1970convergence}
Charles~George Broyden.
\newblock The convergence of a class of double-rank minimization algorithms 1.
  general considerations.
\newblock {\em IMA Journal of Applied Mathematics}, 6(1):76--90, 1970.

\bibitem{fletcher1970new}
Roger Fletcher.
\newblock A new approach to variable metric algorithms.
\newblock {\em The computer journal}, 13(3):317--322, 1970.

\bibitem{goldfarb1970family}
Donald Goldfarb.
\newblock A family of variable-metric methods derived by variational means.
\newblock {\em Mathematics of computation}, 24(109):23--26, 1970.

\bibitem{shanno1970conditioning}
David~F Shanno.
\newblock Conditioning of quasi-newton methods for function minimization.
\newblock {\em Mathematics of computation}, 24(111):647--656, 1970.

\bibitem{powell1994direct}
Michael~JD Powell.
\newblock {\em A direct search optimization method that models the objective
  and constraint functions by linear interpolation}.
\newblock Springer, 1994.

\bibitem{spall1987stochastic}
James~C Spall.
\newblock A stochastic approximation technique for generating maximum
  likelihood parameter estimates.
\newblock In {\em 1987 American control conference}, pages 1161--1167. IEEE,
  1987.

\bibitem{spall1992multivariate}
James~C Spall.
\newblock Multivariate stochastic approximation using a simultaneous
  perturbation gradient approximation.
\newblock {\em IEEE transactions on automatic control}, 37(3):332--341, 1992.

\bibitem{grimsley_AdaptiveVariationalAlgorithm_2019}
Harper~R. Grimsley, Sophia~E. Economou, Edwin Barnes, and Nicholas~J. Mayhall.
\newblock An adaptive variational algorithm for exact molecular simulations on
  a quantum computer.
\newblock {\em Nature Communications}, 10(1):3007, July 2019.

\bibitem{wierichs_GeneralParametershiftRules_2022}
David Wierichs, Josh Izaac, Cody Wang, and Cedric Yen-Yu Lin.
\newblock General parameter-shift rules for quantum gradients.
\newblock {\em Quantum}, 6:677, March 2022.

\bibitem{barkoutsos_QuantumAlgorithmsElectronic_2018}
Panagiotis~Kl. Barkoutsos, Jerome~F. Gonthier, Igor Sokolov, Nikolaj Moll, Gian
  Salis, Andreas Fuhrer, Marc Ganzhorn, Daniel~J. Egger, Matthias Troyer,
  Antonio Mezzacapo, Stefan Filipp, and Ivano Tavernelli.
\newblock Quantum algorithms for electronic structure calculations:
  {{Particle-hole Hamiltonian}} and optimized wave-function expansions.
\newblock {\em Physical Review A}, 98(2):022322, August 2018.

\bibitem{watanabe_OptimizingParameterizedQuantum_2021}
Hiroshi~C. Watanabe, Rudy Raymond, Yu-Ya Ohnishi, Eriko Kaminishi, and
  Michihiko Sugawara.
\newblock Optimizing {{Parameterized Quantum Circuits}} with {{Free-Axis
  Selection}}.
\newblock In {\em 2021 {{IEEE International Conference}} on {{Quantum
  Computing}} and {{Engineering}} ({{QCE}})}, pages 100--111, October 2021.

\bibitem{watanabe_OptimizingParameterizedQuantum_2023}
Hiroshi~C. Watanabe, Rudy Raymond, Yu-Ya Ohnishi, Eriko Kaminishi, and
  Michihiko Sugawara.
\newblock Optimizing {{Parameterized Quantum Circuits With Free-Axis
  Single-Qubit Gates}}.
\newblock {\em IEEE Transactions on Quantum Engineering}, 4:1--16, 2023.

\bibitem{wada_SimulatingTimeEvolution_2022}
Kaito Wada, Rudy Raymond, Yu-ya Ohnishi, Eriko Kaminishi, Michihiko Sugawara,
  Naoki Yamamoto, and Hiroshi~C. Watanabe.
\newblock Simulating time evolution with fully optimized single-qubit gates on
  parametrized quantum circuits.
\newblock {\em Physical Review A}, 105(6):062421, June 2022.

\bibitem{wada_SequentialOptimalSelections_2024}
Kaito Wada, Rudy Raymond, Yuki Sato, and Hiroshi~C. Watanabe.
\newblock Sequential optimal selections of single-qubit gates in parameterized
  quantum circuits.
\newblock {\em Quantum Science and Technology}, 9(3):035030, May 2024.

\bibitem{kurogi_OptimizingParameterizedControlled_2024}
Hiroyoshi Kurogi, Katsuhiro Endo, Yuki Sato, Michihiko Sugawara, Kaito Wada,
  Kenji Sugisaki, Shu Kanno, Hiroshi~C. Watanabe, and Haruyuki Nakano.
\newblock Optimizing a parameterized controlled gate with {{Free Quaternion
  Selection}}, 2024.

\bibitem{slattery_UnitaryBlockOptimization_2022}
Lucas Slattery, Benjamin Villalonga, and Bryan~K. Clark.
\newblock Unitary block optimization for variational quantum algorithms.
\newblock {\em Physical Review Research}, 4(2):023072, April 2022.

\bibitem{arrazola_UniversalQuantumCircuits_2022}
Juan~Miguel Arrazola, Olivia Di~Matteo, Nicol{\'a}s Quesada, Soran Jahangiri,
  Alain Delgado, and Nathan Killoran.
\newblock Universal quantum circuits for quantum chemistry.
\newblock {\em Quantum}, 6:742, June 2022.

\bibitem{feniou_GreedyGradientfreeAdaptive_2023}
C{\'e}sar Feniou, Baptiste Claudon, Muhammad Hassan, Axel Courtat, Olivier
  Adjoua, Yvon Maday, and Jean-Philip Piquemal.
\newblock Greedy {{Gradient-free Adaptive Variational Quantum Algorithms}} on a
  {{Noisy Intermediate Scale Quantum Computer}}, September 2023.

\bibitem{boyd_ComputingZerosMaxima_2006}
John Boyd.
\newblock Computing the zeros, maxima and inflection points of {{Chebyshev}},
  {{Legendre}} and {{Fourier}} series: {{Solving}} transcendental equations by
  spectral interpolation and polynomial rootfinding.
\newblock {\em Journal of Engineering Mathematics}, 56:203--219, November 2006.

\bibitem{bartlett1989alternative}
Rodney~J Bartlett, Stanislaw~A Kucharski, and Jozef Noga.
\newblock Alternative coupled-cluster ans{\"a}tze ii. the unitary
  coupled-cluster method.
\newblock {\em Chemical physics letters}, 155(1):133--140, 1989.

\bibitem{suzuki1976generalized}
Masuo Suzuki.
\newblock Generalized trotter's formula and systematic approximants of
  exponential operators and inner derivations with applications to many-body
  problems.
\newblock {\em Communications in Mathematical Physics}, 51(2):183--190, 1976.

\bibitem{jordan1993paulische}
Pascual Jordan and Eugene~P Wigner.
\newblock {{\"U}ber das Paulische {\"A}quivalenzverbot}.
\newblock {\em Zeitschrift f{\"u}r Physik}, 47(9):631--651, September 1928.

\bibitem{bravyi_FermionicQuantumComputation_2002}
Sergey~B. Bravyi and Alexei~Yu. Kitaev.
\newblock Fermionic {{Quantum Computation}}.
\newblock {\em Annals of Physics}, 298(1):210--226, May 2002.

\bibitem{tranter_BravyiKitaevTransformation_2015}
Andrew Tranter, Sarah Sofia, Jake Seeley, Michael Kaicher, Jarrod McClean, Ryan
  Babbush, Peter~V. Coveney, Florian Mintert, Frank Wilhelm, and Peter~J. Love.
\newblock The {{Bravyi}}--{{Kitaev}} transformation: {{Properties}} and
  applications.
\newblock {\em International Journal of Quantum Chemistry}, 115(19):1431--1441,
  2015.

\bibitem{tranter2018comparison}
Andrew Tranter, Peter~J Love, Florian Mintert, and Peter~V Coveney.
\newblock A comparison of the bravyi--kitaev and jordan--wigner transformations
  for the quantum simulation of quantum chemistry.
\newblock {\em Journal of chemical theory and computation}, 14(11):5617--5630,
  2018.

\bibitem{kottmann_FeasibleApproachAutomatically_2021}
Jakob~S. Kottmann, Abhinav Anand, and Al{\'a}n {Aspuru-Guzik}.
\newblock A feasible approach for automatically differentiable unitary
  coupled-cluster on quantum computers.
\newblock {\em Chemical Science}, 12(10):3497--3508, March 2021.

\bibitem{seeley2012bravyi}
Jacob~T Seeley, Martin~J Richard, and Peter~J Love.
\newblock The bravyi-kitaev transformation for quantum computation of
  electronic structure.
\newblock {\em The Journal of chemical physics}, 137(22), 2012.

\bibitem{ramôa2024reducingresourcesrequiredadaptvqe}
Mafalda Ram{\^o}a, Panagiotis~G. Anastasiou, Luis~Paulo Santos, Nicholas~J.
  Mayhall, Edwin Barnes, and Sophia~E. Economou.
\newblock Reducing the resources required by {{ADAPT-VQE}} using coupled
  exchange operators and improved subroutines.
\newblock {\em npj Quantum Information}, 11(1):86, May 2025.

\bibitem{nooijen2000can}
Marcel Nooijen.
\newblock Can the eigenstates of a many-body hamiltonian be represented exactly
  using a general two-body cluster expansion?
\newblock {\em Physical review letters}, 84(10):2108, 2000.

\bibitem{lee2018generalized}
Joonho Lee, William~J Huggins, Martin Head-Gordon, and K~Birgitta Whaley.
\newblock Generalized unitary coupled cluster wave functions for quantum
  computation.
\newblock {\em Journal of chemical theory and computation}, 15(1):311--324,
  2018.

\bibitem{elfving_HowWillQuantum_2020}
V.~E. Elfving, B.~W. Broer, M.~Webber, J.~Gavartin, M.~D. Halls, K.~P. Lorton,
  and A.~Bochevarov.
\newblock How will quantum computers provide an industrially relevant
  computational advantage in quantum chemistry?, September 2020.

\bibitem{Utkarsh2023ChemistryComb}
Utkarsh Azad.
\newblock Pennylane quantum chemistry datasets.
\newblock \url{https://pennylane.ai/datasets/qchem/h2-molecule},
  \url{https://pennylane.ai/datasets/qchem/h3-plus-molecule},
  \url{https://pennylane.ai/datasets/qchem/lih-molecule},
  \url{https://pennylane.ai/datasets/qchem/h2o-molecule}, 2023.

\bibitem{larocca_ReviewBarrenPlateaus_2024}
Martin Larocca, Supanut Thanasilp, Samson Wang, Kunal Sharma, Jacob Biamonte,
  Patrick~J. Coles, Lukasz Cincio, Jarrod~R. McClean, Zo{\"e} Holmes, and
  M.~Cerezo.
\newblock A {{Review}} of {{Barren Plateaus}} in {{Variational Quantum
  Computing}}, May 2024.

\bibitem{giner_BasissetErrorCorrection_2020}
Emmanuel Giner, Anthony Scemama, Pierre-Fran{\c c}ois Loos, and Julien
  Toulouse.
\newblock A basis-set error correction based on density-functional theory for
  strongly correlated molecular systems.
\newblock {\em The Journal of Chemical Physics}, 152(17):174104, May 2020.

\bibitem{mitarai_QuantumCircuitLearning_2018}
K.~Mitarai, M.~Negoro, M.~Kitagawa, and K.~Fujii.
\newblock Quantum circuit learning.
\newblock {\em Physical Review A}, 98(3):032309, September 2018.

\bibitem{schuld_EvaluatingAnalyticGradients_2019}
Maria Schuld, Ville Bergholm, Christian Gogolin, Josh Izaac, and Nathan
  Killoran.
\newblock Evaluating analytic gradients on quantum hardware.
\newblock {\em Physical Review A}, 99(3):032331, March 2019.

\bibitem{anselmetti_LocalExpressiveQuantumnumberpreserving_2021}
Gian-Luca~R Anselmetti, David Wierichs, Christian Gogolin, and Robert~M
  Parrish.
\newblock Local, expressive, quantum-number-preserving {{VQE}} ans{\"a}tze for
  fermionic systems.
\newblock {\em New Journal of Physics}, 23(11):113010, November 2021.

\bibitem{bravyi_TaperingQubitsSimulate_2017}
Sergey Bravyi, Jay~M. Gambetta, Antonio Mezzacapo, and Kristan Temme.
\newblock Tapering off qubits to simulate fermionic {{Hamiltonians}}, January
  2017.

\bibitem{setia_ReducingQubitRequirements_2020}
Kanav Setia, Richard Chen, Julia~E. Rice, Antonio Mezzacapo, Marco Pistoia, and
  James~D. Whitfield.
\newblock Reducing {{Qubit Requirements}} for {{Quantum Simulations Using
  Molecular Point Group Symmetries}}.
\newblock {\em Journal of Chemical Theory and Computation}, 16(10):6091--6097,
  October 2020.

\bibitem{wiersema_HereComesMathrmSU_2023}
Roeland Wiersema, Dylan Lewis, David Wierichs, Juan Carrasquilla, and Nathan
  Killoran.
\newblock Here comes the {SU}({N}): multivariate quantum gates and gradients.
\newblock {\em {Quantum}}, 8:1275, March 2024.

\bibitem{Barison2021efficientquantum}
Stefano Barison, Filippo Vicentini, and Giuseppe Carleo.
\newblock An efficient quantum algorithm for the time evolution of
  parameterized circuits.
\newblock {\em {Quantum}}, 5:512, July 2021.

\bibitem{Benedetti2021HardwareEfficient}
Marcello Benedetti, Mattia Fiorentini, and Michael Lubasch.
\newblock Hardware-efficient variational quantum algorithms for time evolution.
\newblock {\em Phys. Rev. Res.}, 3:033083, Jul 2021.

\bibitem{Jaderberg_2020}
B~Jaderberg, A~Agarwal, K~Leonhardt, M~Kiffner, and D~Jaksch.
\newblock Minimum hardware requirements for hybrid quantum–classical dmft.
\newblock {\em Quantum Science and Technology}, 5(3):034015, jun 2020.

\bibitem{boyd2006computing}
John~P Boyd.
\newblock Computing the zeros, maxima and inflection points of chebyshev,
  legendre and fourier series: solving transcendental equations by spectral
  interpolation and polynomial rootfinding.
\newblock {\em Journal of Engineering Mathematics}, 56:203--219, 2006.

\bibitem{shannon1949communication}
Claude~Elwood Shannon.
\newblock Communication in the presence of noise.
\newblock {\em Proceedings of the IRE}, 37(1):10--21, 1949.

\bibitem{beck_IntroductionNonlinearOptimization_2014}
Amir Beck.
\newblock {\em Introduction to Nonlinear Optimization: Theory, Algorithms, and
  Applications with {{MATLAB}}}.
\newblock {{MOS-SIAM}} Series on Optimization. {Society for Industrial and
  Applied Mathematics : Mathematical Optimization Society}, Philadelphia, 2014.

\bibitem{bergholm2022pennylaneautomaticdifferentiationhybrid}
Ville Bergholm, Josh Izaac, Maria Schuld, Christian Gogolin, Shahnawaz Ahmed,
  Vishnu Ajith, M.~Sohaib Alam, Guillermo Alonso-Linaje, B.~AkashNarayanan, Ali
  Asadi, Juan~Miguel Arrazola, Utkarsh Azad, Sam Banning, Carsten Blank,
  Thomas~R Bromley, Benjamin~A. Cordier, Jack Ceroni, Alain Delgado, Olivia~Di
  Matteo, Amintor Dusko, Tanya Garg, Diego Guala, Anthony Hayes, Ryan Hill,
  Aroosa Ijaz, Theodor Isacsson, David Ittah, Soran Jahangiri, Prateek Jain,
  Edward Jiang, Ankit Khandelwal, Korbinian Kottmann, Robert~A. Lang, Christina
  Lee, Thomas Loke, Angus Lowe, Keri McKiernan, Johannes~Jakob Meyer, J.~A.
  Montañez-Barrera, Romain Moyard, Zeyue Niu, Lee~James O'Riordan, Steven Oud,
  Ashish Panigrahi, Chae-Yeun Park, Daniel Polatajko, Nicolás Quesada, Chase
  Roberts, Nahum Sá, Isidor Schoch, Borun Shi, Shuli Shu, Sukin Sim, Arshpreet
  Singh, Ingrid Strandberg, Jay Soni, Antal Száva, Slimane Thabet, Rodrigo~A.
  Vargas-Hernández, Trevor Vincent, Nicola Vitucci, Maurice Weber, David
  Wierichs, Roeland Wiersema, Moritz Willmann, Vincent Wong, Shaoming Zhang,
  and Nathan Killoran.
\newblock Pennylane: Automatic differentiation of hybrid quantum-classical
  computations, 2022.

\bibitem{Kandala2017}
Abhinav Kandala, Antonio Mezzacapo, Kristan Temme, Maika Takita, Markus Brink,
  Jerry~M. Chow, and Jay~M. Gambetta.
\newblock Hardware-efficient variational quantum eigensolver for small
  molecules and quantum magnets.
\newblock {\em Nature}, 549(7671):242–246, September 2017.

\bibitem{Spall1998}
James~C. Spall.
\newblock An overview of the simultaneous perturbation method for efficient
  optimization.
\newblock {\em Hopkins APL Technical Digest}, 19(4):482–492, 1998.

\bibitem{Powell1994}
M.J. Powell.
\newblock A direct search optimization method that models the objective and
  constraint functions by linear interpolation.
\newblock {\em Advances in Optimization and Numerical Analysis}, pages 51--67,
  1994.

\bibitem{kim_ScalableErrorMitigation_2023}
Youngseok Kim, Christopher~J. Wood, Theodore~J. Yoder, Seth~T. Merkel, Jay~M.
  Gambetta, Kristan Temme, and Abhinav Kandala.
\newblock Scalable error mitigation for noisy quantum circuits produces
  competitive expectation values.
\newblock {\em Nature Physics}, 19(5):752--759, May 2023.

\bibitem{mckay_Efficient$Z$Gates_2017}
David~C. McKay, Christopher~J. Wood, Sarah Sheldon, Jerry~M. Chow, and Jay~M.
  Gambetta.
\newblock Efficient \${{Z}}\$ gates for quantum computing.
\newblock {\em Physical Review A}, 96(2):022330, August 2017.

\bibitem{shawe-taylor_KernelMethodsPattern_2004}
John {Shawe-Taylor} and Nello Cristianini.
\newblock {\em Kernel {{Methods}} for {{Pattern Analysis}}}.
\newblock Cambridge University Press, Cambridge, 2004.

\bibitem{murphy_ProbabilisticMachineLearning_2022}
Kevin~P. Murphy.
\newblock {\em Probabilistic Machine Learning: {{An}} Introduction}.
\newblock MIT Press, 2022.

\bibitem{nielsen_QuantumComputationQuantum_2010}
Michael~A. Nielsen and Isaac~L. Chuang.
\newblock {\em Quantum Computation and Quantum Information}.
\newblock Cambridge University Press, Cambridge ; New York, 10th anniversary ed
  edition, 2010.

\bibitem{sung_NonGaussianNoiseSpectroscopy_2019}
Youngkyu Sung, F{\'e}lix Beaudoin, Leigh~M. Norris, Fei Yan, David~K. Kim,
  Jack~Y. Qiu, Uwe {von L{\"u}pke}, Jonilyn~L. Yoder, Terry~P. Orlando, Simon
  Gustavsson, Lorenza Viola, and William~D. Oliver.
\newblock Non-{{Gaussian}} noise spectroscopy with a superconducting qubit
  sensor.
\newblock {\em Nature Communications}, 10(1):3715, September 2019.

\bibitem{endo_OptimalParameterConfigurations_2023}
Katsuhiro Endo, Yuki Sato, Rudy Raymond, Kaito Wada, Naoki Yamamoto, and
  Hiroshi~C. Watanabe.
\newblock Optimal parameter configurations for sequential optimization of the
  variational quantum eigensolver.
\newblock {\em Physical Review Research}, 5(4):043136, November 2023.

\bibitem{stein_FourierAnalysisIntroduction_2003}
Elias~M. Stein and Rami Shakarchi.
\newblock {\em Fourier Analysis: An Introduction}.
\newblock Number~1 in Princeton Lectures in Analysis / {{Elias M}}. {{Stein}}
  \& {{Rami Shakarchi}}. Princeton University Press, Princeton Oxford, 15.
  druck edition, 2003.

\bibitem{hastie2009elements}
Trevor Hastie, Robert Tibshirani, Jerome~H Friedman, and Jerome~H Friedman.
\newblock {\em The Elements of Statistical Learning: Data Mining, Inference,
  and Prediction}, volume~2.
\newblock Springer, 2009.

\bibitem{evangelista2019exact}
Francesco~A Evangelista, Garnet~Kin Chan, and Gustavo~E Scuseria.
\newblock Exact parameterization of fermionic wave functions via unitary
  coupled cluster theory.
\newblock {\em The Journal of chemical physics}, 151(24), 2019.

\bibitem{xu2020test}
Luogen Xu, Joseph~T Lee, and JK~Freericks.
\newblock Test of the unitary coupled-cluster variational quantum eigensolver
  for a simple strongly correlated condensed-matter system.
\newblock {\em Modern Physics Letters B}, 34(19n20):2040049, 2020.

\bibitem{chen2021quantum}
Jia Chen, Hai-Ping Cheng, and James~K Freericks.
\newblock Quantum-inspired algorithm for the factorized form of unitary coupled
  cluster theory.
\newblock {\em Journal of Chemical Theory and Computation}, 17(2):841--847,
  2021.

\bibitem{freericks2022operator}
James~K Freericks.
\newblock Operator relationship between conventional coupled cluster and
  unitary coupled cluster.
\newblock {\em Symmetry}, 14(3):494, 2022.

\bibitem{PhysRevB.79.174515}
Jacob Jordan, Rom\'an Or\'us, and Guifr\'e Vidal.
\newblock Numerical study of the hard-core bose-hubbard model on an infinite
  square lattice.
\newblock {\em Phys. Rev. B}, 79:174515, May 2009.

\bibitem{wu2002qubits}
L-A Wu and DA~Lidar.
\newblock Qubits as parafermions.
\newblock {\em Journal of Mathematical Physics}, 43(9):4506--4525, 2002.

\bibitem{ferris2022quantum}
Kaelyn~J. Ferris, A.~J. Rasmusson, Nicholas~T. Bronn, and Olivia Lanes.
\newblock Quantum simulation on noisy superconducting quantum computers, 2022.

\bibitem{armaos_EfficientParabolicOptimisation_2021}
V.~Armaos, Dimitrios~A. Badounas, Paraskevas Deligiannis, Konstantinos Lianos,
  and Yordan~S. Yordanov.
\newblock Efficient {{Parabolic Optimisation Algorithm}} for adaptive {{VQE}}
  implementations, October 2021.

\bibitem{li_EfficientRobustParameter_2024}
Weitang Li, Yufei Ge, Shixin Zhang, Yuqin Chen, and Shengyu Zhang.
\newblock Efficient and {{Robust Parameter Optimization}} of the {{Unitary
  Coupled-Cluster Ansatz}}, January 2024.

\end{thebibliography}

    \clearpage
    \appendix

    \part*{Appendix}\addcontentsline{toc}{part}{Appendix} 
    
    \localtableofcontents
          
    \section{Experimental setup}\label{app:experimental_setup}
    
        The majority of the experiments were implemented in Python using the quantum computing framework \texttt{Pennylane} \cite{bergholm2022pennylaneautomaticdifferentiationhybrid}.
            
        \subsection{Hyperparameter tuning and calibration} %
        \label{sec:methods_hyperparam}
            For both GD and Adam we perform a hyperparameter tuning for the step size. For Adam we set the momentum parameters to $\beta_1=\num{0.9}$ and $\beta_2=\num{0.99}$ and keep them fixed for every step size. We consider the following step sizes: $0.5\times 10^n$ and $2.5\times10^n$ for $n\in\{-4, -3, -2,-1, 0 \}$. Figure \ref{fig:hyperparams_gd_adam} in the appendix shows the performance of the investigated step sizes. SPSA also has two tunable hyperparameters, the learning rate and the perturbation for the gradient approximation. We use the SPSA implementation in \texttt{qiskit} \cite{Kandala2017} and use its calibration function to tune both hyperparameters before starting the VQE optimization at the constant cost of $\num{50}$ energy evaluations \cite{Spall1998}. To account for this calibration phase, we plot the Hartree-Fock energy for the first $\num{50}$ SPSA energy evaluations for each molecule. We find that tuning the hyperparameters of COBYLA has negligible impact on the convergence and, therefore, use the default parameters of the \texttt{SciPy} implementation \cite{Powell1994}. For the BFGS optimizer we used the \texttt{SciPy} implementation which has no hyperparameters. We emphasize again that ExcitationSolve has no hyperparameters that need to be tuned.

        \subsection{Fixed UCCSD ansatz}
        \label{sec:methods_fixed_ansatz}
            We choose the UCCSD ansatz in its first Trotter-approximation and use the STO-3G basis set as provided by the Pennylane datasets \cite{Utkarsh2023ChemistryComb} and the Jordan-Wigner (JW) mapping \cite{jordan1993paulische}. The single- and double excitations are ordered as provided by Pennylane \cite{bergholm2022pennylaneautomaticdifferentiationhybrid} (version \texttt{0.37.0}).
            The system is initialized in the HF state and zero parameters. In each VQE iteration, all parameters are optimized in the order in which they appear in the ansatz (first the double-, then the single-excitations), unless explicitly stated otherwise. The step sizes for each molecule determined through hyperparameter tuning are listed in \Cref{tab:stepsizes}.
            \begin{table}[htb]
                \centering
                \caption{Optimal step sizes for the GD and Adam optimizers.}
                \begin{tabular}[t]{lcc}
                \toprule
                & \textbf{GD} & \textbf{Adam} \\ \midrule
                \ce{H2}  & 0.5           & 0.005         \\ 
                \ce{H3+} & 0.5           & 0.005         \\ 
                \ce{LiH} & 0.25         & 0.005         \\ 
                \ce{H2O} & 0.025        & 0.0025        \\ 
                \bottomrule
                \end{tabular}
                \label{tab:stepsizes}
            \end{table}

        \subsection{Adaptive Ansatz}
        \label{sec:methods_adapt}

            \begin{table}[htb]
                \centering
                \caption{Optimal threshold values in $\si{\Ha}$ for operator selection and VQE convergence. For ExcitationSolve the threshold is an absolute energy difference between evaluations, for GD the threshold is a gradient.}
                \begin{tabular}[t]{lcc|cc}
                \toprule
                & \multicolumn{2}{c}{\textbf{ExcitationSolve}} & \multicolumn{2}{c}{\textbf{ADAPT-VQE}} \\ \midrule
                & parameter optimization & operator selection & parameter optimization  & operator selection\\ \midrule
                \ce{H2}  & $10^{-6}$ &$10^{-6}$ & $2 \times 10^{-13}$ &$2 \times 10^{-13}$ \\ 
                \ce{H3+} & $10^{-6}$ &$10^{-6}$ & $2 \times 10^{-13}$ &$2 \times 10^{-8}$ \\ 
                \ce{LiH} & $10^{-7}$ &$10^{-7}$ & $2 \times 10^{-7}$ &$2 \times 10^{-7}$ \\ 
                \ce{H2O} & $10^{-6}$ &$10^{-6}$ & $2 \times 10^{-8}$ &$2 \times 10^{-8}$ \\ 
                \bottomrule
                \end{tabular}
                \label{tab:thresholds}
            \end{table}
        
            The experimental setting follows the one of the experiments with the fixed UCCSD ansatz, except that the optimization starts with an empty ansatz. Instead, the fermionic excitation operators from the UCCSD ansatz in its first Trotter-approximation constitute the operator pool for ADAPT-VQE. We employ pool draining, which means that once an operator was selected to extend the ansatz, it is removed from the pool and cannot be used again -- thus, the number of ADAPT steps is limited by the size of the pool. We also set a threshold that no more operators are attached when their impact is less than a  set threshold value. This value has to be tuned for each molecule individually to achieve optimal convergence. For ExcitationSolve the threshold is an absolute energy difference, for ADAPT-VQE the threshold is a gradient. The chosen values are shown in \Cref{tab:thresholds}.\\
            For ExcSolve2D, the two operators that have the biggest impact are selected, optimized using 2D ExcitationSolve and then appended to the ansatz. As we already calculated the energies for the individual operators in the ranking, we can reuse those in the optimization. The five single parameter values can be viewed as the shifts of $\theta_i$ where $\theta_j=0$ and vice versa. The value for $\theta_i=\theta_j=0$ appears in both selections, so only nine values can be recycled. This leaves leaves $25-9=16$ energy evaluations to be done. Overall we calculate the energies for 24 sets of $(\theta_i,\ \theta_j)$, but the effort is split between operator ranking and actual optimization of the chosen operators.\\
            After each ADAPT step, all parameters are re-optimized until convergence, the corresponding thresholds can again be found in \Cref{tab:thresholds}, they are the same in 1D and 2D ExcitationSolve. ExcitationSolve follows the parameter order in which they were attached.
            The step sizes for GD in the original ADAPT-VQE counterpart are the same as for the fixed ansatz listed in \Cref{tab:stepsizes}.

        \subsection{ExcitationSolve 2D optimization} \label{sec:method_2D_disscurve}
            In the 2D optimization variant of ExcitationSolve, we first perform a sweep over all parameters and rank them based on the difference between their global energy minimum and the Hartree-Fock~(HF) energy. This assesses the immediate impact of the operators on improving the HF error. During this initial sweep no parameter values are updated. The top two most impactful parameters are first simultaneously optimized in each VQE iteration, as explained in \Cref{sec:multi_param_gen}, while afterwards all other parameters are optimized independently, i.e., using standard 1D ExcitationSolve. In the evaluation and plotting of the experiments, the initial sweep is included in the total number of energy evaluations, treated equivalently to a VQE iteration in the 1D ExcitationSolve optimization.
            
        \subsection{Dissociation experiments}
        \label{sec:methods_strong_corr}
            We use a fixed UCCSD ansatz in its first Trotter-approximation with fermionic excitations with a Hamiltonian representation in the STO-3G basis set, using all available bond lengths in the Pennylane datasets \cite{Utkarsh2023ChemistryComb}. The initial state is again the HF state for all configurations. The hyperparameters for the used optimizers are the same as in \Cref{subsec:fixedAnsatz}.
            Convergence is defined as approaching the FCI energy up to a certain threshold. As not all molecules converge to the same accuracy, the threshold is specified for each molecule individually depending on the lowest accuracy achieved over all bond lengths (see \Cref{tab:convergenceThresholds}).
            
            \begin{table}[htb]
                \centering
                \caption{Convergence threshold values for each molecule}
                \begin{tabular}[t]{ll}
                \toprule
                & \textbf{Threshold}  \\ \midrule
                \ce{H2}  & $\SI{5.78E{-12}}{\Ha}$            \\ 
                \ce{H3+} & $\SI{2.64E{-13}}{\Ha}$           \\ 
                \ce{LiH} & $\SI{2.17E{-5}}{\Ha}$     \\ 
                \ce{H2O} & $\SI{2.27E{-3}}{\Ha}$        \\ 
                \bottomrule
                \end{tabular}
                \label{tab:convergenceThresholds}
            \end{table}
        
        \subsection{Shot noise simulations}
        \label{sec:methods_shot_noise}
            The experimental setup is identical to that of the fixed UCCSD ansatz experiments, except that a finite number of shots is used, resulting in energy evaluations being based on estimates rather than exact values. 
            We find that when defining a fixed shot budget, changing the number of shots per energy evaluation or using more energy values for reconstructing the energy landscape has negligible impact on the convergence of ExcitationSolve. Therefore, we use $5$ energy values, where we do not reuse the energy for the parameter value $\theta=0$ but estimate this energy every time by executing the quantum circuit. %
            For COBYLA and BFGS their default termination schemes in the respective \texttt{scipy} implementations are used. The optimization of gradient descent and ExcitationSolve was automatically stopped if the energy after the current VQE iteration was larger than the energies after the last two VQE iterations. We use the parameter-shift rule for all gradient-based optimizers.

        \subsection{NISQ hardware experiments}\label{sec:method_NISQ}
            
            This section describes the IBM-Q implementation specifics and particularly highlights deviations from the fixed and adaptive ansatz experiments in simulation, as previously detailed in Sec.~\ref{sec:methods_fixed_ansatz} and Sec.~\ref{sec:methods_adapt}, respectively. The execution of quantum circuits on a real NISQ device as opposed to a simulation involves two additional steps, which are the \emph{transpilation} to map the circuit to the hardware specification and constraints, as well as the \emph{error mitigation and suppression} to counteract errors occurring in the execution due to the noisy hardware realization. The specifics are described as follows.
            
            For the transpilation, i.e., the compilation of the quantum circuits from the excitation operator ansatz to the \texttt{ibm\_quebec} native gate set $\left\lbrace \mathrm{ECR}, I, R_Z(\cdot), \sqrt{X}, X \right\rbrace$ subject to the \texttt{ibm\_quebec} coupling map, the \texttt{qiskit} transpiler service was utilized. The transpilation was configured by setting the \texttt{optimization\_level} to the maximum level 3 to achieve most optimized circuits at the cost of longer compilation times as part of the pre-processing on the classical computer.
            Importantly, the transpilation does not only consider the native gate sets and connectivity of the IBM-Q backend into account but optimizes with respect to the gate errors based on the most recent calibration run provided by the \texttt{qiskit} runtime service. Especially due to the latter fact, transpilations of the same circuit or ansatz at different times may lead to different outcomes.
            For all optimizers, close-to-zero parameters are ignored as their expected impact is negligible by removing the associated operator from the circuit before transpilation. This can significantly enhance the energy estimates obtained from the then shallower transpiled circuits. The zero threshold is $\sfrac{1}{8}$, which sends parameters in the interval $[-1/8, 1/8]\approx [-0.04\pi, 0.04\pi]$, i.e., 4\% of the total period of $2\pi$, to zero.
            Despite the maximum optimization level, the \ce{H3+} ansatz was transpiled into exceedingly deep circuits. Therefore, the fermionic double excitations were transpiled manually.
            We quickly illustrate the technique at hand of a
            qubit double excitation from orbitals $p,q$ to $r,s$. We write the generator arising from the JW transformation as
            \begin{align}
                G = \frac{1}{8} X_p X_q X_r Y_s & (I_p I_q I_r I_s + I_p I_q Z_r Z_s - I_p Z_q I_r Z_s - I_p Z_q Z_r I_s \\
                &- Z_p I_q I_r Z_s - Z_p I_q Z_r I_s + Z_p Z_q I_r I_s + Z_p Z_q Z_r Z_s).\nonumber
            \end{align}
            This product decomposition is specifically designed such that the $XXXY$ string commutes with the remaining $Z$-terms. This would not be the case if one were to factor out, e.g,~ $XXYY$. Since the Hartree-Fock ground state is represented by only one computational basis state, the $Z$-terms can be replaced using the eigenvalue relations, therefore reducing the effective generator $G$ to a single Pauli rotation with $G\propto XXXY$. For subsequent excitations, the number of removable strings decreases, until the full 8 strings of $G$ are unavoidable. Thus, the technique is only relevant for the first few excitations. 
            The standard gate decomposition for the single excitations remained. In addition, all Clifford gates in the decomposed circuit were collected and, if at the end of the circuit, absorbed into the Hamiltonian or, otherwise, optimally (heuristically) re-synthesized when acting on at most 3 (more than 3) qubits.
            
            For the error mitigation, the default settings in the \texttt{qiskit} runtime for the highest robustness are activated through the currently maximal \texttt{resilience\_level} of 2. This involves readout error mitigation, Pauli twirling, and zero noise extrapolation. In addition, probabilistic error amplification (PEA) as a more recent and sophisticated noise amplification technique in ZNE \cite{kim_ScalableErrorMitigation_2023} is performed with the noise factors $\lbrace 1, 1.5, 2 \rbrace$. Error suppression is achieved by activating dynamical decoupling, which aims to protect idling qubits through noise-canceling pulse sequences (set to a pulse sequence of two Pauli-X with opposite phases). Idling qubits are particularly common in the transpiled excitation operator circuits due to the high number of two-qubit (ECR) gates with execution times that typically are significantly higher than single-qubit gates. Unless otherwise stated, $8192$ shots are spent per energy evaluation.

            To limit error propagation, ExcitationSolve is based on five energy evaluations as opposed to of four in simulation to incorporate an energy evaluation of the unshifted parameters instead of re-using the previously determined optimal energy, which match in theory.
            Since (approximately) non-unique minima can occur in the analytic energy reconstructions, ExcitationSolve is adapted to pick the closer one. 
            This choice can be justified with the assumption of studying weakly correlated systems at the molecule equilibrium geometries. 
            Although $Z$-rotations are implemented virtually on IBM devices \cite{mckay_Efficient$Z$Gates_2017} and thus do not contribute to hardware noise independent of the angle magnitude, selecting the smaller angle reduces circuit depth by skipping the transpilation into additional physical gates in case of parameters close to zero (according to the zero threshold).
            Furthermore, the coefficients of the two lower frequencies in the analytic energy reconstructions for ExcitationSolve are thresholded at \num{5e-2} (refit under the corresponding frequency coefficients being set to zero) and if both are set to zero, the period in which the optima are analyzed is halved.
            The step size in GD is reduced by $\sfrac{1}{5}$ compared to the ones used in simulation. This more conservative choice showed a decrease in the sensitivity to noisy gradients and the associated risk of divergence in preliminary experiments. To increase the comparability, ExcitationSolve and GD are based on the same parameter equidistant shifts for the analytic energy reconstruction and (four-term) parameter-shift rule, respectively. For the sake of clarity in presenting the experimental results, the number of energy evaluations per parameter for GD is matched with ExcitationSolve to five.
        
            The implementation of the \ce{LiH} experiments on the IBM-Q device differs from the simulated ones in two size-reducing aspects to compensate for noise further as follows. Hereby, some excitation operators can be removed or at least some spin orbitals (qubits) they act on can be removed, which implicitly decreases the depth of the transpiled circuits. First, certain (occupied) orbitals are \emph{frozen}, which means that these orbitals will be fully occupied. Thus, these orbitals can be excluded from the VQE calculation because they contribute with a constant energy, which can be efficiently computed from the Hamiltonian $H$ classically. For \ce{LiH}, the lowest two, so-called \emph{core}, spin orbitals are frozen. Second, \emph{qubit tapering} \cite{bravyi_TaperingQubitsSimulate_2017,setia_ReducingQubitRequirements_2020} further decreases the qubit requirements below one qubit per spin orbital as per the Jordan-Wigner (JW) mapping \cite{jordan1993paulische}, which results in six final qubits and ten UCCSD excitation operator \ce{LiH}. Neither technique is employed for \ce{H2} and \ce{H3+} such that the ansatz exactly matches the full UCCSD one of the simulations.
    
    \clearpage
    \section{Supplementary experimental results and analyses}

    \label{app:supp_experiments}
    
        The following material provides further details and insight into the performed experiments and analysis of ExcitationSolve.

    \subsection{Hyperparameter tuning results}
\begin{figure}[!hpt]
                \centering
                \begin{subfigure}[t]{0.49\textwidth}
                    \centering
                    \includegraphics[width=\textwidth]{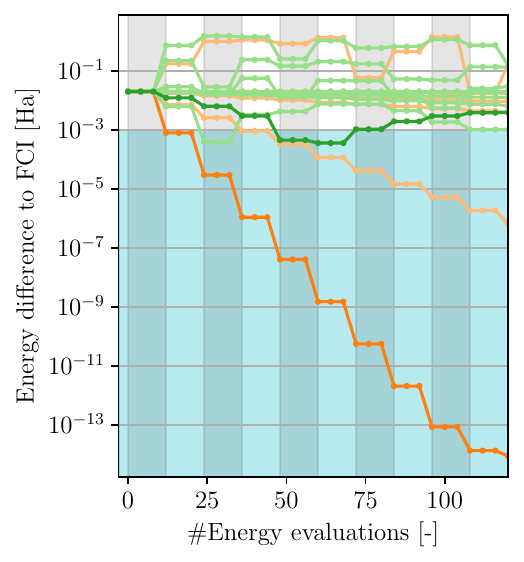}
                    \vspace{-0.75cm}
                    \caption{\ce{H2}, 4 qubits.}
                    \label{fig:hyperparam_H2}
                \end{subfigure}
                \hfill
                \begin{subfigure}[t]{0.49\textwidth}
                    \centering
                    \includegraphics[width=\textwidth]{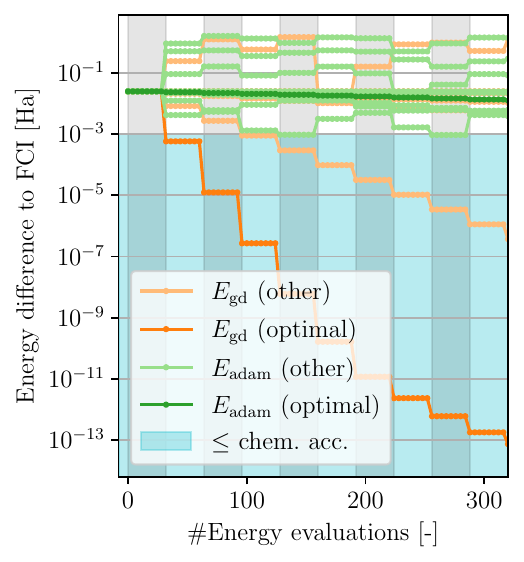}
                    \vspace{-0.75cm}
                    \caption{\ce{H3+}, 6 qubits.}
                    \label{fig:hyperparam_H3plus}
                \end{subfigure}
                \hfill
                \begin{subfigure}[t]{0.49\textwidth}
                    \centering
                    \includegraphics[width=\textwidth]{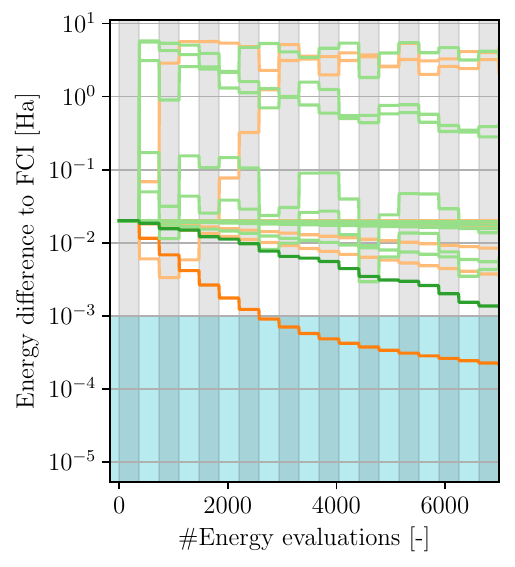}
                    \vspace{-0.75cm}
                    \caption{\ce{LiH}, 12 qubits.}
                    \label{fig:hyperparam_LiH}
                \end{subfigure}
                \hfill
                \begin{subfigure}[t]{0.49\textwidth}
                    \centering
                    \includegraphics[width=\textwidth]{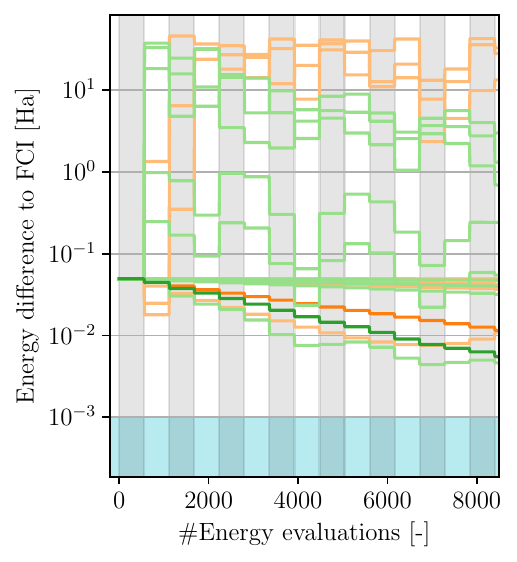}
                    \vspace{-0.75cm}
                    \caption{\ce{H2O}, 14 qubits.}
                    \label{fig:hyperparam_H2O}
                \end{subfigure}
                \caption{\textbf{Comparison of optimizer step sizes.} The optimizers under consideration are Gradient descent (yellow) and Adam (green). The plots show the error of the VQE with respect to the FCI solution
                ${|E_{\text{VQE}}-E_{\text{FCI}}|}$ over the number of energy evaluations for all tested step sizes. The optimal step size is highlighted. The light blue region signifies the chemical accuracy ($\SI{e-3}{\Ha}$) and the alternating vertical shading marks each iteration over all parameters.}
                \label{fig:hyperparams_gd_adam}
            \end{figure}

        As found in preliminary experiments, the performance of the gradient-based optimizers is susceptible to a well-tuned step size hyperparameter. We consider the following step sizes: $0.5\times 10^n$ and $2.5\times10^n$ for $n\in\{-4, -3, -2,-1, 0 \}$. The detailed results for the different runs for the performed hyperparameter tuning as outlined in Sec.~\ref{sec:methods_hyperparam} are presented here in Fig.~\ref{fig:hyperparams_gd_adam}.

    \subsection{ADAPT-VQE resource comparison}\label{app:adapt_vqe_resource_comp}
\begin{figure}
            \centering
            \begin{tikzpicture}%
                \node[inner sep=0pt] (gridplot) at (0,0)
                {\includegraphics[width=1\textwidth]{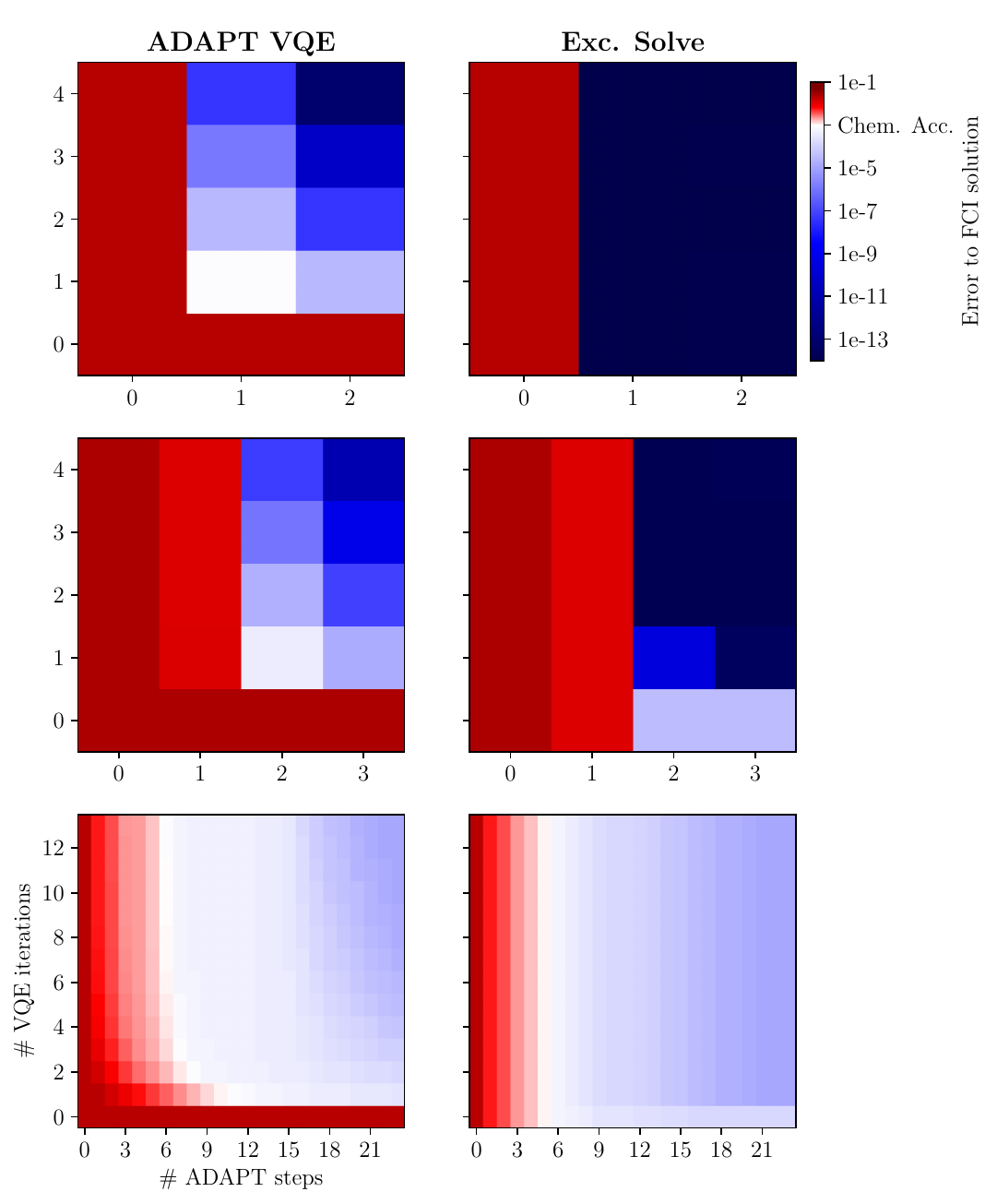}};
                \node[] at (-9,6.5) {\textbf{\ce{H2}}};
                \node[] at (-9,0) {\textbf{\ce{H3+}}};
                \node[] at (-9,-6.5) {\textbf{\ce{LiH}}};
            \end{tikzpicture}
            \caption{\textbf{Evaluation of ExcitationSolve for ADAPT-VQE.} ExcitationSolve (right) in an adaptive setting compared with the original ADAPT-VQE (left) based on GD for molecules (top to bottom) \ce{H2}, \ce{H3+}, \ce{LiH}. On both axes are the resources spent on the calculation: The number of ADAPT steps signals how many operators have been appended to the ansatz, the number of VQE iterations indicates how often each of the parameters has been optimized in each ADAPT step. The color code signals how close the result is to the exact FCI solution.}
            \label{fig:Adapt_resources}
        \end{figure}

        \Cref{fig:Adapt_resources} displays how well ExcitationSolve and ADAPT-VQE converge to the ground state given a fixed amount of computational resources for the following molecules: \ce{H2}, \ce{H3+}, \ce{LiH}. On the x-axis is the number of operators that have been attached to the ansatz. As we employ pool-draining, this number is limited by the number of operators in the pool for each molecule. The y-axis shows how often each parameter is re-optimized after each ADAPT-step. Here the threshold is set by a convergence criterion that the change in energy between re-optimizations must be larger than \SI{E{-6}}{\Ha}. An analysis for \ce{H2O} has been spared due to exceeding reasonable computational time.
        
        Apart from the overall faster convergence that can be inferred from Fig.~\ref{fig:Adapt_Benchmark_combined}, ExcitationSolve particularly shines regarding the initialization/warm-start strategy: ExcitationSolve initializes the newly attached operator in its optimal configuration, so it immediately has an impact without any additional energy evaluation overhead. In contrast, the standard ADAPT-VQE initialization sets the parameter to zero, which does not have any direct impact on the energy but heavily relies on the subsequent VQE re-optimization. Indeed, we find that while the GD based algorithm always needs to re-optimize its parameters, further VQE iterations rarely have a substantial impact when using ExcitationSolve.
        Importantly, this detailed analysis also reveals that ExcitationSolve is capable of finding shorter ansätze, in the case of the more complex molecule \ce{LiH}, than standard ADAPT-VQE (compare the horizontal transition through the white color, marking chemical accuracy, in the bottom two plots of Fig.~\ref{fig:Adapt_resources} in a region where the VQE re-optimization is converged.) This hints towards the advantage of the globally-informed operator selection criterion with ExcitationSolve as opposed to the local gradient-based criterion in the original ADAPT-VQE.
        
        \subsection{ADAPT-VQE impact analysis of operator selection criterion vs parameter optimizer} \label{app:mixing}
\begin{figure}[!hpt]
            \centering
            \begin{subfigure}[t]{0.49\textwidth}
                \centering \includegraphics[width=\textwidth]{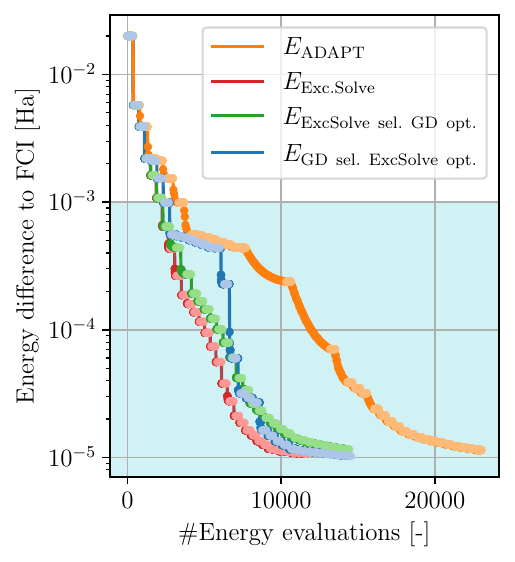}
                \vspace{-0.75cm}
                \caption{\ce{LiH}, $12$ qubits. 
                }
                \label{fig:adapt_mixed_optimizers_LiH}
                \vspace{0.4cm}
            \end{subfigure}
            \hfill
            \begin{subfigure}[t]{0.49\textwidth}
                \centering \includegraphics[width=\textwidth]{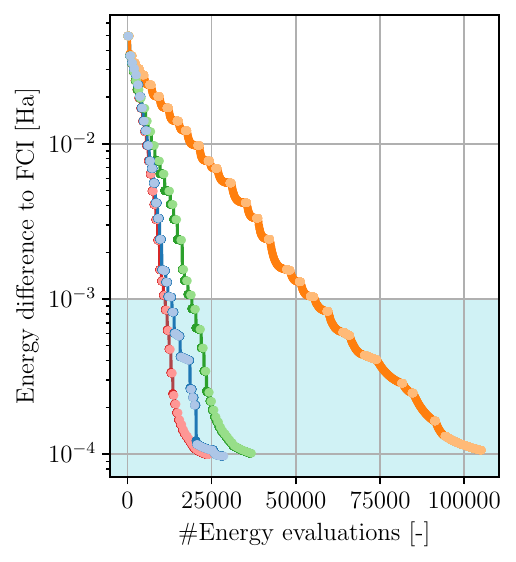}
                \vspace{-0.75cm}
                \caption{\ce{H2O}, $14$ qubits.
                }
                \label{fig:adapt_mixed_optimizers_H2O}
            \end{subfigure}
            \hfill
            
            \caption{\textbf{Comparison of the impact of optimizer and selection method for adaptive ansätze.} In addition to the standard ExcitationSolve and ADAPT-VQE optimization, one setting uses ExcitationSolve only for the operator selection but GD for the parameter optimization (green), and, vice versa, the original (gradient-based) ADAPT-VQE operator selection and ExcitationSolve for the parameter optimization in another setting (blue). Experiments are for the molecules a) \ce{LiH} and b) \ce{H2O}. The light blue region signifies the chemical accuracy ($\SI{e-3}{\Ha}$).}
            \label{fig:adapt_mixed_optimizers}
        \end{figure}

            To separate the respective impacts of the optimizers and the operator selection criteria on the convergence, in Fig.~\ref{fig:adapt_mixed_optimizers} we compare adaptive ansatz optimizations with combinations of GD as parameter optimizer and ExcitationSolve as operator selector, and, vice versa, with pure gradient- and ExcitationSolve-based ADAPT-VQE. It is apparent that while using ExcitationSolve for both the optimization and selection is still the most efficient method. Already replacing only one part with ExcitationSolve leads to a great speed-up over the original (purely gradient-based) ADAPT-VQE. The reason that convergence using GD is much faster when using ExcitationSolve as operator selector already is its initialization strategy. 
            Initializing the newly selected operator with a parameter $\theta_i\neq0$ that is often near-optimal with respect to all parameters $\bm{\theta}$ naturally reduces the number of GD steps required to reach convergence. This is in excellent agreement with the observation in Fig.~\ref{fig:Adapt_resources}, which shows that little re-optimization is required when using ExcitationSolve for the selection and initialization. 
            
            For \ce{LiH} in Fig.~\ref{fig:adapt_mixed_optimizers_LiH}, one can clearly identify a region (shortly after chemical accuracy was reached) in which the gradient-based selection criterion overestimates the number of operators required to reach a certain energy threshold -- independent of the optimization strategy. This results in slower convergence in the start. However, once this \enquote{plateau} is breached, using ExcitationSolve as the optimizer provides a significant advantage over GD. 
        
            In both cases of \ce{LiH} in Fig.~\ref{fig:adapt_mixed_optimizers_LiH} and \ce{H2O} in Fig.~\ref{fig:adapt_mixed_optimizers_H2O}, in comparison with the original ADAPT-VQE, we find that using gradient-selection along with ExcitationSolve optimization yields faster convergence but the opposite approach leads to an improved operator selection quality and, hence, fewer operators in the ansatz (cf.~Table~\ref{tab:num_ops_adapt}).
            In any case, it is therefore advisable to employ ExcitationSolve for both methods to profit from both the convergence speed-up and the reduction in circuit depth.
        
        \begin{table}[!hpt]
            \centering
            \caption{\textbf{Number of operators selected for all four combinations in adaptive optimization impact study for \ce{LiH} and \ce{H2O}.} Combinations are labeled selector/optimizer. Indicates numbers of operators to reach full convergence and, in parentheses, chemical accuracy.}
            \begin{tabular}{ccccc}
                \toprule
                Molecule & ExcSolve/ExcSolve & ExcSolve/GD & GD/ExcSolve & GD/GD \\ \midrule
                LiH & 30 (6)               & 31 (6)         & 34    (6)      & 34 (6)   \\
                H2O & 42 (20)                & 42 (20)         & 48 (22)         & 48 (22)  \\
                \bottomrule
            \end{tabular}
            \label{tab:num_ops_adapt}
        \end{table}
        
        \subsection{Dissociation curve energy errors}
        \label{app:diss_curves_energy_errors}
\begin{figure}[ht]
            \centering
            \includegraphics[width=0.9\textwidth]{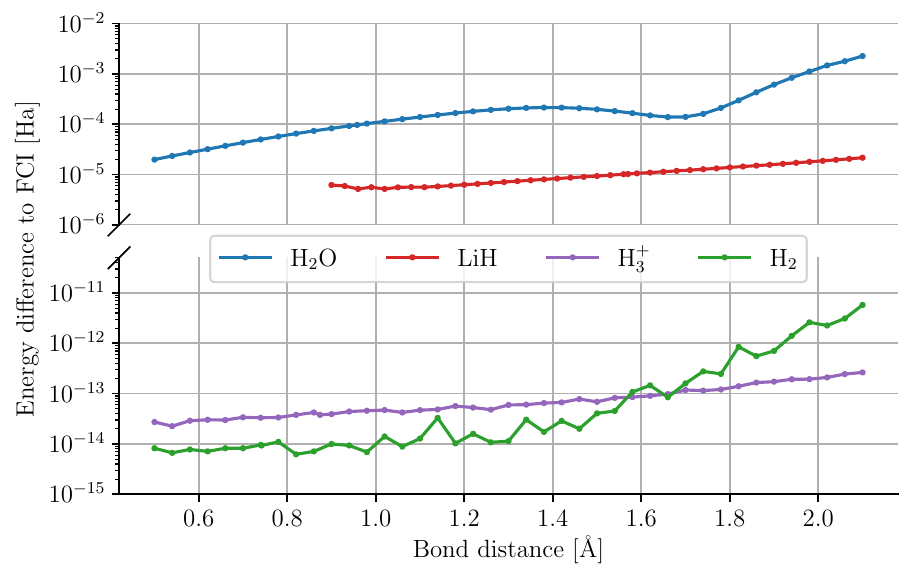}
            \caption{Final energy differences to FCI when convergence is reached for each molecule and bond distance shown in Figure~\ref{fig:combined_diss_curves}.}
            \label{fig:diss_curve_final_vqe_error}
        \end{figure}

        Here we present the final energy differences to FCI when convergence is reached for our dissociation curves from Section~\ref{subsec:StrongCorrelations}. The energy differences are the same for all optimizers but depend on the molecule. The final energy differences for each molecule and bond distance are shown in Figure~\ref{fig:diss_curve_final_vqe_error}. For \ce{H2} and \ce{H3+} the final energy differences are below $\SI{e-11}{\Ha}$ and $\SI{e-12}{\Ha}$, respectively. For \ce{LiH} and \ce{H2O} the final energy differences vary from about $\SI{5e-6}{\Ha}$ to about $\SI{3e-5}{\Ha}$ and from about $\SI{1e-5}{\Ha}$ to about $\SI{3e-3}{\Ha}$ based on the bond distance, respectively. For the bond distances \SI{2.02}{\angstrom} and \SI{2.06}{\angstrom} in \ce{H2O} we ignored the cases where ExcitationSolve got stuck in local minima when computing the final energy differences. These two cases are separately discussed in Appendix~\ref{app:local_min_diss_curves}.
        
        \subsection{Analysis of dissociation curve experiments on local minima avoidance} \label{app:local_min_diss_curves}
        
\begin{figure}[!hpt]
                \centering
                \begin{subfigure}[t]{0.49\textwidth}
                    \centering
                    \includegraphics[width=\textwidth]{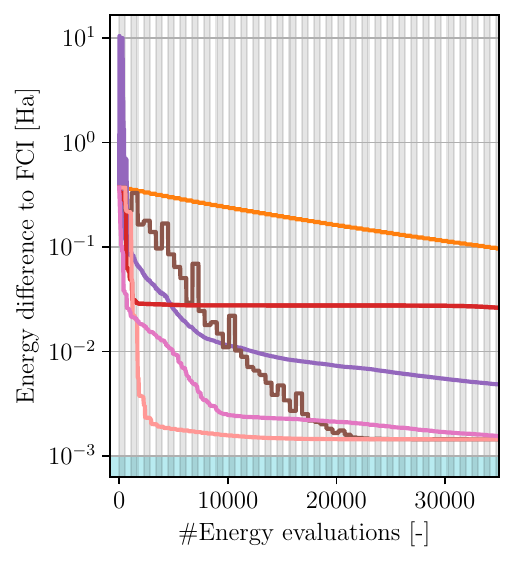}
                    \vspace{-0.75cm}
                    \caption{\ce{H2O}, 14 qubits, bond distance \SI{2.02}{\angstrom}.}
                    \label{fig:H2O_idx39}
                \end{subfigure}
                \hfill
                \begin{subfigure}[t]{0.49\textwidth}
                    \centering
                    \includegraphics[width=\textwidth]{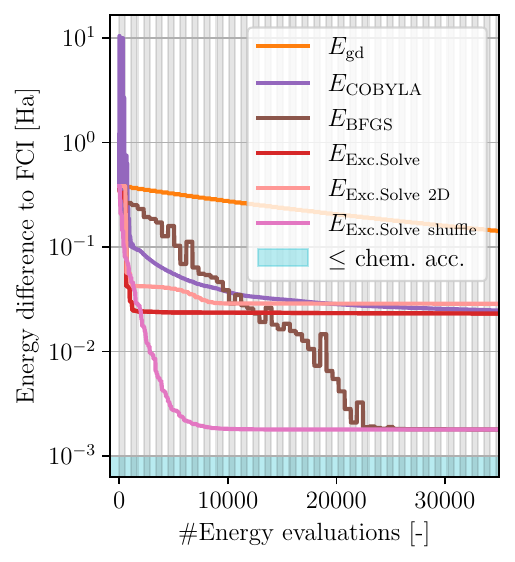}
                    \vspace{-0.75cm}
                    \caption{\ce{H2O}, 14 qubits, bond distance \SI{2.06}{\angstrom}.}
                    \label{fig:H2O_idx40}
                \end{subfigure}
                \caption{\textbf{Optimization of \ce{H2O} for two specific bond distances.} The optimizers under consideration are ExcitationSolve (red), COBYLA (purple), Gradient descent (yellow) and BFGS (brown). The plots show the error of the VQE with respect to the FCI solution
                ${|E_{\text{VQE}}-E_{\text{FCI}}|}$ over the number of energy evaluations for the bond distances marked separately in Fig.~\ref{fig:diss_curve_H2O_gd_cobyla} where parameter shuffling is used. The light blue region signifies the chemical accuracy ($\SI{e-3}{\Ha}$) and the alternating vertical shading marks each iteration over all parameters.}
                \label{fig:H2O_remaining_bls}
            \end{figure}

            Particular attention should be drawn to the two data points of the \ce{H2O} dissociation curve (Fig.~\ref{fig:diss_curve_H2O_gd_cobyla}) at the bond distances of $\SI{2.02}{\angstrom}$ and $\SI{2.06}{\angstrom}$. The convergence analysis for these data points can be inferred from Fig.~\ref{fig:H2O_remaining_bls}. As detailed in Sec.~\ref{sec:methods_fixed_ansatz}, all parameters have been optimized in their order of appearance in the UCCSD ansatz, i.e., a fixed order. Figures \ref{fig:H2O_idx39} and \ref{fig:H2O_idx40} depict cases in which this order causes ExcitationSolve to get stuck either in local minima or extremely flat parts of the optimization landscape, thus failing to converge within a reasonable number of energy evaluations. Meanwhile, Fig.~\ref{fig:H2O_idx39} also provides further evidence for the utility of 2D optimization, which successfully converges to the global minimum within the expressivity of the ansatz. While the 2D optimization can be utilized to avoid local minima, our heuristic to simultaneously optimize the two most impacting parameters may not succeed in some rare instances. We have found one example in Fig.~\ref{fig:H2O_idx40}, where both the 1D and 2D optimizers get stuck. Fortunately, randomly shuffling the parameter order in each VQE iteration while performing solely 1D optimization achieves convergence for both cases. Randomly shuffling the parameter order means that we randomly change the order in which we optimize the parameters at the beginning of each VQE iteration, while all parameters are optimized using ExcitationSolve 1D optimization. We note that this shuffling does not guarantee convergence to the optimal parameters in general and here we show only one instance of shuffling where it succeeded. We did not find a systematic way of deciding when to apply shuffling to avoid local minima and leave this for future work. In the case where the 2D optimizer already converges (Fig.~\ref{fig:H2O_idx39}), it is still notably faster than the shuffled 1D approach. 
            
        \subsection{Shot noise} \label{subsec:Noise}
        
        We repeat the experiments from Sec.~\ref{subsec:fixedAnsatz} with shot noise instead of exact state vector simulations. Due to the large number of shots needed to achieve chemical accuracy and the increasing computation time for larger molecules, we restrict ourselves to the molecules \ce{H2} and \ce{H3+}. We perform $\num{e7}$ shots for all optimizers and molecules in the results we show here since the overall qualitative behavior of the optimizers was rather independent of the number of shots. 
        Implementation details can be found in Sec.~\ref{sec:methods_shot_noise}.
        Note that, unlike in the results presented in the main text, the parameter order considered here comprises of first the single then the double excitations in the UCCSD ansatz, which is slightly sub-optimal in the first iteration as the single excitations cannot cause any change and results in a shifted energy reduction pattern in the plots.
        
\begin{figure}[ht]
            \centering
            \begin{subfigure}[t]{0.49\textwidth}
                \centering
                \includegraphics[width=\textwidth]{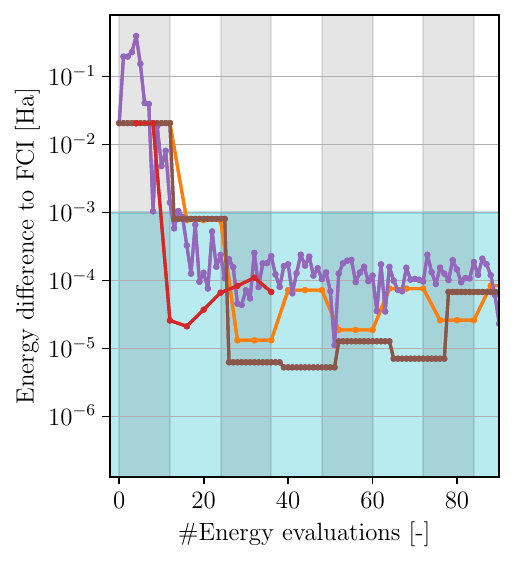}
                \vspace{-0.75cm}
                \caption{\ce{H2}, $4$ qubits. %
                }
                \label{fig:shots_H2}
            \end{subfigure}
            \hfill
            \begin{subfigure}[t]{0.49\textwidth}
                \centering
                \includegraphics[width=\textwidth]{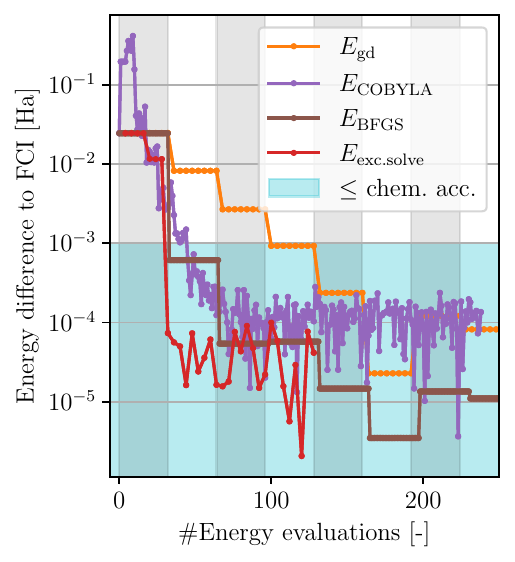}
                \vspace{-0.75cm}
                \caption{\ce{H3+}, $6$ qubits.
                }
                \label{fig:shots_H3}
            \end{subfigure}
            
            \caption{\textbf{Comparison of optimizers under the influence of shot noise.} The optimizers under consideration are ExcitationSolve (red), COBYLA (purple), Gradient descent (yellow) and BFGS (brown) with $\num{e7}$ shots each. The plots show the error of the VQE with respect to the FCI solution
            ${|E_{\text{VQE}}-E_{\text{FCI}}|}$ over the number of energy evaluations for the molecules \ce{H2}~(Fig.~\ref{fig:fixed_ansatz_H2_gd_adam_spsa_cobyla}), \ce{H3+}~(Fig.~\ref{fig:fixed_ansatz_H3_gd_adam_spsa_cobyla}). The light blue background signifies when chemical accuracy has been reached. Vertical lines mark when one iteration over all parameters has been completed.
            }
            \label{fig:combined_shots}
        \end{figure}

        Figure \ref{fig:combined_shots} presents the results.
        For optimizers that do not update parameters at each energy evaluation, we repeatedly plot the latest updated energy which results in energy plateaus which seem not affected by noise. 
        For example, GD has these plateaus are during the calculation of the gradient. We find that the shot noise has a significant impact on all used optimizers and limits the achievable accuracy. The difference between the optimizers is less pronounced than in the state vector simulations form Sec.~\ref{subsec:fixedAnsatz}. Overall, the results are similar to the state vector simulations, only that the maximum achievable accuracy of every optimizer is limited by the shot noise. %
        We see that ExcitationSolve reaches its maximum accuracy within similar number of energy evaluation as in the state vector simulations (cf.~Sec.~\ref{subsec:fixedAnsatz}). This includes \ce{H2}, where ExcitationSolve achieves its maximum accuracy within one VQE iteration. Most importantly, ExcitationSolve reaches its maximum accuracy faster than all other optimizers. With this, we note that the convergence speed of ExcitationSolve is robust against noise.

        \subsection{NISQ robustness of ExcitationSolve adaptive operator ranking}\label{app:nisq_adapt}
        
\begin{figure}[ht]
            \centering \includegraphics[width=\textwidth,clip,trim={0 280 0 250}]{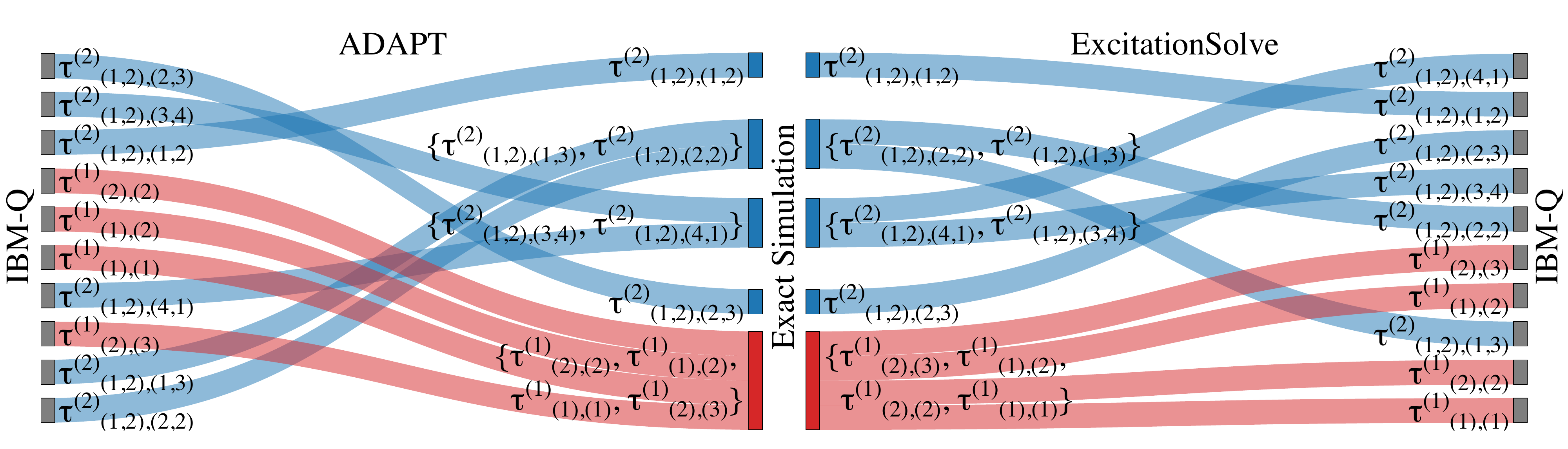}
            \caption{\textbf{Benchmarks on NISQ (\texttt{ibm\_quebec}) quantum processor for adaptive ansätze.}
            The excitation operator rankings for adaptive ansätze on the initial HF states of \ce{LiH} (frozen core, tapered) calculated on the \texttt{ibm\_quebec} device are visualized. The operator scores and resulting noisy rankings for ExcitationSolve (right) and ADAPT-VQE (left) are compared by visualizing their permutations with the true rankings in the middle. The true scores are exactly simulated, where red (blue) indicates zero (non-zero) scores.
            }
            \label{fig:ibmq_LiH}
        \end{figure}

            To study the NISQ robustness of the ExcitationSolve operator ranking for adaptive ansatz optimization compared to ADAPT-VQE, we analyze the permutations from the true operator rankings via slope charts in Fig.~\ref{fig:ibmq_LiH}, which connect matching operators in different rankings.
            There, it becomes apparent that the operator ranking is reproduced more accurately on the IBM-Q device via the ExcitationSolve than the gradient-based ADAPT-VQE scores. 
            Since the rankings produced by both methods match under exact simulation, this implies that the ExcitationSolve scores are more robust against noise in reproducing the true ranking. 
            Operators with zero ExcitationSolve scores cannot promote an energy decrease and should not be selected. However, ADAPT-VQE frequently mixes such operators with contributing (non-zero score) operators in the ranking from IBM-Q evaluations. In the ExcitationSolve ranking, such a confusion only occurs once and otherwise provides a clear separation between contributing and non-contributing operators.
            Furthermore, ExcitationSolve picks an operator ranked higher in simulation as the top choice than ADAPT-VQE and puts the exact top operator second instead of third. 
            Overall, ExcitationSolve seems more likely to append operators that can contribute with a higher energy decrease than the original ADAPT-VQE when evaluated on noisy hardware.
            
        \FloatBarrier %
        
    \clearpage
    \section{ExcitationSolve algorithmic details}\label{app:algorithms}
    
        Algorithmic details for ExcitationSolve for the application to both fixed and adaptive ansätze are provided in the following in form of pseudo code. For implementations of ExcitationsSolve and its variants/extensions in \texttt{Python}, refer to the code availability statement in the main text. We also include additional extensions and methodologies for ExcitationSolve in this appendix that are not present in the experimental studies.
    
        \subsection{ExcitationSolve for fixed ansätze}\label{app:algorithms:fixed}
        \FloatBarrier
        
            Algorithm \ref{algo:adapt_excitation_solve} outlines the ExcitationSolve optimization algorithm for fixed ansätze.
            Here, the $k$ iterations reflect the number of parameter updates that have been performed by ExcitationSolve. Hence, the parameters of a new iteration $k$ are initialized by the parameters of the previous iteration $k-1$, i.e., ${\bm\theta}^{(k)} \leftarrow {\bm\theta}^{(k-1)}$. Importantly, only the line highlighted in purple requires quantum hardware (QC), while everything else is computed efficiently on a classical device. Note that the energy associated with the unshifted current parameter value $\theta_j^{(k)}$ is re-used from the previous iteration or, in the first iteration, from the initial HF energy (CC). The order in which the $N$ parameters are iterated over in the \textit{for-each} loop can be chosen freely. To clearly indicate that the parameter indices and the sweep order are independent, we use two distinct iterators, $j$ and $k$.

            \begin{algorithm}[htb]
                \SetAlgoLined
                \KwHardware{{\color{purple}Quantum Computer (QC)}, Classical Computer (CC)}
                \KwIn{Initial parameters $\bm{\theta}^{(0)} = \bm{0}$, HF/init.~energy $E^{(0)} = f(\bm{\theta}^{(0)})$, HF/init.~state $\ket{\psi_0}$, fixed ansatz $U(\cdot)$ \textit{(excitation operators as in Sec.~\ref{sec:excitation_ops})}}
                \KwOut{Optimized parameters $\bm{\theta}^*$ and energy $E^*$}
                $k = 0$\;
                \Repeat{\textnormal{Convergence (threshold energy reduction $|E^{(k-N)} - E^{(k)}| \leq \epsilon$)} $\rightarrow \bm{\theta}^* = \bm{\theta}^{(k)}, E^* = E^{(k)}$}{
                    \ForEach{\textnormal{Parameter} $\theta_j$}{
                        $k \leftarrow k + 1$\;
                        Keep other parameters ${\theta}_{l\neq j}^{(k)}$ fixed\;
                        Re-use optimal energy from previous iteration $k-1$ as energy evaluation in current iteration $E_0^{(k)} = E^{(k-1)}$ for un-shifted parameter position $\theta_{j,0}^{(k)} = \theta_j^{(k)}$\;
                        {\color{purple} Determine energies $E_1^{(k)},E_2^{(k)},E_3^{(k)},E_4^{(k)}$ at four additional parameter positions $\theta_{j,1}^{(k)},\theta_{j,2}^{(k)},\theta_{j,3}^{(k)},\theta_{j,4}^{(k)}$, e.g., equidistant positions $\theta_{j,l}^{(k)} = \theta_j^{(k)} + {2\pi l}/{5}$ for $l = 1,\ldots, 4$ \textit{(see Eq.~(\ref{eq:energy_func_expectation}) via QC)}\;}
                        Reconstruct energy landscape in parameter $\theta_j$ by solving linear equation system of five ($l = 0, \ldots, 4$) equations $f_{\bm{\theta}^{(k)}}(\theta_{j,l}^{(k)}) \overset{!}{=} E_l^{(k)}$ \textit{(see Eq.~(\ref{eq:analytic_form_fourier}))}\;
                        Determine global minimum of reconstruction $E^{(k)} = \min_{\theta_j} f_{\bm{\theta}^{(k)}}(\theta_j)$ and update parameter $\theta_j^{(k)} \leftarrow \argmin_{\theta_j} f_{\bm{\theta}^{(k)}}(\theta_j)$ \textit{(see companion matrix method in Sec.~\ref{sec:classical_min_analytic_energy_function})}\;
                    }
                }
                \caption{ExcitationSolve optimization algorithm for fixed ansätze.\label{algo:excitation_solve}}
            \end{algorithm}
            
            \FloatBarrier
        
        \clearpage
        
        \subsection{ExcitationSolve for ADAPT-VQE (adaptive ansätze)}\label{app:algorithms:adaptive}

            Algorithm \ref{algo:adapt_excitation_solve} details the application of ExcitationSolve to ADAPT-VQE (adaptive ansätze).
            Here, ADAPT iteration $\ell$ denotes how many operators have been appended to the ansatz, while the number of update steps for re-optimizing all parameters in-between ADAPT iterations is omitted by calling ExcitationSolve for a fixed ansatz (Algorithm \ref{algo:excitation_solve}). The index $m$ describes the index of the operators in the operator pool $\mathcal{P}$. Importantly, the usage of the quantum device solely happens in Algorithm \ref{algo:excitation_solve} when invoked as sub-routines.

            \begin{algorithm}[htb]
                \SetAlgoLined
                \KwHardware{{\color{purple}Quantum Computer (QC)}, Classical Computer (CC)}
                \KwIn{Initial (empty) parameters $\bm{\theta}^{(0)} = \varnothing$, Empty ansatz $U^{(0)}(\cdot) = I$, HF/init.~energy $E^{(0)} = f(\bm{\theta}^{(0)})$, HF/init.~state $\ket{\psi_0}$, Pool of excitation operators $\mathcal{P}$ \textit{(excitation operators as in Sec.~\ref{sec:excitation_ops})}}
                \KwOut{Optimized parameters $\bm{\theta}^*$ and energy $E^*$}
                $\ell = 0$\;
                \While{True}{
                    $\ell \leftarrow \ell + 1$\;
                    \ForEach{\textnormal{Operator in pool} $U_m(\cdot) \in \mathcal{P}$}{
                        New candidate ansatz by appending operator $U^{(\ell-1)}(\bm{\theta}^{(\ell-1)}) \circ U_m(\theta_{m})$\;
                        Evaluate operator candidate via minimum energy and optimal parameter $E_m^{(\ell)}, \theta_m^{(\ell)} \leftarrow$ inner loop in Algorithm \ref{algo:excitation_solve} {\color{purple} incl. QC} (fix previous parameters $\bm{\theta}^{(\ell-1)}$)\;
                    }
                    \If{\textnormal{Convergence (threshold energy reduction $|E^{(\ell-1)} - \min_m E_m^{(\ell)}| \leq \epsilon$)}}{
                        $\bm{\theta}^* = \bm{\theta}^{(\ell-1)}, E^* = E^{(\ell-1)}$\;
                        break\;
                    }
                    Select operator with strongest energy reduction $m^* = \argmin_m E_m^{(\ell)}$ to extend ansatz $U^{(\ell)}(\bm{\theta}^{(\ell)}) = U^{(\ell-1)}(\bm{\theta}^{(\ell-1)}) \circ U_m^*(\theta_{m^*}^{(\ell)})$ and optimally initialize the new parameter $\bm{\theta}^{(\ell)} = \bm{\theta}^{(\ell-1)} \cup ( \theta_{m^*}^{(\ell)})$\;
                    Re-optimize all parameters via ExcitationSolve under fixed ansatz\\$E^{(\ell)}, \bm{\theta}^{(\ell)} \leftarrow$ Algorithm \ref{algo:excitation_solve} {\color{purple} incl. QC}\;
                }
                \caption{ExcitationSolve for ADAPT-VQE (adaptive ansätze). \label{algo:adapt_excitation_solve}}
            \end{algorithm}
            
            \FloatBarrier
        
    \subsection{ExcitationSolve for multiple occurrences of multiple parameters}\label{sec:multi_occ_multi_param}
    
    After having explored the two cases of multiple distinct parameters and multiple occurrences of a single parameter in the main text, it remains to explore the most general case: multiple occurrences of multiple parameters (different parameters may appear different number of times). The result is a straightforward conclusion of both previous results. Each unique parameter $\theta_i$ introduces one dimension and the order of the Fourier series in the corresponding dimension is given by the respective number of occurrences. Let $\bm{\tilde\theta} \subseteq \bm \theta$ be the subset of simultaneously varied parameters and $S_i$ be the number of occurrences of a parameter $\theta_i \in \bm{\tilde \theta}$. The energy landscape can then be expressed as
        
        \begin{equation}
            f_{\bm \theta}(\bm \theta_{\mathcal \bm{\tilde \theta}}) = \bm c \cdot \left[\bigotimes_{\theta_i \in \bm{\tilde \theta}} \left(\cos(\theta_i), \cos(2\theta_i), \dots, \cos(2S_i\theta_i), \sin(\theta_i), \sin(2\theta_i), \dots, \sin(2S_i\theta_i), 1\right)^\top\right],
        \end{equation}
        where $\bm c$ is a real-valued vector with dimension $\prod_{\theta_i \in \bm{\tilde \theta}} (4S_i+1)$.
            
    \subsection{Reconstruction strategies for noise robustness} \label{sec:noise_robustness}

    As the analytic energy landscape is resembled by a second-order Fourier series as in Eq.~\eqref{eq:analytic_form_fourier}, five energy evaluations set the minimum requirement to uniquely determine the five coefficients. However, the energy landscape reconstruction in ExcitationSolve can be readily extended beyond five energy evaluations. Assuming that the energy evaluations are inexact (e.g.~due to device- or shot-noise)%
    , this approach can make ExcitationSolve more robust against noise. %
    
    The then overdetermined linear equation system can then be solved in two ways: Using the least-squares method or a discrete Fourier transform.
    From a statistical perspective, the least-squares method is solving the regression problem in a second-order Fourier basis expansion \cite{shawe-taylor_KernelMethodsPattern_2004}. Then, in terms of maximum-likelihood optimality, the least-squares estimation yields the optimal result under a normally distributed noise assumption \cite{murphy_ProbabilisticMachineLearning_2022}. This assumption is approximately fulfilled for pure shot-noise with practical shot numbers \cite{nielsen_QuantumComputationQuantum_2010}, yet only sometimes observed for hardware noise \cite{sung_NonGaussianNoiseSpectroscopy_2019}.
    The discrete Fourier transform truncated at the second order could be utilized because higher frequencies cannot be contained in the energy landscape as in Eq.~\eqref{eq:analytic_form_fourier} but are solely subject to noise. 
    Both approaches can in fact be seen as equivalent if the parameter-shifts are equidistant, which is necessary for the Fourier transform and generally suggested \cite{endo_OptimalParameterConfigurations_2023}. The equivalence can be supported by the least-squares guarantee when solving the regression problem \cite{murphy_ProbabilisticMachineLearning_2022} and the \emph{best approximation} principle of the truncated Fourier transform \cite{stein_FourierAnalysisIntroduction_2003}. Both arguments are made under the $L^2$ norm and exhibit the geometrical interpretation of orthogonal projections on the feasible function space \cite{hastie2009elements,stein_FourierAnalysisIntroduction_2003}.
    
    From a practical perspective, the possibility of using more than five energy values raises the following question: Given a fixed shot budget $T$, should we rather spend more shots per energy evaluation or query more energy values for the best energy reconstruction? 
    The corresponding variance of the energy estimate when allocating $t$ shots per evaluation for any parameter position $\theta_j \in [-\pi, \pi]$ is given by ${(\Delta H)^2}/{t}$ \cite{nielsen_QuantumComputationQuantum_2010} where (one-shot) observable variance $(\Delta H)^2$ depends on the prepared quantum state and, consequently, on the choice of parameters $\bm{\theta}$. To simplify the subsequent discussion, we assume $(\Delta H)^2$ to be constant and incorporate it into a proportionality constant. For the total shot budget $T$, we perform $t=T/K$ shots for each of the $K$ energy evaluations\footnote{For integer devision, assume compatible $T, K, t\in \mathbb{N}_{> 0}$.}. Then, each energy evaluation is estimated with a noise variance of $\sigma^2 \sim \tfrac{1}{T/K}$, which leads to an average variance of these $K$ estimates of $\sigma^2_T \sim \sigma^2 / K$. Therefore, this average estimate variance becomes $\sigma^2_T \sim 1 / T$, i.e., inverse-proportional to the total shot budget $T$ and, importantly, independent of $K$. In conclusion, the distribution of shots per energy evaluation under a fixed shot budget will not quantitatively change the information extracted from the quantum computer and, thus, does not impact the quality of the reconstruction.
    On the other hand, if we relax the assumption of a constant observable variance $(\Delta H)^2$, which is certainly expected in practice, we cannot make general statements about a trade-off between the shot count and number of energy evaluations. Otherwise, $(\Delta H)^2$ must be known (again of a finite Fourier series form), however, its estimation could pose a significant challenge in practice due to the likely high number of terms in $H^2$. 
    It must be emphasized that our analysis is intentionally kept simple and focuses solely on the average statistical error of the energy evaluation estimates, which serve as the basis for the reconstruction. A detailed analysis of how the error propagates through this reconstruction to parameter regions away from the sampling points, and particularly its dependence on the number and spacing of these points for the reconstruction, is beyond the scope of this work. For a more rigorous treatment, we refer the reader to Ref.~\cite{endo_OptimalParameterConfigurations_2023}, which, for example, identifies three equiangular (i.e., equally spaced) sampling points on the unit circle as an optimal configuration for Rotosolve.

    \clearpage    
    \section{Proofs}
    \label{app:proofs}
    
        \subsection{Analytic energy function in single parameter} \label{app:analy_energy_func_single_param}
        
        We present three proofs that show that the energy function in Eq.~\eqref{eq:energy_func_expectation} has the analytical form of a second-order Fourier series as in Eq.~\eqref{eq:analytic_form_fourier}. 
        First, a constructive proofs is presented, which also shows the explicit connection of the coefficients $a_1, a_2, b_1, b_1, c$ in Eq.~\eqref{eq:analytic_form_fourier} and (expectation values of) observables of variationally prepared states.
        \begin{proof}
            The excitation operators $U(\theta)$ as defined in Eq.~\eqref{eq:excitation_operator_herm_form} have a Hermitian generator $G$ with the property $G^3 = G$. The exponential series simplifies to the Euler formula %
            \begin{equation}\label{eq:euler_formula}
                U(\theta) = \exp( -i \theta G  ) = I + \left(\cos(\theta) - 1\right)G^2 - i\sin(\theta) G.
            \end{equation} %
            because $G^3 = G$ \cite{anselmetti_LocalExpressiveQuantumnumberpreserving_2021}. This property can also be motivated through a hidden SU(2) symmetry associated with operators of the type $G^3=G$, as already studied for excitation operators in Refs.~\cite{evangelista2019exact, xu2020test, chen2021quantum, freericks2022operator}.
            When varying the parameter $\theta_j$ of a single excitation operator, while leaving all the other parameters $\theta_{i < j}$ and $\theta_{i > j}$ of preceding and succeeding excitation operators, respectively, in the circuit $U(\bm{\theta}) = \prod_k U(\theta_k)$ fixed. Hence, these operators can be subsumed in the input state 
            \begin{equation} %
                \left(\prod_{i < j} U(\theta_i)\right) \ket{\psi_0} =: \ket{\psi'}
                 \label{eq:transformed_state}
            \end{equation}
            and observable
            \begin{equation}
                 \left(\prod_{i > j} U(\theta_i) \right)^\dagger H \left(\prod_{i > j} U(\theta_i) \right) =: H',
            \end{equation}
            respectively, allowing us to re-phrase the energy function of Eq.~\eqref{eq:energy_func_expectation} through the following expectation value
            \begin{equation}\label{eq:energy_func_single_param_expectation}
                f_{\bm{\theta}}(\theta) = \braket{\psi' | U^\dagger(\theta) H' U(\theta) | \psi'}.
            \end{equation}
            The dependence of the state $\ket{\psi'}$ and observable $H'$ on the remaining parameters is omitted, as well as the index of the varied parameter $\theta_j = \theta$, for the sake of clarity.
            Inserting now the Euler formula in Eq.~\eqref{eq:euler_formula} into Eq.~\eqref{eq:energy_func_single_param_expectation} yields
            \begin{align}
                f_{\bm{\theta}}(\theta) &= \Braket{\left(I + \left(\cos(\theta) - 1\right)G^2 + i\sin(\theta) G \right) H' \left(I + \left(\cos(\theta) - 1\right)G^2 - i\sin(\theta) G \right)} \\
                &= \phantom{+} \Braket{\left\lbrace H', G^2\right\rbrace}\left(\cos(\theta) - 1\right) 
                    + \Braket{G^2 H' G^2}\left(\cos(\theta) - 1\right)^2 \nonumber\\
                    &\mathrel{\phantom{=}} + \Braket{i\left[ GH'G, G \right]}\left(\cos(\theta) - 1\right)\sin(\theta) \nonumber\\
                    &\mathrel{\phantom{=}} + \Braket{i\left[ G, H' \right]} \sin(\theta) 
                    + \Braket{GH'G} \sin^2(\theta) \nonumber\\
                    &\mathrel{\phantom{=}} + \Braket{H'} \\
                &= \phantom{+} \left(\Braket{\left\lbrace H', G^2\right\rbrace} -2\Braket{G^2 H' G^2} \right)\cos(\theta) \nonumber\\
                &\mathrel{\phantom{=}} + \Braket{i\left[ GH'G, G \right]} \sin(\theta)\cos(\theta) \nonumber\\
                &\mathrel{\phantom{=}} + \left(\Braket{i\left[ G, H' \right]} - \Braket{i\left[ GH'G, G \right]} \right)\sin(\theta) \nonumber\\
                &\mathrel{\phantom{=}} + \Braket{G^2 H' G^2}\cos^2(\theta) \nonumber\\
                &\mathrel{\phantom{=}} + \Braket{G H' G}\sin^2(\theta) \nonumber\\
                &\mathrel{\phantom{=}} + \Braket{H'} -  \Braket{\left\lbrace H', G^2\right\rbrace} + \Braket{G^2 H' G^2},
            \end{align}
            where all expectation values above are to be understood with respect to $\ket{\psi'}$, i.e., $\braket{\cdot} = \braket{\psi' | \cdot | \psi'}$. Considering the three trigonometric identities, \textit{Pythagorean trigonometric identity} $\cos^2(\theta) + \sin^2(\theta) = 1$, \textit{double-angle-formula} $\sin(\theta)\cos(\theta) = \sin(2\theta)/2$, and \textit{power-reduction-formula} $\sin^2(\theta) = \left(1 - \cos(2\theta) \right)/2$, we obtain 
            \begin{align}
                f_{\bm{\theta}}(\theta) &= \phantom{+} \underbrace{\left(\Braket{\left\lbrace H', G^2\right\rbrace} -2\Braket{G^2 H' G^2}\right)}_{= \, a_1} \cos(\theta) \nonumber \\
                &\mathrel{\phantom{=}}+ \underbrace{ \tfrac{1}{2}\left(\Braket{G^2 H' G^2} - \Braket{G H' G} \right)}_{= \, a_2} \cos(2\theta) \nonumber\\
                &\mathrel{\phantom{=}} + \underbrace{\left(\Braket{i\left[ G, H' \right]} - \Braket{i\left[ GH'G, G \right]} \right)}_{= \, b_1} \sin(\theta) \nonumber\\
                &\mathrel{\phantom{=}} + \underbrace{\tfrac{1}{2}\Braket{i\left[ GH'G, G \right]}}_{= \, b_2} \sin(2\theta) \nonumber \\
                &\mathrel{\phantom{=}} + \underbrace{\Braket{H'} -  \Braket{\left\lbrace H', G^2\right\rbrace} + \tfrac{1}{2}\left( \Braket{G H' G} + 3\Braket{G^2 H' G^2}\right)}_{=\,c}. \label{eq:proof_analytic_form_fourier_explicit}
            \end{align} %
            Recognizing that Eq.~\eqref{eq:proof_analytic_form_fourier_explicit} precisely matches the form of a second-order Fourier series for the energy function in a single parameter as in Eq.~\eqref{eq:analytic_form_fourier} concludes the proof.
        \end{proof}
        
        Second, we provide an alternative proof of this connection between Eq.~\eqref{eq:energy_func_expectation} and Eq.~\eqref{eq:analytic_form_fourier}, given the theory of general parameter-shift rules \cite{wierichs_GeneralParametershiftRules_2022}, which links the eigenvalues of $G$ to the frequencies present in the energy function.
        \begin{proof}[Proof (alternative I)]
            Given the eigenvalues $\lbrace \omega_i \rbrace$ of the Hermitian generator $G$ of a single parameterized operator $U(\theta) = \exp(i\theta G)$, Ref.~\cite{wierichs_GeneralParametershiftRules_2022} determines that the energy function Eq.~\eqref{eq:energy_func_expectation} can be written in the form of a finite Fourier series 
            \begin{equation}\label{eq:x}
                f_{\bm{\theta}}(\theta_j) = a_0 + \sum_{\ell = 1}^R a_\ell \cos\left(\Omega_\ell\theta_j\right) + \sum_{\ell = 1}^R b_\ell \sin\left(\Omega_\ell\theta_j\right) \tag{Ref.~\cite{wierichs_GeneralParametershiftRules_2022}, Eq.~(6)}
            \end{equation}
            of the order $\max_\ell\{\Omega_\ell\}$. Here, they introduce the $R$ \emph{unique positive differences} $\lbrace \Omega_\ell \rbrace := \left\lbrace \omega_k - \omega_h \mid \omega_k > \omega_h \right\rbrace $.
            In the case of the excitation operators, we exploit the fact that a Hermitian generator with the property $G^3 = G$ must have eigenvalues $\omega_k \in \lbrace -1, 0, 1 \rbrace$ \cite{anselmetti_LocalExpressiveQuantumnumberpreserving_2021}. If all three different possible eigenvalues are contained in the spectrum, $R=2$ unique positive differences $\lbrace \Omega_\ell \rbrace = \left\lbrace 1, 2 \right\rbrace $ are present, which proves that Eq.~\eqref{eq:energy_func_expectation} is a second-order Fourier series Eq.~\eqref{eq:analytic_form_fourier} when varied in a single parameter $\theta_j$. We now prove that the spectrum of the Hermitian generators of excitation operators $G=i \tau_{\bm o, \bm v}^{(m)}$ does indeed contain all three possible eigenvalues $\{-1,0,1\}$. For the eigenvalues $\omega = \pm 1$, we may construct the corresponding eigenstates explicitly as 
            \begin{equation}
                \ket{\pm} = \frac{1}{\sqrt{2}}\left(\ket{0_{\bm v_1}0_{\bm v_2}\dots 0_{\bm v_m} 1_{\bm o_1}1_{\bm o_2}\dots 1_{\bm o_m}} \pm i \ket{1_{\bm v_1}1_{\bm v_2}\dots 1_{\bm v_m} 0_{\bm o_1}0_{\bm o_2}\dots 0_{\bm o_m}}\right).
                \label{eq:givens_2D_subspace}
            \end{equation}
            Then, we have
            \begin{align}
                G \ket{\pm} &= i \tau_{\bm o, \bm v}^{(m)} \ket{\pm} = i \left(a^\dagger_{v_1} a^\dagger_{v_2} \dots a^\dagger_{v_m}~a_{o_m} \dots a_{o_2} a_{o_1} - \text{H.c.}\right)\ket{\pm} \nonumber \\
                &= \frac{i}{\sqrt{2}} \left(\ket{1_{\bm v_1}1_{\bm v_2}\dots 1_{\bm v_m} 0_{\bm o_1}0_{\bm o_2}\dots 0_{\bm o_m}} \mp i \ket{0_{\bm v_1}0_{\bm v_2}\dots 0_{\bm v_m} 1_{\bm o_1}1_{\bm o_2}\dots 1_{\bm o_m}}\right) \nonumber \\
                &= \pm \frac{1}{\sqrt{2}} \left(\ket{0_{\bm v_1}0_{\bm v_2}\dots 0_{\bm v_m} 1_{\bm o_1}1_{\bm o_2}\dots 1_{\bm o_m}} \pm i \ket{1_{\bm v_1}1_{\bm v_2}\dots 1_{\bm v_m} 0_{\bm o_1}0_{\bm o_2}\dots 0_{\bm o_m}}\right) \nonumber \\
                &= \pm 1 \ket{\pm}
            \end{align}
            One can easily verify that a quantum state $\ket \psi$ that is an arbitrary superposition of any basis states apart from $\ket{\pm}$ gives rise to $G\ket \psi = 0\ket \psi$. Consequently, for the generator $G$ of an $m$-electron excitation, we find two unique eigenstates $\ket \pm$ corresponding to the eigenvalues $\omega_\pm = \pm 1$, as well as a $(4^m-2)$-dimensional eigenspace corresponding to the eigenvalue $\omega=0$. All results hold equivalently for qubit-excitations.
        \end{proof}
        
        Last, we provide a third proof, which shows how our results can be unified with the SMO method \cite{nakanishi_SequentialMinimalOptimization_2020}. The idea is mostly based on the work from Ref.~\cite{kottmann_FeasibleApproachAutomatically_2021}.
        
        \begin{proof}[Proof (alternative II)]
            We once again assume a Hermitian generator with the property $G^3=G$. The generator is then decomposed into the sum of two commuting self-inverse generators $G_\pm$, that is
            
            \begin{equation}
                G = \frac{1}{2}\left(G_++G_-\right),
            \end{equation}
            where
            \begin{equation}
                G_\pm := G \pm (G^2-1).
            \end{equation}
            We first verify that $G_\pm$ are indeed self-inverse:
            
            \begin{align}
                G_\pm^2 &= \left[G \pm (G^2-1)\right]^2 \nonumber \\
                &= G^2 \pm 2 G (G^2-1) + (G^2-1)^2 \nonumber \\
                &= G^2 \pm 2 \underbrace{(G^3-G)}_{=0} + \underbrace{G^4}_{=G^2}-2G^2 + 1 \nonumber \\
                &= 1.
            \end{align}
            The commutation of $G_+$ and $G_-$ is a trivial result, since any operator commutes with any power of itself. The unitary $U(\theta)=\exp(-i\theta G)$ can thus be exactly decomposed as 
            
            \begin{equation}
                U(\theta) = U_-(\theta) U_+(\theta) = \exp\left(-\frac{i}{2}\theta G_-\right) \exp\left(-\frac{i}{2}\theta G_+\right)
                \label{eq:excitation_gate_decomposition}
            \end{equation}
            According to the SMO case describing multiple occurrences of the same parameter (c.f.~Eq.~\eqref{eq:smo_variant_3}), the energy landscape of any observable varied by an operation assuming the form in Eq.~\eqref{eq:excitation_gate_decomposition} gives rise to a second-order Fourier series.
            
        \end{proof}

        \subsection{\texorpdfstring{$G^3=G$ and $G^2\neq I$}{} for generators of excitation operators}
        \label{app:excitation_operators}
        
        In this section, we derive that fermionic- and qubit-excitation generators fulfill the property $G^3=G$, which is the foundation of ExcitationSolve. We start from the $m$-electron excitation generators introduced in Sec.~\ref{sec:excitation_ops}, namely:
        
        \begin{equation}
            \tau^{(m)}_{\bm{o}, \bm{v}} =  
            \prod_{l=1}^m a^\dagger_{v_l} a_{o_l} - \text{H.c.},
            \tag{Eq.~\eqref{eq:excitation_generator} revisited}
        \end{equation}
        where the fermionic creation and annihilation operators obey the canonical anti-commutation relations $\{a_i, a^{\dagger}_j\}=\delta_{ij}$ and $\{a^{\dagger}_i, a^{\dagger}_j\}=\{a_i, a_j\} = 0$.
        For the second power of $\tau^{(m)}_{\bm{o},\bm{v}}$, we obtain
        
        \begin{align}
            {\tau^{(m)}_{\bm{o}, \bm{v}}}^2 &=  \prod_{l=1}^m a^\dagger_{v_l} a_{o_l} a^\dagger_{v_l} a_{o_l} + \prod_{l=1}^m a^\dagger_{o_l} a_{v_l} a^\dagger_{o_l} a_{v_l} - \prod_{l=1}^m a^\dagger_{v_l} a_{o_l} a^\dagger_{o_l} a_{v_l} - \prod_{l=1}^m a^\dagger_{o_l} a_{v_l} a^\dagger_{v_l} a_{o_l}.
        \end{align}
        A direct implication of the anti-commutation relations is that $a^{\dagger 2}_i=a_i^2 = 0$ and $[a_i^{(\dagger)}, a_j a_j^\dagger]=0$. Using this we find that
        
        \begin{align}
            {\tau^{(m)}_{\bm{o}, \bm{v}}}^2 &= - \left(\prod_{l=1}^m a^\dagger_{v_l} a_{v_l} a_{o_l} a^\dagger_{o_l}  + \text{H.c.}\right),
            \label{eq:excitation_generator_power_2}
        \end{align}
        which clearly is not an identity operator.
        Next, for the third power, we similarly obtain from Eq.~\eqref{eq:excitation_generator} and \eqref{eq:excitation_generator_power_2} that
        
        \begin{align}
            {\tau^{(m)}_{\bm{o}, \bm{v}}}^3 &= - \left(\prod_{l=1}^m a^\dagger_{v_l} a_{v_l} a^\dagger_{v_l} a_{o_l} a^\dagger_{o_l} a_{o_l} - \text{H.c.} \right).
        \end{align}
        Utilizing that $a^\dagger_i a_i a^\dagger_i = a^\dagger_i (1-a^\dagger_i a_i) = a^\dagger_i$ and similarly $a_i a_i^\dagger a_i = a_i$, we finally arrive at
        \begin{equation}
             {\tau^{(m)}_{\bm{o}, \bm{v}}}^3 = -\left(\prod_{l=1}^m a^\dagger_{v_l} a_{o_l} - \text{H.c.}\right) = - \tau^{(m)}_{\bm{o}, \bm{v}}.
        \end{equation}
        
        Now we are presented with an anti-Hermitian operator of the form $G^3=-G$ generating the excitation operator $\exp{(\theta G)}$. To fit it within the convention of writing gates in terms of their Hermitian generator, we redefine $G:=i\tau^{(m)}_{\bm{o}, \bm{v}}$ and therefore obtain $G^3=G$. This logic can also be easily inferred from the following equation:
        
        \begin{equation}
            U^{(m)}_{\bm{o},\bm{v}}(\theta) = \exp(\theta~\tau^{(m)}_{\bm{o},\bm{v}}) = \exp(-i^2\theta~\tau^{(m)}_{\bm{o},\bm{v}}) = \exp(-i\theta~\underbrace{i\tau^{(m)}_{\bm{o},\bm{v}}}_{=G}).
        \end{equation}
        The same properties can easily be shown for qubit-excitation generators. In QEB-ansätze, the fermionic creation and annihilation operators $a^\dagger$ and $a$ in Eq.~\eqref{eq:excitation_generator} are replaced by qubit creation- and annihilation operators $Q^\dagger =\sigma^-$ and $Q=\sigma^+$\cite{yordanov2021qubitexcitationbased}, giving rise to the qubit-excitation generator. These operators fulfill the commutation relations $[Q_i, Q_j^\dagger] = \delta_{ij}(1-2Q_i^\dagger Q_i)$ with the (qubit-) occupation number $n_i=Q_i^\dagger Q_i$ being restricted to $0$ or $1$. These are the same algebraic properties as known for hard-core bosons \cite{PhysRevB.79.174515} or parafermions \cite{wu2002qubits}, allowing for a mapping-independent interpretation of qubit-excitations\footnote{In the literature, the generated qubit-excitation gates are also sometimes referred to as Givens rotations \cite{arrazola_UniversalQuantumCircuits_2022, ferris2022quantum}, as they can be visualized as a rotation in a two-dimensional subspace. More details about this subspace can be inferred from Eq.~\eqref{eq:givens_2D_subspace} in Appendix \ref{app:analy_energy_func_single_param}.}.
        The steps of the proof are the same, apart from skipping the sign argument due to the non-local commutation relations between qubit-creation/annihilation operators. 
        
        \subsection{General Fourier Series for Multi-Parameter Optimization}
        \label{app:analy_energy_func_multi_param}
        
        In Appendix \ref{app:analy_energy_func_single_param}, we have derived an analytical expression for the energy functional in a single parameter, which takes the form of a second-order Fourier series (c.f.~Eq.~\eqref{eq:analytic_form_fourier}). In the following, we prove inductively that an $D$-dimensional multi-parameter optimization landscape assumes the form of a $D$-dimensional second-order Fourier series. 
        
        \begin{proof}
            The base case, that is $D=1$, has already been proven in Appendix \ref{app:analy_energy_func_single_param}. For the induction step, we assume that, without loss of generality, the $(D+1)$-th parameter acts on the quantum state after the previous $D$ parameters. We define the ordered index sets $\mathcal M^{(D)}=\{j, \dots, k\}$, which contains the $D$ simultaneously optimized parameters in ascending order, and $\mathcal{M}^{(D+1)} = \{j, \dots, k, l\}$, which further includes the index $l$ of the $(D+1)$-th parameter $\theta_l$. To establish the induction hypothesis and carry out the induction step, we first define the effective initial state
    
            \begin{equation} %
                \ket{\psi'} :=\left(\prod_{i < j} U(\theta_i)\right) \ket{\psi_0},
                \tag{Eq.~\eqref{eq:transformed_state} revisited}
            \end{equation}
            and the effective Hamiltonian
            \begin{equation}
                 H^{(D)} := \left(\prod_{i > \max \mathcal{M}^{(D)}} U(\theta_i) \right)^\dagger H \left(\prod_{i > \max \mathcal{M}^{(D)}} U(\theta_i) \right).
            \end{equation}
            We further denote the effective unitary, including all operations sandwiched by the first and last variational (not fixed) unitary, for $D$ parameters as
            \begin{equation}
                U^{(D)} := \prod_{\substack{i \geq \min \mathcal{M}^{(D)} \\ i \leq \max \mathcal{M}^{(D)}}} U(\theta_i).
            \end{equation}
            Following these definitions, we may express the induction hypothesis as
            \begin{equation}
                f_{\bm \theta}(\bm{\theta}_{\mathcal M^{(D)}}) = \bra{\psi'}U^{\dagger(D)} H^{(D)} U^{(D)}\ket{\psi'} =
                \bm c^{(D)} \cdot \left[\bigotimes_{i\in \mathcal M^{(D)}}\begin{pmatrix}
                    \cos(\theta_i)  \\
                    \cos(2\theta_i) \\
                    \sin(\theta_i) \\
                    \sin(2\theta_i) \\
                    1
                \end{pmatrix}\right],
                \label{eq:induction_hypothesis}
            \end{equation}
            where $H^{(D)}$ is some arbitrary Hamiltonian since $H$ is arbitrary. 
            Next, we abbreviate all the fixed operations between $k$ and $l$ as
            \begin{equation}
                V^{(D+1)} := \prod_{\substack{i > \max \mathcal{M}^{(D)} \\ i < \max \mathcal{M}^{(D+1)}}} U(\theta_i).
            \end{equation}
            The energy landscape of the $(D+1)$-parameter case can then be written as 
            \begin{align}
                f_{\bm \theta}(\bm \theta_{\mathcal M^{(D+1)}}) 
                &=  \braket{\psi'| U^{\dagger (D+1)} H^{(D+1)} U^{(D+1)}|\psi'}
                \\
                &= \braket{\psi'| U^{\dagger (D)}V^{\dagger (D+1)} U^\dagger (\theta_l) H^{(D+1)} U(\theta_l) V^{(D+1)} U^{(D)}|\psi'}
            \end{align}
            Using the same reasoning as in Appendix \ref{app:analy_energy_func_single_param}, that is the Euler formula in Eq.~\ref{eq:euler_formula} and the trigonometric identities, we find that
                
            \begin{align}
                f_{\bm{\theta}}(\bm \theta_{\mathcal M^{(D+1)}}) &= \phantom{+} 
                \bra{\psi'} U^{\dagger (D)}\underbrace{V^\dagger\left(\left\lbrace H, G^2\right\rbrace -2G^2 H G^2\right) V}_{H_1} U^{(D)}\ket{\psi'}\cos(\theta_l) \nonumber \\
                &\mathrel{\phantom{=}}+ 
                \tfrac{1}{2}\bra{\psi'} U^{\dagger (D)}\underbrace{V^\dagger\left(G^2 H G^2 - G H G\right)V}_{H_2} U^{(D)}\ket{\psi'} \cos(2\theta_l) \nonumber\\
                &\mathrel{\phantom{=}} +
               \bra{\psi'} U^{\dagger (D)}\underbrace{V^\dagger\left(i\left[ G, H \right]- i\left[GHG, G \right]\right)V}_{H_3} U^{(D)}\ket{\psi'}  \sin(\theta_l) \nonumber\\
                &\mathrel{\phantom{=}} + 
                \tfrac{1}{2}\bra{\psi'} U^{\dagger (D)}\underbrace{V^\dagger i\left[ GHG, G \right]V}_{H_4} U^{(D)}\ket{\psi'} \sin(2\theta_l) \nonumber \\
                &\mathrel{\phantom{=}} + 
                \bra{\psi'} U^{\dagger (D)}\underbrace{V^\dagger\left(H- \left\lbrace H, G^2\right\rbrace + G H G + 3 G^2 H G^2\right)V}_{H_5} U^{(D)}\ket{\psi'}, \label{eq:proof_multi_param_induction_step}
            \end{align} %
            where we abbreviated $H^{(D+1)}=H$ and $V^{(D+1)}=V$. Notice that all of the expectation values $\bra{\psi'}U^{\dagger(D)} H_i U^{(D)}\ket{\psi'}$ for $i=1,\dots, 5$ must assume a $D$-dimensional second-order Fourier series according to the induction hypothesis in Eq.~\eqref{eq:induction_hypothesis} (the coefficients $\bm c^{(D)}$ differ across the different effective Hamiltonians $H_i$). Finally, we conclude that the energy landscape can be rewritten as
            \begin{equation}
                f_{\bm \theta}(\bm \theta_{\mathcal M^{(D+1)}}) = 
                \bm c^{(D+1)} \cdot \left[\bigotimes_{i\in \mathcal M^{(D+1)}}
                \begin{pmatrix}
                    \cos(\theta_i)  \\
                    \cos(2\theta_i) \\
                    \sin(\theta_i) \\
                    \sin(2\theta_i) \\
                    1
                \end{pmatrix}\right],
            \end{equation}
            thus completing the proof.
            
        \end{proof}
        
        \subsection{Fourier series for multiple occurrences of a single parameter} \label{app:multi_occ_single_param}
        
        In this Appendix, we derive an expression for the energy functional in a single parameter which occurs $S$ times in the circuit. We will inductively prove that the energy landscape is given by finite Fourier series of order $2S$:
        
        \begin{equation}
            f_{\bm\theta}(\theta) = \sum_{s = 1}^{2S} a_s \cos(s \theta_j) +  \sum_{s = 1}^{2S} b_s \sin(s \theta_j) + c.
            \tag{Eq.~\eqref{eq:energy_multi_occ} revisited}
        \end{equation}
        
        \begin{proof}
            Once again, the base case $S=1$ has already been proved in Appendix \ref{app:analy_energy_func_single_param}. Assuming that Eq.~\eqref{eq:energy_multi_occ} holds for some $S$, we consider the case with $S+1$ occurrences. Following exactly the same steps as in the previous proof of the multi-parameter case (Appendix \ref{app:analy_energy_func_multi_param}), but redefining $\mathcal M^{(S)}$ such that it corresponds to the equal parameters, i.e.~$\bm \theta_{\mathcal M^{(S)}} = \theta$, we find that the energy landscape is given by
            
            \begin{equation}
                f_{\bm \theta}^{(S+1)}(\theta) = a_1^{(s)}(\theta) \cos(\theta) + a_2^{(s)}(\theta) \cos(2\theta) + b_1^{(s)}(\theta) \sin(\theta) + b_2^{(s)}(\theta) \sin(2\theta) + c^{(s)}(\theta),
            \end{equation}
            
            where the parameterized coefficients $a_1^{(s)}(\theta), a_2^{(s)}(\theta), b_1^{(s)}(\theta), b_2^{(s)}(\theta)$ and $c^{(s)}(\theta)$ obey the induction assumption from Eq.~\eqref{eq:energy_multi_occ}. To obtain the order of the Fourier series, we need to compute the highest possible frequency $\omega_\mathrm{max}^{(S+1)}$ obtained from reducing the trigonometric form of the expressions above. For that purpose, we employ the following trigonometric identities:
            
            \begin{equation}
                \begin{split}
                    \sin(ax)\sin(bx) &= \frac{1}{2} \left[\cos((a-b)x) - \cos((a+b)x)\right], \\
                    \sin(ax)\cos(bx) &= \frac{1}{2} \left[\sin((a-b)x) + \sin((a+b)x)\right], \\
                    \cos(ax)\cos(bx) &= \frac{1}{2} \left[\cos((a-b)x) + \cos((a+b)x)\right].
                \end{split}
            \end{equation}
            
            The highest frequency is thus obtained as the sum of the largest frequency for $S$ occurrences, i.e.~$\omega_\mathrm{max}^{(S)}=2S$, and the additional double frequency $2$ of the $(S+1)$-th occurrence, giving rise to $\omega_\mathrm{max}^{(S+1)} = 2(S+1)$, and thus
            
            \begin{equation}
                f_{\bm \theta}^{(S+1)}(\theta) = \sum_{s = 1}^{2(S+1)} a_s \cos(s \theta_j) +  \sum_{s = 1}^{2(S+1)} b_s \sin(s \theta_j) + c.
            \end{equation}

        \end{proof}

    \clearpage
    \section{Comprehensive overview of standard approaches in variational quantum algorithms}\label{app:methods}

        \subsection{Gradients via parameter-shift rules for excitation operators}\label{sec:param_shift_rules}
                For gradient-based optimization of parameterized quantum circuits, analytical gradients can be computed for specific types of parameterized operators and gates through so-called \emph{parameter-shift rules}. As the name suggests, the (partial) derivative of a function $f_{\bm{\theta}}$ w.r.t. parameter $\theta_j$ is composed of energy function evaluations at shifts of parameter $\theta_j$. For excitation operators, fulfilling the generator property $G^3 = G$ without being self-inverse, i.e., $G^2 \neq I$, Ref.~\cite{anselmetti_LocalExpressiveQuantumnumberpreserving_2021} states the four-term paramter-shift rule relying on the energy values of four parameter shifts $\pm \alpha, \pm \beta$ as 
                \begin{equation}\label{eq:four_term_param_shift_rule}
                    f_{\bm{\theta}}'(\theta_j) = 
                    \frac{\partial}{\partial \theta_j} f(\bm{\theta}) = 
                    d_1 \left(f_{\bm{\theta}}(\theta_j + \alpha) - f_{\bm{\theta}}(\theta_j - \alpha) \right) - d_2 \left(f_{\bm{\theta}}(\theta_j + \beta) - f_{\bm{\theta}}(\theta_j - \beta) \right)
                \end{equation}
                with, for example,
                \begin{equation}
                    d_1 = \tfrac{1}{2}, \qquad 
                    d_2 = \tfrac{\sqrt{2}-1}{4}, \qquad
                    \alpha = \tfrac{\pi}{2}, \qquad
                    \beta = \pi.
                \end{equation}
                Other choices of $\alpha, \beta, d_{1,2}$ are possible subject to conditions \cite{anselmetti_LocalExpressiveQuantumnumberpreserving_2021}.
                
                Variations of parameter-shift rules exist in which the quantum circuit is dressed by additional gates. This leads to a decrease in the number of required shifts and, hence, energy evaluations on the quantum device to \emph{two} if the wave function (i.e., quantum state) is \textit{real}. For excitation operators%
                , this was derived in Ref.~\cite{kottmann_FeasibleApproachAutomatically_2021}. As both the four-term parameter-shift rule \cite{anselmetti_LocalExpressiveQuantumnumberpreserving_2021} and ExcitationSolve rely on energies of pure parameter shifts, the four-term parameter-shift rule is considered for a fair comparison between ExcitationSolve and gradient-based optimizers.  
                While this approach requires circuits differing from the ones necessary to evaluate the energy, the cost of the distinct circuit component may be upper-bounded by the cost of a regular excitation \cite{kottmann_FeasibleApproachAutomatically_2021}. Therefore, we may for the sake of simplicity still compare the costs in terms of energy evaluations, even though the shifts for the real wave functions are strictly speaking not energy evaluations as in ExcitationSolve or other optimizers (e.g. Table~\ref{tab:resource_comparison}).

            \subsection{Quantum-aware optimization for rotations: Rotosolve and SMO}\label{sec:rotosolve_smo}
                
                The Rotosolve \cite{ostaszewski_StructureOptimizationParameterized_2021} optimization method describes a coordinate descent approach, i.e., only a single parameter $\theta_j$ is updated in each iteration while the other parameters $\theta_{i\neq j}$ are held fixed. 
                Multiple extensions of Rotosolve/SMO have been proposed such as Free-Axis Selection (Fraxis) \cite{watanabe_OptimizingParameterizedQuantum_2021,watanabe_OptimizingParameterizedQuantum_2023,wada_SimulatingTimeEvolution_2022}, Free Quaternion Selection (FQS) \cite{wada_SequentialOptimalSelections_2024,kurogi_OptimizingParameterizedControlled_2024} and the Unitary Block Optimization Scheme (UBOS) \cite{slattery_UnitaryBlockOptimization_2022}. On the other hand, Rotosolve has never been extended to excitation operators as studied here despite several attempts \cite{armaos_EfficientParabolicOptimisation_2021,li_EfficientRobustParameter_2024} using polynomial fits, unaware of the correct analytical form as a second-order Fourier series. For each update step, the entire energy function along the current parameter is reconstructed, which has the form of a simple cosine curve
                \begin{equation}\label{eq:rotosolve_analytic_form}
                f_{\bm{\theta}}(\theta_j) = A \cos(\theta_j - \Phi) + c,    
                \end{equation}
                and the parameter is set to the then classically and analytically determined minimum of the reconstruction. To determine the coefficients $A, \Phi, c$ the energy is evaluated on the quantum computer for three suitable shifts of the parameter $\theta_j$. Note that one evaluation can be saved by reusing the energy value from the previous iteration. Importantly, the applicability of Rotosolve is limited to rotations, i.e., parameterized operators of the form $\exp(i\theta_j G /2)$ with $G^2 = I$, and, moreover, all parameters must be independent of each other, meaning that each $\theta_j$ must only occur once in the variational quantum circuit. 
                
                While Rotosolve was independently proposed under the name Sequential Minimal Optimization (SMO) in the first variant in Ref.~\cite{nakanishi_SequentialMinimalOptimization_2020}, SMO comes in two further variants:
                The second variant of SMO is a multi-parameter generalization, concerning the simultaneous optimization of a \emph{subset} of parameters $\bm{\theta}_{\mathcal{M}}$ where $\mathcal M$ denotes the index set of the $|\mathcal{M}|=D$ parameters to be optimized. A multi-parameter generalization was also mentioned in Ref.~\cite{parrish_JacobiDiagonalizationAnderson_2019}. The $D$-dimensional energy function reconstruction includes $3^D$ coefficients $\bm{c}$ and has the analytical form of
                \begin{equation}
                    f_{\bm \theta}(\bm{\theta}_{\mathcal M}) = 
                    \bm c \cdot \left[\bigotimes_{i\in \mathcal M}\begin{pmatrix}
                        \cos(\theta_i)  \\
                        \sin(\theta_i) \\
                        1
                    \end{pmatrix}\right].
                    \label{eq:analytic_form_fourier_d_dim_smo}
                \end{equation}
                The third variant of SMO lifts the requirement that each parameter must occur once in the variational quantum circuit, which SMO and Rotosolve impose otherwise. If a parameter $\theta_j$ occurs $S$ times, the energy function along parameter $\theta_j$ is no longer a simple cosine function but incorporates $S$ frequencies, i.e., obeys the form of a Fourier series of order $S$ as
                \begin{equation}\label{eq:smo_variant_3}
                    f_{\bm \theta}(\theta_j) = \sum_{s = 1}^S a_s \cos(s \theta_j) +  \sum_{s = 1}^S b_s \sin(s \theta_j) + c.
                \end{equation}
                Thus, determining $2S + 1$ coefficients require $2S + 1$ energy evaluations on the quantum computer to obtain the reconstruction to optimize $\theta_j$ (Again, one evaluation can be skipped by reusing the final energy of the previous iteration.)

        \subsection{ADAPT-VQE}\label{sec:method_adapt_vqe}
                In ADAPT-VQE, we optimize adaptive ansätze in VQE by alternately growing of the ansatz and optimization of the parameters as introduced in Ref.~\cite{grimsley_AdaptiveVariationalAlgorithm_2019}. Each Adapt(VQE)-Step consists of two parts: First, a suitable operator is appended to the ansatz from an operator pool, e.g., the pool of all single and double fermionic excitation operators. Second, all parameters are re-optimized while keeping the ansatz fixed, which equals a standard VQE run with a warm-start, i.e., the parameter values from the previous Adapt-Steps are used as initial guesses (while the newly added operator is initialized with its parameter set to zero.)
                
                For selecting a new operator from the pool, a scoring criterion assesses the quality of each operator candidate. The original ADAPT-VQE \cite{grimsley_AdaptiveVariationalAlgorithm_2019} obeys a gradient-based criterion where the operator is selected that admits the highest magnitude of its partial derivative in zero. This is the operator with the strongest \emph{local} impact on the energy. While the partial derivative could be computed through the (four-term) parameter-shift rule as in Eq.~\eqref{eq:four_term_param_shift_rule}, the partial derivative in zero constitutes a special case such that it can be alternatively expressed through the expectation value of a commutator
                \begin{equation}\label{eq:adapt_commutator_grad}
                    \left. \frac{\partial f(\theta_1, \theta_2, \ldots, \theta_N, \theta_{N+1})}{\partial \theta_{N+1}} \right|_{\theta_{N+1} = 0} = \quad \Braket{i\left[ H, G \right]}
                \end{equation} %
                where $H$ and $G$ are the Hamiltonian and generator of the tested excitation operator as in Eq.~\eqref{eq:excitation_operator_herm_form}, respectively. The expectation is taken over the state $\ket{\psi^{(N)}}$, which is prepared by the previous $N$ parameters in the current ansatz before being extended. %

\end{document}